\documentstyle{elsart}
\input epsfig
\begin{document}


\newcommand{\Bl}{ \left( }
\newcommand{\Br}{ \right)}

\begin{frontmatter}

\title{Dynamics of galaxies with triaxial haloes}

\author[sussex,technion]{A.A. El-Zant}
\author[sussex,hamburg]{B. Hassler}
\address[sussex]{Astronomy Centre, University of Sussex, Brighton, BN1 9QH, UK}
\address[technion]{Physics Department, Technion --- Israel Institute of Technology,
Haifa 32000, Israel}
\address[hamburg]{DESY, Hamburg, Germany}


\begin{abstract}
We study the stability of trajectories near the disk plane of
galaxy models with a triaxial dark matter halo component. We also examine
the effect of weak discreteness noise, rapidly rotating bar perturbations and
weak dissipation on these trajectories. 
The latter effect is studied  both by  adding 
a dissipative component to the force law and by using particle simulations. 
If the matter distribution is triaxial and has a constant density 
core, dissipation leads to  inflow of material inside the core 
radius to the centre (since no non-self-intersecting 
closed orbits exist in the central areas). This leads to the formation of central masses
which in turn destabilise the  trajectories of any stars  formed in these regions.
In particular, even if the gas  settles by 
dissipation into a flat disk, stars formed in that disk will later form a bulge like 
distribution the extent of which would be related to the core radius of the halo and the
original asymmetry in the plane.
The process of gas inflow is regulated by the fact that too large and condensed a central mass 
leads to the creation of stable closed loop orbits in the central area around
which the gas can move. This would appear to stop 
the accumulation of central mass {\em before}
it becomes large enough for rapid loss of halo triaxiality. 
It was found that weak discreteness noise
can increase the fraction of such trajectories significantly 
and can therefore have important consequences for the modelling of galaxies.
The addition of rapidly rotating bar perturbations also increases the degree of instability 
dramatically. So if bars can form in triaxial haloes they are likely to be quickly
destroyed leaving a bulge like structure behind (which may explain the absence of bars in surveys
of high redshift galaxies). 
In they do survive their main effect on the gas dynamics is to create attractors other than the 
centre around which the gas can move.  
We discuss some possible consequences of the aforementioned
effects. In particular, it is suggested 
that the halo core radius and initial 
asymmetry may be important in determining 
the relative disk-halo contribution to
the rotation curve of a galaxy --- and hence its Hubble type.

\end{abstract}

\end{frontmatter}

\section{Introduction}
\label{tax:wyax}

Numerical simulations of
dissipationless collapse starting from cosmological initial   
conditions consistently predict triaxial final states for cold
dark matter (CDM) halos (e.g., Dubinsky \& Carlberg 1991; Warren et al. 1992; 
Cole \& Lacey 1996). 
Considering the fact that collapsing
spherical objects are unstable towards non-spherical perturbations (Lin, Mestel
\& Shu 1965), and that violent relaxation is not completely effective in washing out
initial asymmetries (Aarseth \& Binney 1978), these results are perhaps not that surprising. 
In fact, given
the centrally concentrated  radial density distribution
of the resulting halos, 
the observed asymmetries should be considered as lower limits, 
since, for such structures, many of the trajectories will be chaotic. 
As these trajectories explore their phase space, the 
equilibrium figure will tend to relax towards more symmetric shape. This 
process  would be expected to be strongly amplified by discreteness noise
(Merritt \& Vallury 1996 and Section~\ref{tax:noise} of the present paper) which is much larger
in numerical simulations than in CDM system.

Using parameters obtained from numerical simulations (of Dubinsky \& Carlberg 1991),
Kuijken \& Tremaine (1991) estimated
that a disk equipotential axis ratio of $\sim 0.9$ can be inferred. 
Since then, there has been a number of papers suggesting that such  
a value is incompatible with observations. These include arguments 
that the scatter in the Tully-Fisher relation would be too large (Franx \& de Zeeuw 1992).  As was
pointed out  by Rix \& Zaritsky (1995) however, the fact that the Tully-Fisher
relationship is used for distance measurement might make the 
sample used  particularly biased 
against galaxies with large-scale distortions.
These authors use the shapes of the isodensity contours from
near-infrared images of external galaxies
to obtain an axis ratio of $\sim 0.95$ for the
disk isopotentials ---
this is consistent with a halo potential axis ratio $\sim 0.9$,
if the disk and halo contribute more or less equally to the potential.
Another argument for axisymmetric haloes  makes use of the velocity
dispersion of gas clouds  in the Milky Way. 
From such considerations Blitz \& Spergel (1991) deduce 
that the potential axis ratio is near 1 (axisymmetric) unless the Sun 
lies near a symmetry axis of the potential.
Kuijken \& Tremaine (1994), on the other hand, study in detail  both local
and global photometric and kinematic properties of our galaxy, and
deduce that the Sun lies near the minor axis of the Milky Way (give
or take 10 degrees), and that the data is consistent with a disk 
equipotential axis ratio of $\sim  0.9$ (meaning a halo potential axis
ratio nearer to 0.8). 
Less controversial are measurements of the
flatness of the halo. These can be obtained  from observations
of polar ring galaxies.
 They suggest a density axis ratio of about a half
(e.g., Sackett \& Sparke 1990; Sackett et al. 1994). In general, therefore,
one should expect the dark halos of galaxies to be very aspherical ---
in contrast to how they are usually modelled in simple treatments.

Whatever the shapes of haloes today, one may like to speculate
about their evolution towards the present state and their influence
on the early dynamics of disk galaxies.
The fact that the flatness of haloes appears to be larger than 
the asymmetries in the disk plane, and that some studies constrain this
asymmetry to be rather small, led some investigators to suggest that
the dissipational formation of a disk in the centre of a triaxial halo causes the
latter to evolve to more oblate nearly axisymmetric state: dissipation
might lead to the formation of central concentration, which in turn might
destroy the triaxiality (we consider this in detail later in this
paper). However this is expected to be a slow process which should
occur over many dynamical times. Indeed, triaxial figures of equilibrium
with mild central density cusps have been successfully built in a 
self consistent manner (Merritt \& Fridman 1996: see also Schwarschild 1993
for an application to the singular logarithmic potential), and numerical simulations
seem to show that there is a threshold for the ratio of the central mass
to that of the triaxial figure ($\sim$ a few percent) below which rapid loss of  triaxiality 
does not occur (Merritt \& Quinlan 1997).

Studies of dissipational galaxy formation
(e.g., Katz \& Gunn 1991) appear to contradict  this. It is  found that,
in the presence of dissipation, the resulting haloes become nearly
 axisymmetric, and that this happened in the initial formation stages.
However, modelling dissipational galaxy formation is not an exact science
since the exact role and relative importance of the various  hydrodynamic 
and thermodynamic effects is still unclear. 
Even more difficult to
predict is the effect of star formation and the accompanying energy feedback. 
These effects, 
 if underestimated,
can cause over-dissipation leading to the buildup of  central
concentrations far greater than what can be realistically expected.  
In addition, the
small number of particles used by Katz \& Gunn (about 4000),
 means that the discreteness noise 
is large. As mentioned in the opening paragraph, this can lead to artificial relaxation.  
A more controlled study was undertaken by Dubinsky (1994) who considers
the slow growth of a Kuzmin Toomre disk initially placed 
at the centre of a live halo and the motion of which is evolved 
as a particle in the simulation. His conclusion is that, in the presence 
of the disk, the halo orbits change adiabatically, in a way as to cause
 the halo to become  more oblate and less asymmetric in the disk 
plane. Nevertheless, the halo retains some of its triaxiality.

Indirect  evidence that some disk galaxies might have had a triaxial halo
at one point in their history but may have evolved towards axisymmetry
relates to how one interprets  the counter-rotating galactic disks 
that are observed
(particularly the case of  the $ {\rm S0}$ galaxy NGC 4550: Rubin et al. 1992;
Rix et al. 1992). According to Evans \& Collett (1994), these symmetric
populations result from stars which were on box orbits when the potential was
triaxial, and follow loop orbits when the potential is axisymmetric.
 The stars on the box orbits in the original triaxial potential are either
formed on these eccentric orbits or subsequently
 ``heated'' into box orbits. In fact, if the inner parts
of galaxies are non-axisymmetric and are dominated by a nearly harmonic constant
density core, it would be natural for stars confined within that core to move
on box orbits. Indeed, the high velocity dispersions found in the inner regions
of galactic disks (Lewis \& Freeman 1989) suggest that a large
fraction of stars in these regions
either once were or still are on box orbits. Unless the harmonic core is very
large,  this population is likely to be confined to the inner few kpc.
Moreover, we shall see in this paper, in the presence of a central mass, 
this group of
stars can be transformed into a bulge population. This may explain why large 
scale counter-rotating disks are not commonly observed (Kuijken et al. 1996). 

The work of Dubinsky (1994) and Evans \& Collett (1994) are examples 
of the evolutionary effects 
that disks can induce on galactic halos and vice-versa. The two effects
described above however depend on the properties of the regular orbits
in the potential: how they can be arranged in different ways.  A richer
variety of interesting phenomena  are related to changes in the qualitative
structure of orbits --- that is the transition of orbits from regular
to chaotic and vice-versa. In this case, a whole set of new phenomena
may appear. Strongly chaotic orbits (visiting most of their allowed energy subspace) 
will usually have a time averaged density in the 3 dimensional
configuration space that is much more round and isotropic than the 
underlying density distribution. The existence of a large number of these
in a certain potential therefore implies that the velocity and density fields
will evolve towards more isotropic distributions. 
An example of this effect is now well documented in the case of galactic bars. When a central mass
is present, the once regular trajectories trapped
around the $x_{1}$ closed periodic orbits (which support the asymmetric 
structure of the bar) become chaotic as the latter are destabilised --- leading
to the dissolution of the bar and the growth of bulge like structures
(e.g., Hasan et al. 1993; Pfenniger \& Friedli 1991; Friedli \& Benz 1993;
Norman et al. 1995). Similar effects are also thought to be important in the
evolution of elliptical galaxies, where it is thought that the presence of 
a central mass or a significant density cusp causes these galaxies to lose
their triaxiality (Norman et al. 1985; Gerhard \& Binney 1985; Merritt
\& Fridman 1996; Merritt \& Quinlan 1997).

We wish here to study similar effects in
models of disk galaxies with triaxial halos. In this situation, 
both  processes described above may act. 
One expects that 
the effect of chaotic behaviour will  modify the disk stellar 
distribution, while at the same time causing
the halo to become more axisymmetric.  
The   model 
characteristics that are expected to be important in the determining the extent of  
chaotic behaviour (the degree of asymmetry and central concentration)
are the same in this new context (Section~\ref{statprop}  and~\ref{dishac}), though the interpretations 
may be different. 
In addition, also interesting here are situations in which both a non-rotating triaxial 
halo and a rapidly rotating bar are present (Section~\ref{tax:rapbars}). 
We will also be examining the effect of weak 
discreteness noise (Section~\ref{tax:noise}) and weak dissipation 
(Section~\ref{tax:disp}  and~\ref{tax:disin}).
We will find that the latter property has the interesting  effect of triggering significant
inflow of material to the central regions of our models.

We will be concerned with  the stability of motion near the disk
plane of fully three dimensional models. 
By symmetry, all orbits have to
pass at least once by the disk plane. Trajectories that remain confined
near the disk plane however are usually regular.
Those exploring the full configuration space bounded
by the (approximately ellipsoidal) zero velocity  surface are strongly 
chaotic (an exception  was found in the case  of models
with rotating bars, but here one is mainly interested in the bars'
effect on orbits in the plane in any case).
Thus, testing for vertical stability near the disk plane, at the same
time gives an idea of the fraction of chaotic orbits in a given model,
while also determining the fraction of trajectories that would be vertically
unstable from an initial distribution  started near the disk plane.
These two related effects determine the plausibility of maintaining a triaxial
halo, and a flat distribution of stars in the inner hotter areas of the disk.
While a search confined to the vicinity of the disk plane is obviously not
exhaustive, it should be representative since the three dimensional
regular orbits are usually parented by those in the symmetry planes 
(e.g., Binney \& Tremaine 1987 (BT))
and strongly chaotic orbits should pass at any given point in the disk plane with a vertical
velocity close to zero (Poincar\'e recurrence theorem). 
On the other hand, the reduction in the number of phase space  parameters to
be searched, can be used in exploring  the extensive set of models which
we will now describe.

\section{The models}
\label{tax:mod}
\subsection{The halo}
\label{tax:halo}

A well known  simple  expression for a triaxial halo with an azimuthally
averaged rotation curve that flattens at large radii is  
\begin{equation}
\Phi_{H} = \frac{1}{2} v^{2}_{0} \log \Bl R^{2}_{0}+x^{2} + p y^{2} + q z^{2} \Br ,
\label{tax:log}
\end{equation}
where  $v_{0}$ is the maximum  rotation velocity  and $R_{0}$ is the
radius of a core where the potential is nearly harmonic.
The parameters $p$ and $q$ are related to the potential axis ratios by
$p=(a/b)^{2}$ and $q=(a/c)^{2}$, with $a>b>c$
for a triaxial halo.

 We will usually fix $v_{0}$ to
a value close to 200 Km/s. This is consistent with the value 
attributed to the halo contribution in the Schmidt (1985) model
 of the Galaxy (see also van der Kruit 1986). In this model, the value
for $R_{0}$ is found to be about 2.8 kpc. 
 $R_{0}$  also can be  determined by decomposing
the rotation curve into disk, halo and bulge contributions
 using the ``maximum disk'' procedure (see, e.g., van der 
Kruit 1989).
It
is still not clear whether real galaxies satisfy this assumption,
although  some authors do argue in that direction (e.g., Freeman 1993). 
In this  
decomposition method  $R_{0}$ can be very large
(Verdes-Montenegro et al. 1996 and Botteno 1996 found it 
to be 11 and 30 kpc respectively). 
By spanning a range of possible core radii however (not only those consistent 
with maximum disk), it is possible to find smaller values.
For example for  NGC 2903, Hofner \& Sparke (1994)
find  values between 2.4 to 8 kpc. 
 Numerical simulations based on CDM models on the other hand predict no core 
radius at all, but a density increasing as $1/r$ in the central
areas. 
It has been
argued however (Flores et al. 1993) 
 that  small
(or zero) values for the core radius would mean that the density distribution
is halo dominated even in the inner regions of bright  galaxies, making it
difficult to accommodate for the luminous contributions of galactic disks
(which can explain the observed rotation in the inner regions of such galaxies: e.g.,
Persic, Salucci \& Stel 1996; Tully \& Verheijen 1997. See however Navarro 1996).
In addition, dynamical arguments concerning the braking of rotating bars, 
 suggest that long lived bars must reside in nearly maximal disk
density distributions  (Debattista \& Sellwood 1996). 
Moreover, for low surface brightness galaxies, where the halo does indeed
dominate over all radii, such centrally concentrated halo densities as
those predicted by simulations would be
incompatible with the observed rotation curves (Flores \& Primack 1994).
Thus, despite the success of CDM models in reproducing other observed features
of dark haloes (like correlations between total mass and scale lengths),
observational evidence  suggests that dark haloes should
have substantial harmonic cores (Burkert 1997; Persic, Salucci \& Stel 1996).

The importance of a harmonic core lies in the fact that within it
only box orbits can exist. Dissipation therefore implies that the only stable attractor
is the centre (since there will be no stable closed loop orbits to oscillate 
against: Section~\ref{tax:dispdisk}). 
This will imply that, as a disk settles down in such a potential, there
will be a 
 building up of central mass concentration, which in turn would destabilise  both the
halo box orbits and those stars formed in the plane where the disk is settling which,
at least in the central areas, 
will
also mostly have to be on box
 orbits. On the other hand, in  the case when the core radius is very small, stable 
loop orbits exist everywhere, and gas can settle around them. 
Nevertheless,
stars born in rapidly evolving non-equilibrium situations, or subsequently heated
by discreteness effects (e.g., encounters with gas clouds)
from their nearly circular orbits,  can have eccentric 
trajectories --- and in cusped potentials these are usually chaotic. 
 The very high velocity dispersion found for old stars in the inner
disk by Lewis \& Freeman (1989) suggests that one of the above scenarios should hold.  

In our models we  have tried values between $R_{0}=2$ kpc
and $R_{0}=6$ kpc which span a  range of observationally allowable values. 
We also did some runs with $R_{0} \sim 0$ with and axisymmetric halos --- on the 
assumption that such highly cusped structures would not remain triaxial.

The concept of triaxiality can be made more formal by using the ``triaxiality parameter''
(Warren et al. 1992; Franx et al. 1991)
\begin{equation}
T=\frac{ a^{2} - b^{2}}{ a^{2}- c^{2}}.
\end{equation}
If $T \sim 0$ then the halo figure shape is nearly oblate, while when
$T \sim 1$ the halo it is essentially prolate.
 We  choose two values for the
triaxiality parameter in our models: $T=0.85$, corresponding to a strongly
 prolate shape
 close to the most probable values found by Warren et al. (1992)
for the equidensities, and $T=0.65$ --- close to the value found by Dubinsky
(1994)  for his equidensities (we assume here that the triaxiality of the
potential distribution follows that of the density, at least in the case when
the asymmetry is not too large). Although a value of $T=0.65$ still describes
a shape that is more prolate than oblate, we will refer to halos that have
$T=0.65$ as ``oblate''  to distinguish them from  
those with $T=0.85$ which will be referred to as  ``prolate''.

\subsection{The disk}
Although an exponential disk is most realistic in modelling the galactic
disk density distribution, we have decided to use 
a Miyamoto-Nagai (Miyamoto
\& Nagai 1975) disk because it enables
one to model simultaneously a bulge and disk distribution without
any additional components. Is is also  easily handled 
computationally  with
the force obtained in closed form by taking the derivatives 
of the potential given by
\begin{equation}
\Phi_{D}= -  \frac{GM_{D}} {\sqrt{ x^{2}+y^{2}+ 
\Bl a_{d}+\sqrt{b_{d}^{2}+z^{2}} \Br^{2}} },
\label{tax:disk}
\end{equation}
where $GM_{D}$ is the mass of the disk in units in which the gravitational 
constant is unity. The parameters $a_{d}$ and $b_{d}$ determine the 
central concentration
and  the flatness of the disks respectively. 
In our models the mass of
the disk was usually about $3 \times 10^{10}$
or $6 \times 10^{10}$ solar 
masses,  and the scale length $a_{d}$ was
usually taken to be  either 
1.5 kpc or 3 kpc while the scale height $b_{d}$ was usually either 0.5 kpc
or 1 kpc. 

\subsection{Models with disk and halo}
The choice of simple analytic formulae for our potentials 
ensured that the force calculations  are
not CPU-time consuming. This enabled us to explore  large parts
of the parameter spaces.
In Table~\ref{tax:tabmods}  we list the different disk and halo parameters
that were used.
These are given in units of kpc, kpc/Myr and $G=1$ --- which implies that the
unit of mass is about $ 2 \times 10^{11} M_{\odot}$. {\em Any numerical values for
the different dynamical 
variables will be assumed throughout the rest of this paper to be 
expressed in terms of the above units unless it is explicitly stated otherwise}.

\begin{table}
\begin{tabular}{l|l|l|r|r|l}
\multicolumn{6}{c}{ }\\
\hline
Model & $V_{0}$ & $R_{0}$ & $GM_{D}$ & $a_{d}$ & $b_{d}$ \\
\hline
Model 1  & 0.2 &   2 & 0.15 & 3 & 1 \\
\hline
Model 2  & 0.19 &  6 & 0.3 & 3 & 1\\ 
\hline
Model 3  & 0.205 & 6 & 0.15 & 1.5 & 0.5\\
\hline
Model 4  & 0.19 &  6 & 0.3 & 7.8  &  0.2 \\ 
\hline
Model 5  & 0.18 &  4 & 0.3 & 3 & 1\\
\hline
Model 6  & 0.1 &   4 & 0.6 & 3 & 1 
\end{tabular}
\caption{\label{tax:tabmods} Parameters of the different disk-halo models that
were studied. The first three models are discussed in detail in this paper}
\end{table}

\begin{figure}
\begin{center}
\epsfig{file=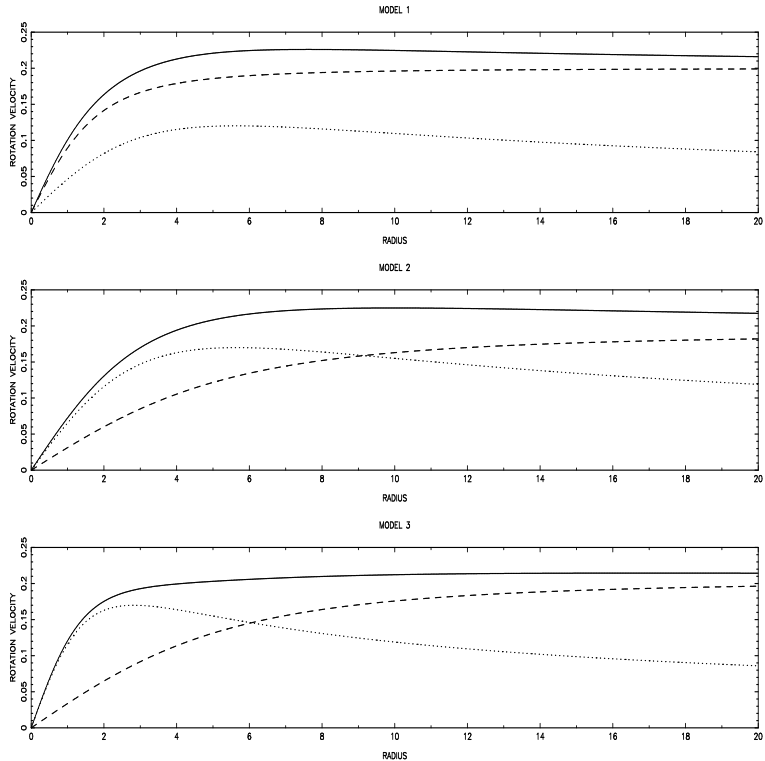,width=10.0cm,height=10.0cm,angle=-90}
\end{center}
\caption{\label{RADPLOT}
 Rotation curves of the three  disk-halo models studied in this papers.
The dotted lines correspond to disk contributions, the dashed lines to the halo 
contributions while the combined disk-halo rotation curves are shown by the solid lines}
\end{figure}

 We model a sequence of galaxies 
with different disk-halo contributions to the rotation curve: from those
for which the halo dominates everywhere (e.g. Model 1), which is likely
to be the case in low surface brightness galaxies, 
to others where this
component dominates only in the outer region (e.g., Model 2 or Model 3). 
Clearly,  in the latter case,
the disk will have to be either more centrally concentrated  or more massive.
These two situations are represented by Models 3 and 2 respectively. 
The first case is a typical near maximal disk situation, while the second case
represents a situation where both components are comparable (as suggested by Navarro 1996).
The three
first models are therefore representative of the range of objects 
of interest. Nothing qualitatively interesting arose from the examination
of the other models and so they will not be discussed any further.

The rotation curves  of the three chosen models are shown in Fig.~\protect\ref{RADPLOT}.
All model parameters were fixed so that the rotational velocity at
$20$ kpc was $\sim 0.22$ kpc/Myr ($\sim$ 220 km/s). Although according to
Persic, Salucci \& Stel (1996) different galaxy types are 
characterised by different
maximal rotational velocities, it facilitates the comparison of dynamical
properties if all models have the same rotation speed at a given radius
(and hence roughly the same total mass distribution and dynamical time-scale). 
In practice, the variation of total mass and mass to light ratios
is rather small over a wide range of Hubble types, and the scatter in galaxy properties 
is such that it is easy to find  galaxies of 
different types with the same maximal rotation speed (Roberts \& Haynes 1994).

\begin{figure}
\begin{center}

\epsfig{file=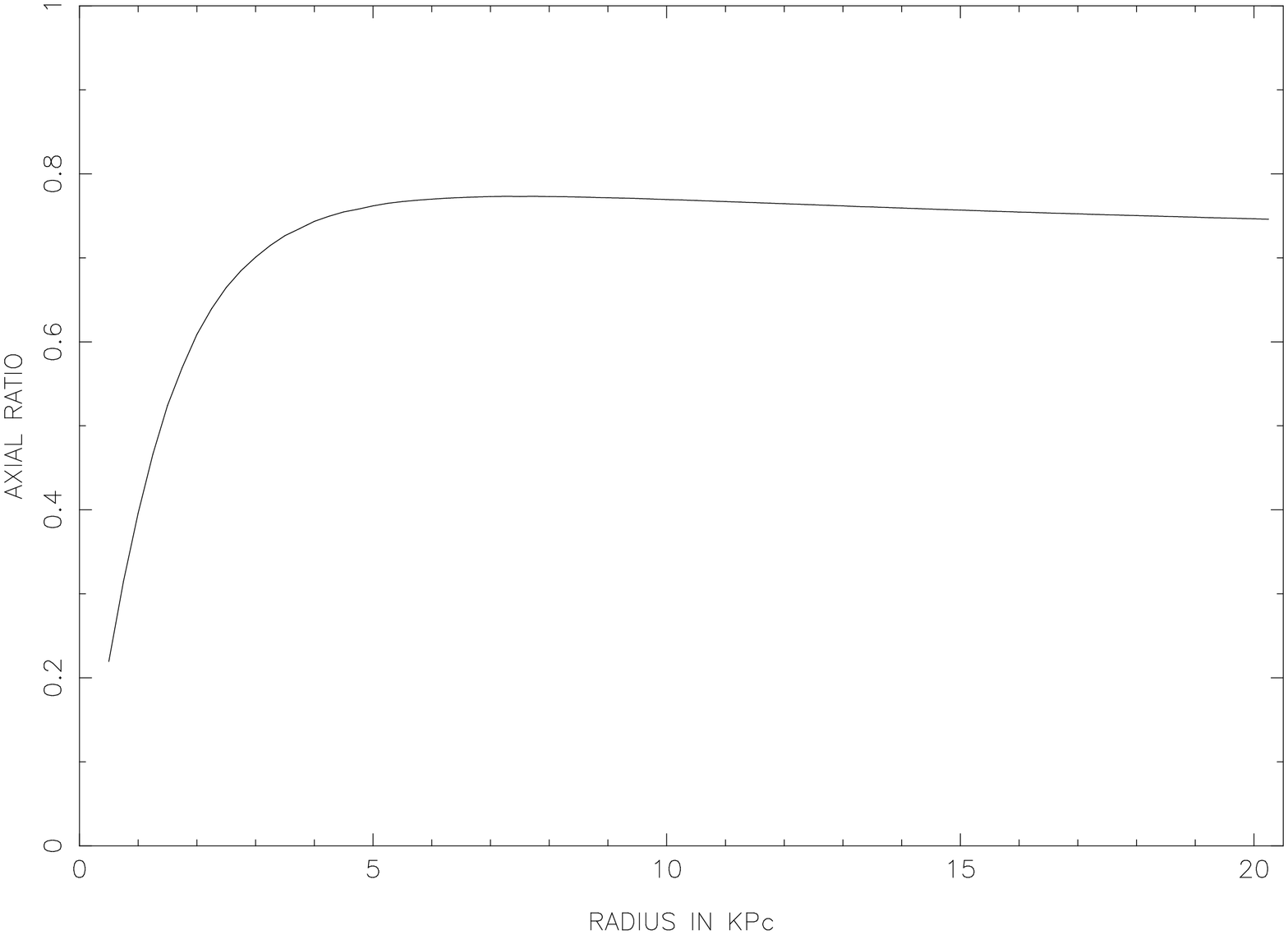,width=12.0cm,height=3.0cm,angle=-90}

\vspace{0.2in}

\epsfig{file=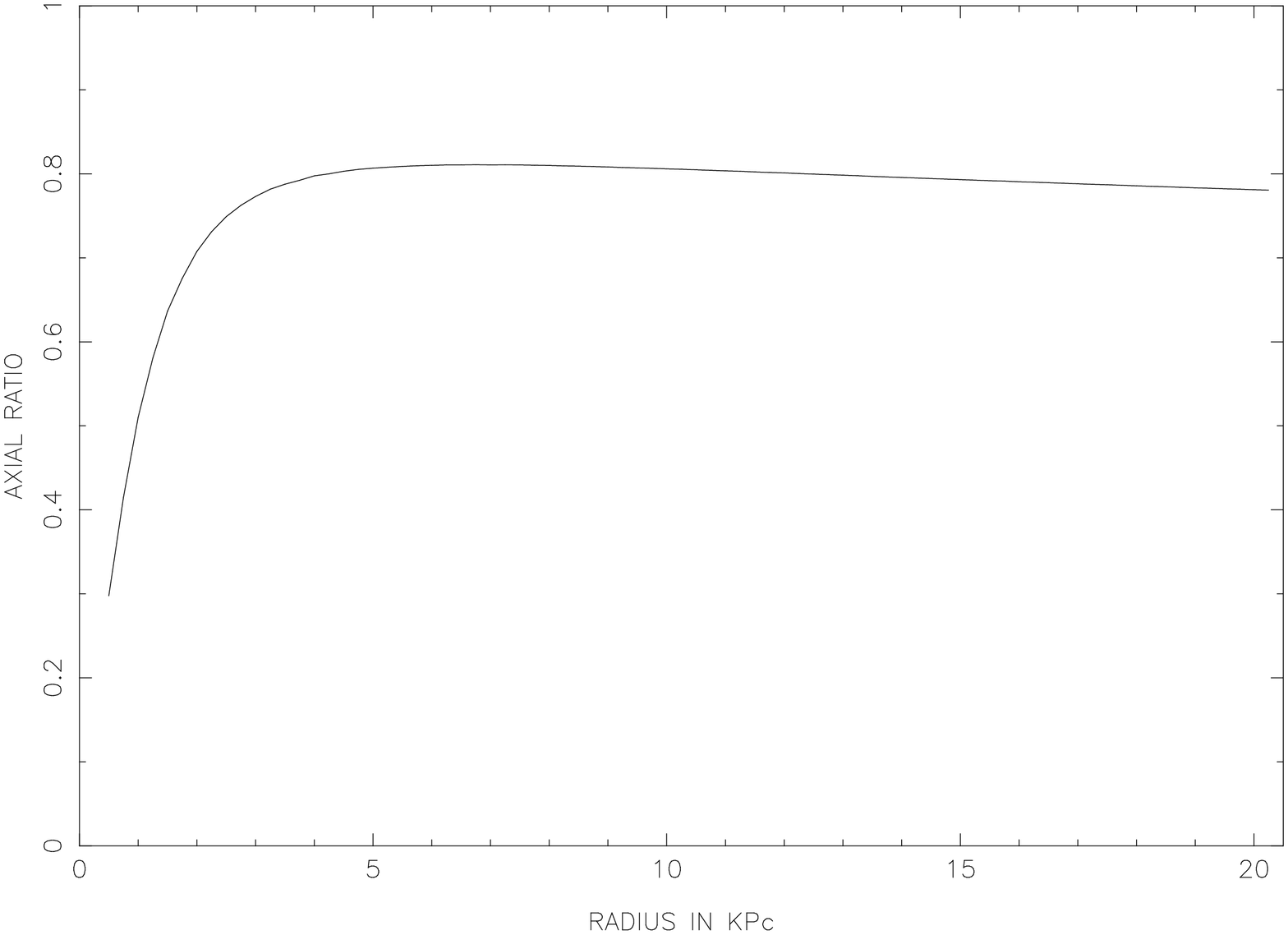,width=12.0cm,height=3.0cm,angle=-90}

\vspace{0.2in}

\epsfig{file=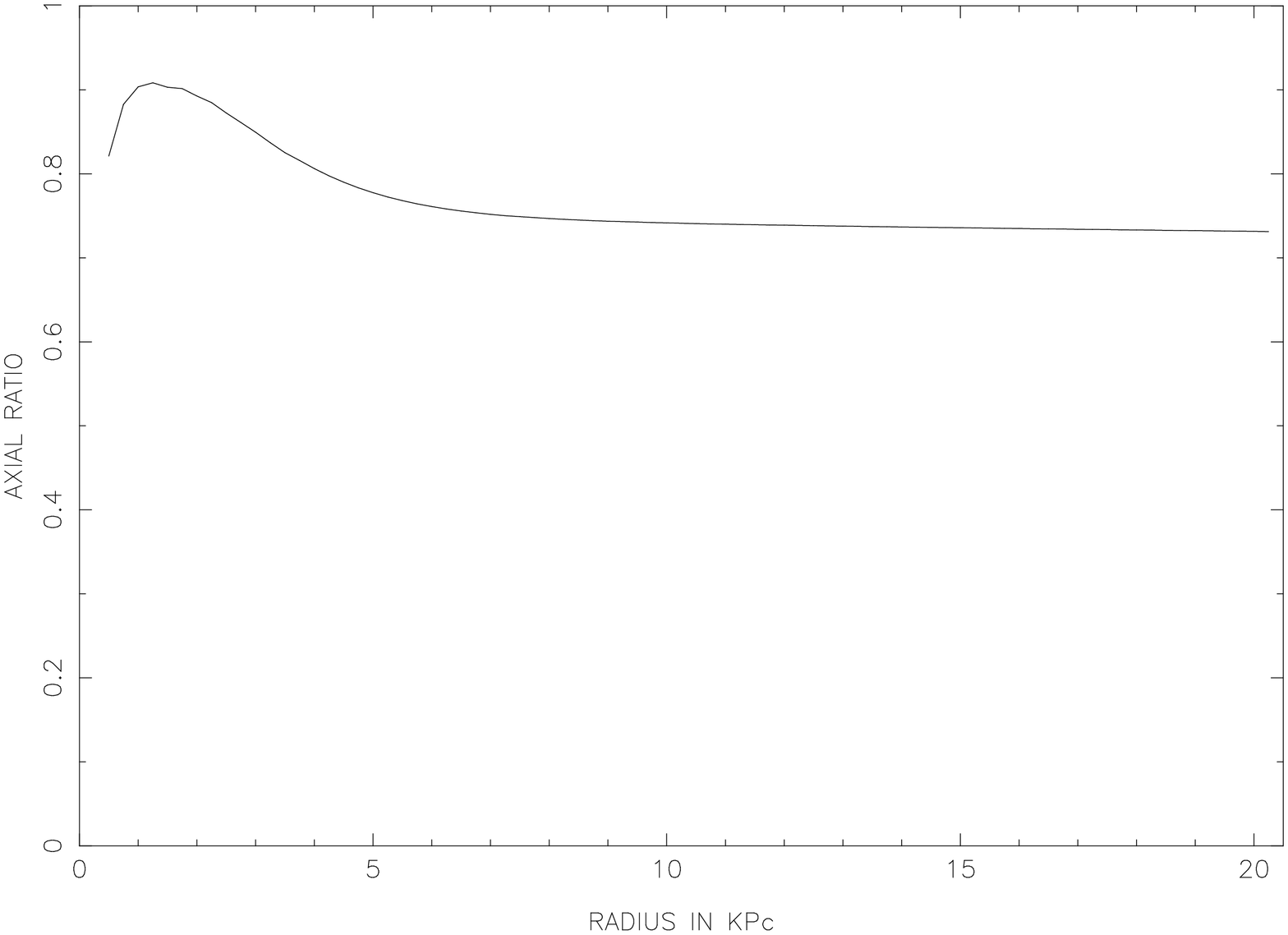,width=12.0cm,height=3.0cm,angle=-90}

\end{center}
\caption{\label{AXRDMOD1}
 Variation with radius of the axis ratios of the closed loop orbits in the disk planes of
Model 1 (top), Model 2 (centre) and Model 3 (bottom)
when the halo potential axis ratio in that plane
is $b/a=0.7$}
\end{figure}

\begin{figure}
\begin{center}

\epsfig{file=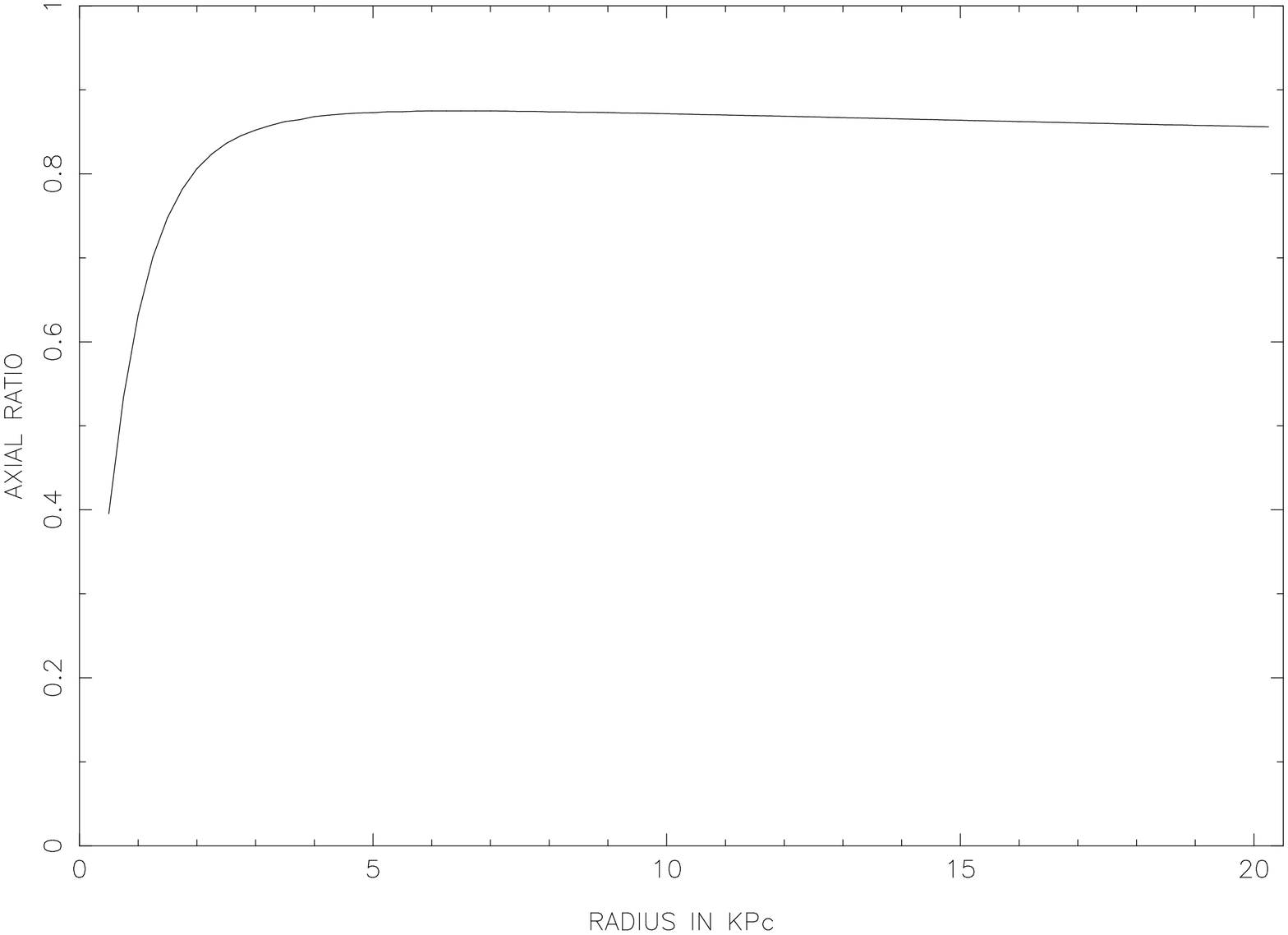,width=12.0cm,height=3.0cm,angle=-90}

\vspace{0.2in}

\epsfig{file=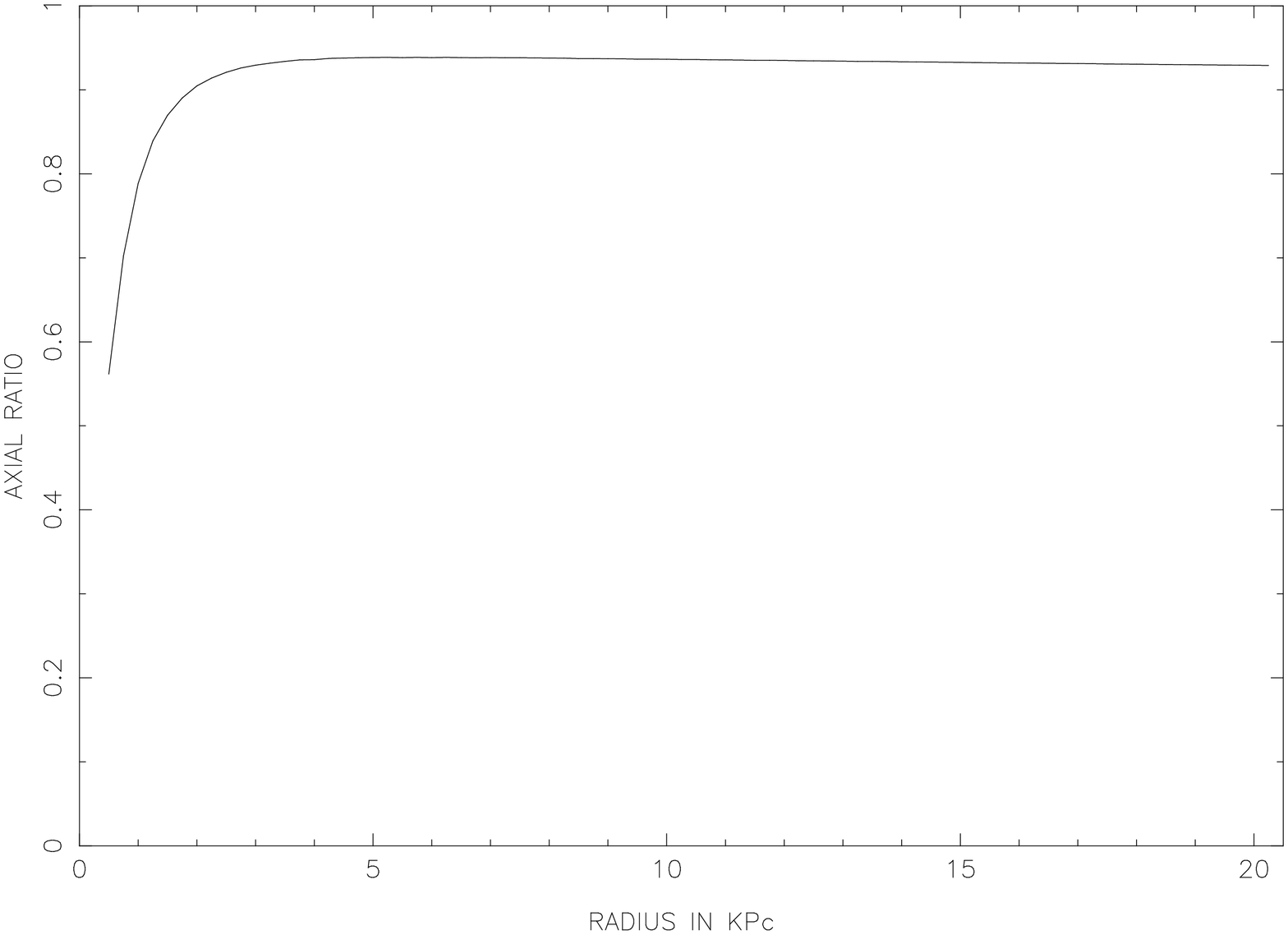,width=12.0cm,height=3.0cm,angle=-90}

\end{center}
\caption{\label{8AXRDMOD2}
 Variation with radius of the axis ratios of the closed loop orbits in the plane of
Model 2 when the potential axis ratio in that plane is $b/a=0.8$ (top)
and when it takes the value $b/a=0.9$ (bottom)}
\end{figure}

To completely fix the system parameters,
there remains now the task of determining the values taken by  
the potential axis ratio of the halo 
in the disk principal plane. This  fixes $p$ and, since  we have chosen two
definite values for the triaxiality parameter $T$ (see Section~\protect\ref{tax:halo}),
$q$ is then automatically fixed.
This was done by looking at the axis ratios of the 
closed periodic orbits, which is the main  observable measure of the total
asymmetry in the disk-halo system that is produced by a given asymmetry in the halo
(e.g., Rix 1995). 

The variations of the axis ratio of the closed 
loops with radius for the three main models with halo potential axis ratio
$b/a=0.7$
are shown in Fig.~\protect\ref{AXRDMOD1}, while the plots in
Fig.~\protect\ref{8AXRDMOD2} show the variation of this parameter as a function radius for
Model 2 with potential axis ratios $b/a=0.9,0.8$. Clearly, in general, an
axis ratio  of $b/a=0.9$ is compatible with all current observational 
constraints (Section~\protect\ref{tax:wyax}) on the eccentricity of galactic disks,
an axis ratio of $b/a=0.8$ is more or less compatible with the results of 
Kuijken \& Tremaine (1991,1994), 
while an axis ratio of $b/a=0.7$ is  more eccentric than  what is compatible 
with current observations (except perhaps for certain regions of Model 3) 
--- it is not ruled out however that such eccentricities
were present at earlier stages of galaxy evolution. 
For the prolate case ($T=0.85$)
all three values of $b/a=(0.9,0.8,0.7)$ were used, 
for the oblate ($T=0.65$) case only the values 0.9 and 0.8 were used. 
In the latter case, the value $0.7$ was not used because it corresponds 
to a density distribution that was seriously dented 
and because one expects
the oblate shapes, which form after dissipation is complete, to be less asymmetric 
(Dubinsky 1994).

Although the logarithmic halo is known to have wild dents in its density distribution
when the asymmetry is too great, the models we have chosen have a well behaved density 
distribution that is positive everywhere. The conditions for this to be the case 
as well as equidensity contours for some of the models studied here are given in the 
Appendix.

\subsection{Central mass}

This was taken to be a Plummer sphere  with
core radius $b_{p}=0.01$ kpc. The potential for this object is
\begin{equation}
\Phi_{C}=- \frac{GM_{C}}{\sqrt{b_{p}^{2}+x^{2}+y^{2}+z^{2}}}.
\end{equation}

Many values of the mass $GM_{C}$ were experimented with 
but only runs corresponding to models with no central mass, 
or with a ``weak'' central mass of $GM_{C}=0.0005$ 
(about $10^{8} M_{\odot}$  
which amounts to $\sim 0.05 \%$ the total mass at 20 kpc),
or a ``strong'' central mass of $GM_{C}=0.01$ (which is about $2 \times 10^{9} M_{\odot}$, 
  $\sim 1\%$ the total mass at 20 kpc)
 are discussed in detail since they convey the essential results.

\subsection{Naming disk-halo-central mass systems}
\label{tax:naming}

Although we have tried to keep this to a minimum, the following nomenclature
was sometimes used in referring to a certain run. If the model used was Model 1
for example we refer to it as M1, this is followed by either the letter P
(for prolate models) or O (for oblate ones) and then by one of the following
numbers 1 (for spherical halo) or by 9 , 8 or 7 for halo potential axis ratio 
$b/a=0.9,0.8,0.7$ . If no central mass was present, this is followed by C1,
if $GM_{C}=0.0005$ it is followed by C2 while if $GM_{C}=0.01$ this is followed 
by C3. Thus a complete description of a run  is referred to by a series of symbols 
like M1P7C2 (which in this case describes a system with the general 
characteristics of Model 1
and when the halo is prolate, $b/a=0.7$ and $GM_{C}=0.0005$). 
If a disk is not present in the given system then -D is added (e.g.,
M1--DP7C2).

\subsection{Bars}
\label{tax:bars}
In some  runs a  rapidly rotating
barred perturbation was added in the disk plane. Two types of bar models were 
considered: a Ferrers ellipsoid of order two (BT; Pfenniger 1984a)
and a homogeneous rectangular box (Macmillan 1958).
 The density of the Ferrers bar is constant along  contours given by
\begin{equation}
m^{2}=x^{2}/a_{b}^{2}+y^{2}/b_{b}^{2}+z^{2}/c_{b}^{2},
\end{equation}
where $a_{b}>b_{b}>c_{b}$ are the semi-axes of the density distribution.
Inside the mass distribution ($m<1$) the density varies as
\begin{equation}
\rho=\rho_{c} (1-m^{2})^{2},
\end{equation}
where the central density $\rho_{c}$ is determined by the total mass of the bar
$GM_{B}$.
The density is zero for  $m \geq 1$ .

The semi-axes are taken as $a=6$, $b=1.5$ and $c=0.5$ (as in Pfenniger 1984a). 
The rotation speed of the bars was always chosen so that
the bar major axis  coincided with the distance from
the center to the Lagrangian point ${\rm L_{1}}$
 near corotation (calculated while assuming an  axisymmetric halo
 and not including the bar's
own contribution to the potential). 
This corresponded to an angular speed of
about 0.0372, 0.0361, 0.0343 in Models 1,2 and 3 respectively.  
Unless otherwise stated, the bar mass was taken 
to be $GM_{B}= 0.02$ ($\sim 4 \times 10^{9}$), which amounts to   
 $13 \%$ of the disk's total mass in Models 1 and 3 and
 $6.5 \%$ of that mass in Model 2.

\section{Stability of periodic orbits}

\subsection{Order and chaos in triaxial galaxy models and the 
role of periodic solutions}
\label{tax:staper}

It is well known (e.g., LL; BT) that the stability properties of periodic orbits
are of central importance in determining the qualitative phase space
structure of Hamiltonian systems.
For regular trajectories are parented by stable periodic orbits while
at least some periodic orbits have to be unstable for a system to 
support chaotic trajectories (i.e., be non-integrable).
If periodic solutions form a dense set of 
 unstable trajectories, then a system can be said to be uniformly
hyperbolic and {\em almost all} trajectories are chaotic. Separable systems 
on the other hand have a finite number of stable and unstable closed orbits    
with the remainder being marginally stable. 
General systems will lie on the transition between these two extremes
(for an interesting original perspective on  the nature of this transition see  Aubry 1985).

It is intuitively clear from the above remarks that as the density of periodic 
solution increases (for a given period length), it is less likely that we have the
remarkable situation of a separable system where all but a finite number of 
closed solutions are marginally stable. Indeed, it can be shown (e.g., Gutzwiller 1990) 
that the number of periodic solutions in a chaotic system increases much faster 
(exponentially) with the period than that in a regular system.
The density of periodic orbits will depend on the nonlinearity of  a
given system. For example, in a linear system the frequencies of the independent
oscillations do not depend on the amplitudes. In this case therefore, the ratios
of the oscillation frequencies in the different degrees of freedom  are constant and
are likely to be irrational. Therefore, without getting into the details, one can guess that
 the more the oscillation frequencies
vary with amplitude, the denser is the set of resonant solutions  and the more likely
the system is to be chaotic. In particular, inside the nearly harmonic cores of systems 
studied here, we expect most orbits to be regular and parented by the axial orbits 
(e.g., BT). Instability in this ``box'' orbit family will appear 
only when non-linearity
starts becoming unimportant and initially around the lower order resonances (in particular
we will see that the 2:1 resonance is highly effective). 
The addition of a central mass causes the oscillation frequencies along the 
principal axes to vary strongly with amplitude. This creates  a dense set of unstable
periodic orbits (represented by gaps along an axial orbit) and the axial orbit
is destabilised almost everywhere. When a sufficiently strong  central mass is added, 
the boxes, which are parented by the axial 
orbits, are therefore destroyed and replaced by other families of orbits.

Even in  a situation where the 
frequencies of oscillations in the different 
degrees of freedom vary (may be even strongly) with amplitude it is
possible that their ratios remain
constant. For example, in power law potentials, the ratio of the frequency of circular
motion to that of a perturbation normal to it (in the plane and in the vertical direction)
is constant. In such a situation one again expects the trajectories parented by the stable
closed nearly circular orbits to be regular quasiperiodic. In potentials with a harmonic core 
and which are superpositions of two power law terms (as is the case with our disk-halo 
systems) the scale free nature of the potential is destroyed and the frequency ratios
are only nearly constant. However these will vary slowly 
enough for the above statement to be
essentially true --- especially far outside the core (precisely 
where motion on loop orbits is predominant). 
Even more generally, the frequency ratios of nearly circular orbits
 do not vary much with radius for any potential in which the 
density decreases outwards (e.g., BT)  and therefore this situation is likely to
be representative of  most nonrotating triaxial potentials. 
The only way one breaks this symmetry
is by adding a {\em rotating} perturbation --- in which case it is possible to have 
a concentration of resonances and instability in the nearly circular $x_{1}$ orbit
family --- an effect which is again enhanced by the presence of a central mass concentration
(see Pfenniger \& Norman 1990, especially the appendix). 

Finally,  for a  system  to support chaotic orbits, it cannot have too much symmetry.
This can be either trivial symmetry as in axisymmetry leading to angular momentum
conservation  or more subtle symmetries in special integrable
potentials. For a generic potential that contains no special symmetries  one will
expect that the more asymmetric  the potential
(as measured by the flatness and the triaxiality parameter), the further it will
be from being integrable. In the case when the potential is inherently time dependent, 
 time symmetry may also be broken and the Hamiltonian may not be 
conserved. In general it will be the combination of the factors mentioned above that will
determine how widespread the chaotic behaviour is in any one of our models.

\begin{figure}





\caption{\label{MAP1}
 Poincar\'e maps of the ensemble of orbits 
in the disk plane of Model 1 when $b/a=0.8$ .  Initial
conditions are given by Equations~(\protect\ref{tax:inihx}) to~(\protect\ref{tax:inihz})  
(except that in the disk plane $z=0$ for all $i$ and $j$). 
Three different values for 
the central mass are chosen: $GM_{C}=0.0$ (top), $GM_{C}=0.0005$ (centre) and
$GM_{C}=0.01$ (bottom). THIS FIGURE WAS TOO LARGE TO BE INCLUDED} 
\end{figure}

We illustrate  these ideas with $y - \dot{y}$
Poincar\'e maps of trajectories in the $x-y$ plane of  Model 1.
The initial conditions chosen are those described in Section~\protect\ref{dishac}.
These span a whole range of energies. 
Since Poincar\'e maps however are most effective when taken at definite energies
(in that case, for example, the projections of different KAM tori do not
overlap), we group the various orbits into six different energy levels. 
We then displace the $y$ coordinates of the orbits according to the  energy
interval they lie in: orbits in the first (lowest energy) interval do not have their coordinates
displaced while orbits in the subsequent groups have their coordinates displaced 
by -5 kpc, -11 kpc, -20 kpc, -34 kpc or -50 kpc respectively.
We vary  the central mass while keeping the axis ratio
of the halo constant with a value $b/a=0.8$. The integrations are done for  50 000 Myr.

The results for central masses of $GM_{C}=0$, $GM_{C}=0.0005$ and $GM_{C}=0.01$
are shown in Fig.~\protect\ref{MAP1}. When no central mass is present,
most orbits belong to either the box  or the loop orbit families.
In the first energy level (nearest to the centre)
all orbits are box orbits while  at the second, third, and fourth energy levels 
the loop orbits start 
appearing in progressively larger fractions. The last two energy levels,
 contain a few higher order islands corresponding to stable ``boxlet'' and ``looplet''
 orbits (see ,e.g., Schwarzschild 1993 and the references therein) and the first 
appearance of weakly chaotic trajectories.

A central mass of 0.0005 leads to the destruction of all box orbits. Most resulting
chaotic orbits are ``strongly stochastic''  --- not venturing only in the areas where the 
numerous higher order islands are present. Increasing the central mass further to 
0.01 destabilizes many of these islands however and the chaotic orbits in these
regions are free to roam through a still larger area of the phase space. This is
especially so in the inner regions where the nonlinearity of the potential is strongest.
 In the outer regions some regular trajectories are found. These 
are parented either by higher order
periodic orbits or by the closed loops (which
remain regular since there is no rotating perturbation).

\begin{figure}
\hspace{0.2cm}




\hspace{0.2cm}
\caption{\label{MAP1D}
First two energy levels of Poincar\'e maps which are
similar  in all respects to those shown in Fig.~\protect\ref{MAP1} except that
the following values for the central mass are adopted. $GM_{C}=0$ (top left),
$GM_{C}=1 \times 10^{-6}$ (top right), $GM_{C}= 1 \times 10^{-5}$ (centre left),
$GM_{C}= 5 \times 10^{-5}$ (centre right), $GM_{C}=0.0001$ (bottom left) and
$GM_{C}=0.0005$ (bottom right). THIS FIGURE WAS TOO LARGE TO BE INCLUDED} 
\end{figure}

A more detailed view is provided in  Fig.~\protect\ref{MAP1D}
where we have reproduced the Poincar\'e maps of the first two energy levels for a
series of central mass concentrations of values up to $GM_{C}=0.0005$. 
It  is clear that a central mass $GM_{C}=1 \times 10^{-5}$
(which is about $10^{-5}$ the mass of the system at 20 kpc)
 is already sufficient 
to destroy all box orbits in the central regions (although in that case most are
still replaced by regular boxlets). Although by $GM_{C}=0.0005$ most trajectories in
the inner region form part of a connected `stochastic sea', significant remaining 
structure is still evident. Over shorter time-scales this structure is much more evident
but its destruction is often speeded up by additional perturbations. It is therefore
important in determining a system's response to these. We will discuss this
 in  a little more detail in Section~\protect\ref{tax:disp}.

\subsection{Statistical stability properties of periodic orbits in disk halo models}
\label{statprop} 
\subsubsection{Method}

A periodic orbit is an orbit whose trajectory returns to exactly the same point in phase space
and appears as a set of isolated  points on a Poincar\'e map.
 Therefore, if one chooses an arbitrary point ${\bf x}$ 
on a periodic orbit, under the mapping transformation ${\bf A}$ one has
\begin{equation}
{\bf A}^{k} ({\bf x}) = {\bf x},
\label{tax:per}
\end{equation}
where $k$, the winding number, is an integer depending on  the ratio of oscillation
frequencies in the different  degrees of freedom,
 and the plane
chosen for the Poincar\'e mapping. An orbit  with a given value of $k$ will return
to its original location on the map after $k$ iterations. It is the low $k$ orbits
which determine the basic phase space characteristics (e.g., LL).

We have made extensive searches for closed periodic orbits in the disk plane
up to order four in $k$. 
For each orbit we checked the stability  to perturbations in and
normal to the disk plane.
We did not follow in detail the main orbital families and their bifurcations.
In a far from integrable system this is a fairly intricate task 
due to the large number of orbital families and accompanying bifurcations.
It also 
misses  ``irregular'' orbits (Barbanis \&
Contopoulos 1995) which do not bifurcate from other families. The above procedure
is especially complex when one has to deal with several bifurcation parameters 
--- the flatness, the triaxiality, the energy and the central mass.
In such a situation, it might seem that the best chance of obtaining
a representative sample of the types of closed orbits populating phase space,
is by undertaking  a systematic search of that space. This
search was made possible by the use of the globally convergent 
search technique (see below) for solving Eq.~\protect\ref{tax:per}. 
Whereas in the standard method one finds the
main families at one energy by taking a Poincar\'e map
and then simply following their characteristic curves using local search routines,
using a globally convergent routine means that one can search the whole phase space. 
This method  is of course more time consuming
and the grid of trial initial conditions is necessarily coarse. 
In addition, the routine may fail for some initial
conditions. Nevertheless, due to the 
advantages stated above, we have found it preferable.

The initial conditions for the searches were chosen as follows. Five values of the
energy were chosen. These corresponded to the values of the potential energies 
on the $x$-axis at 1 and
21 kpc and three intermediate values sampled at regular intervals. 
 The initial positions
for the searches  --- which are taken along the $x$-axis  --- are chosen to be between
1  and 15 kpc (inclusive) and at intervals of 2 kpc.  The initial $x$ velocities
are then varied at intervals of 0.06 kpc/Myr.
 One stops incrementing the $x$ velocities when the corresponding 
$y$ velocities, which are obtained from
\begin{equation}
\dot{y}=\sqrt{ 2 ( E - V) - (\dot{x})^{2}},
\label{tax:codi}
\end{equation}
become imaginary.

The number of iterations required by the  routine to solve  Eq.~\protect\ref{tax:per}
of $n$ variables grows  as $n^{2}$. It is therefore important to minimize the number
of variables involved. On the Poincar\'e section on which the periodic solutions are
searched there are three variables. It is possible to explicitly
eliminate one more variable 
by using  condition~(\protect\ref{tax:codi}). However, during its search far from a root, the
routine is likely to venture into regions of the phase space where this relation
implies complex $\dot{y}$ causing frequent failure. An alternative strategy
(implicitly using the conservation of energy) was therefore adopted. 
At each iteration, the $x$ coordinate was
preset to a given value (the one from which the search originally started) and
the search was conducted over the two velocity coordinates which were left to
vary until the final values were identical to the original preset values
 (within a relative error of $\sim 10^{-10}$)
for two successive iterations. Because energy is conserved during the integration 
and  $y=0$  on the map, the $x$ coordinate is then automatically determined
to within a sign. This is fixed by requiring the mapping to contain the condition of 
$x \geq 0$ in addition to the standard condition $\dot{y} \geq 0$ (the symmetry of
the potential means that this constraint causes no loss of generality).
This procedure of course means that the final energy of the periodic orbit is
different from that of the initial search energy. It is usual however that this
procedure finds the roots nearest to the search energy, so that a more or less 
uniform distribution of energies emerges. This however was not always found to be the case
and some orbits were found that extended to very large radii.

Periodic orbits are located by finding zeros of the function
\begin{equation}
{\bf F(x)} = {\bf A}^{k}({\bf x}) - {\bf x}.
\end{equation}
For this purpose, we have used the {\em globally} convergent
hybrid method of Powell (1968) as implemented in the NAG routine C05NBF. One starts
from an initial guess to the solution which need not necessarily be close to
an actual periodic orbit as in the Newton-Raphson method (e.g., Parker \& Chua 1989).
This latter procedure is only applied after a scaled gradient minimizing technique 
has identified such a neighbourhood from which it is not
likely to fail. It then converges quadratically in this neighbourhood (i.e., at each
iteration, the number of digits to which the solution is accurate is doubled).

\begin{figure}
\epsfig{file=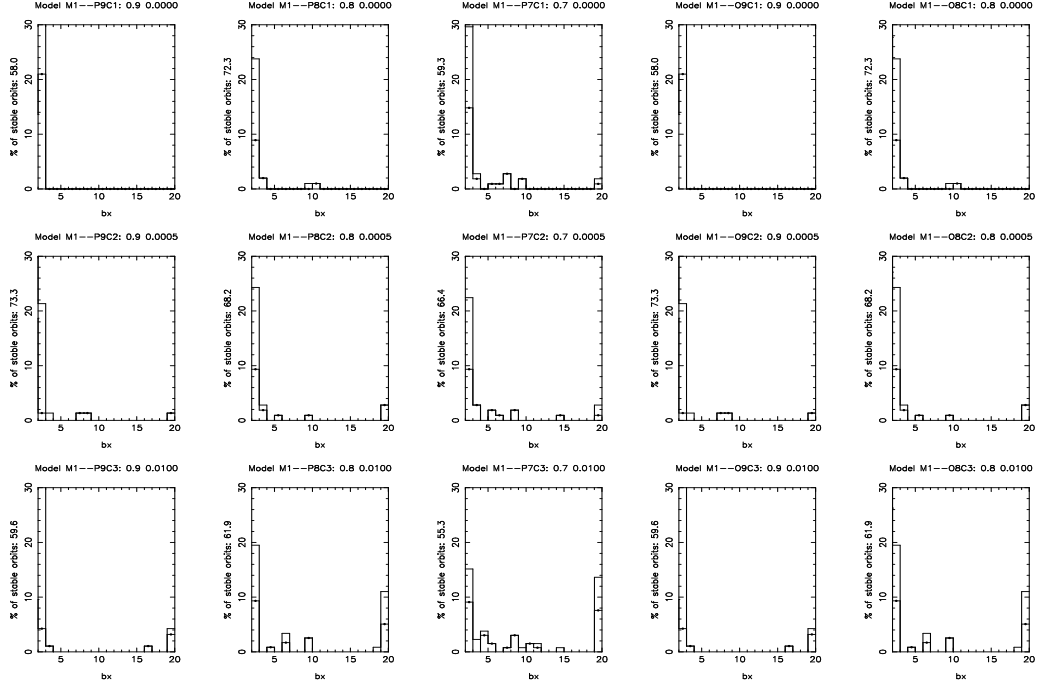,width=13.6cm,height=9.0cm,angle=-90}
\caption{\label{TM1--bx}
Distribution of horizontal stability parameter $b_{x}$ for the periodic orbits of
Model 1 of winding number up to 4. The different runs are labeled according to the
convention described in Section~\protect\ref{tax:naming}. Orbits with $\parallel b_{x} \parallel \leq 2.00005$
are labeled stable and the fraction of such orbits is 
indicated on each diagram}
\end{figure}

The linear stability of periodic orbits
 is determined by the derivatives of the map~(\protect\ref{tax:per}).
In a Poincar\'e section, at constant energy, two coordinates are eliminated (which
in our case happen to be $y$ and $\dot{y}$).
For orbits in the symmetry plane of a galactic potential,
 the variational equations are decoupled in $x-\dot{x}$ and $z-\dot{z}$ 
independent components  (since in that case mixed derivatives of the potential
vanish: H\'enon 1973). 
The stability of periodic orbits near the disk plane can then be expressed in
terms of the stability parameters $b_{x}= \partial A_{x} /\partial x +\partial 
A_{\dot{x}}/\partial \dot{x}$ and $b_{z}= \partial A_{z} /\partial z +\partial 
A_{\dot{z}}/\partial \dot{z}$. These
were approximated  using 
\begin{equation}
\frac{\partial A_{i}}
{\partial x_{j}}= \frac{ A_{i} (x_{1},..., x_{j} + \Delta x_{j},...)
- A_{i} (x_{1},...,x_{j} - \Delta x_{j},...)}{2 \Delta x_{j}}.
\label{tax:derv}
\end{equation} 
For each derivative, several values of the increment were tried (so as 
to ensure that the result  did not sensitively depend on 
the increment, for as is well known, taking numerical derivatives is
a numerically unstable operation which has to be performed with care)
and an error estimate was derived. The increments were  chosen so as to 
ensure that the relative error was less than $10^{-4}$.

In each of the models many hundreds of orbits were recovered but most were found to
duplicate each other. After removing the duplicates one hundred or so orbits were
found  in each model. For these orbits, we bin  the horizontal and 
vertical stability indices  depending on their absolute values. The first bin
contains all stable (and very close to stable) values 
$\parallel b_{x} \parallel , \parallel b_{z} \parallel \leq 2.0005$. The subsequent
bins cover the interval from $\parallel b_{x} \parallel , \parallel b_{z} \parallel
= 2.0005$ to $\parallel b_{x} \parallel, \parallel b_{z} \parallel = 20.0$ in intervals
of one (except for the second bin which covers the slightly smaller interval from 
values of 2.0005 to 3). The number of orbits is then normalized to a hundred.

\subsubsection{Results}

\begin{figure}
\epsfig{file=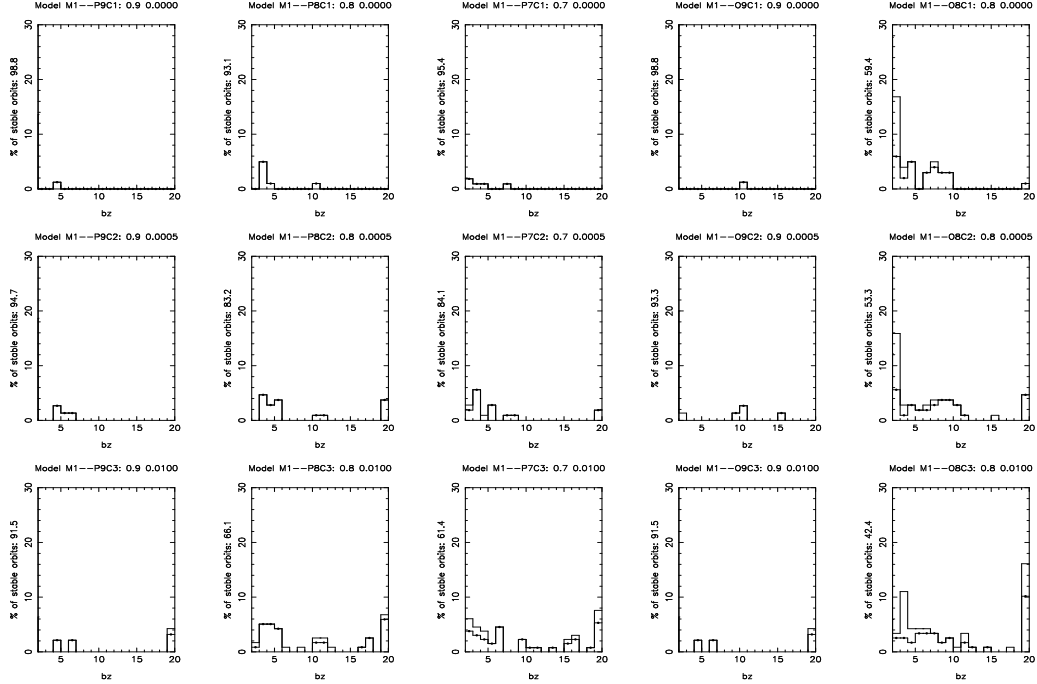,width=13.6cm,height=9.0cm,angle=-90}

\caption{\label{TM1--bz}
Distribution of vertical stability parameter $b_{z}$ for the periodic orbits of
Model 1 of winding number up to 4. The different runs are labeled according to the
convention described in Section~\protect\ref{tax:naming}. Orbits with $\parallel b_{z} 
\parallel \leq 2.00005$
are labelled stable and the fraction of such orbits is indicated on each diagram}
\end{figure}

The resulting  histograms for values of  $b_{x}$ in Model 1 are shown in
Fig.~\protect\ref{TM1--bz}. As would be expected, since one is testing stability 
in the plane, the results are unaffected by the  flatness of the halo and
depend only on the asymmetry in the $x-y$ plane and on the existence and 
strength of the central mass. Also as expected from    the discussion 
above (Section~\protect\ref{tax:staper}), 
the fraction of unstable orbits increases with both of these 
quantities. For as can be seen from this figure, the periodic orbits found
for Model M1--P9C1 for example are all stable, or very close to being so (i.e,
close to marginal stability). When a central mass $GM_{C}=0.0005$ is present, 
the fraction of stable periodic orbits actually increases. This is due to many
previously marginally (un)stable orbits becoming stable. A corresponding number
of orbits become highly unstable ($\parallel b_{x} \parallel > 3$). Increasing
the central mass to $GM_{C}=0.01$ increases the number of unstable orbits
--- both the ones near marginal stability and ones with largest absolute values
of $b_{x}$. This happens at the expense of the fraction of stable orbits which
decreases through second order bifurcations. 

The above effects are more pronounced for models with smaller values of
$a/b$.  However,  there is always  a significant fraction 
of stable periodic orbits. These include the closed loops and the 
boxlets  which do not pass near the centre. It is interesting to
note however that many of the most unstable orbits in some of the runs (e.g., 
M1--P8C3 and M1--P7C3) are tubelets which do not pass by the centre.
This last result suggests that although  one may
think of the reason orbits in centrally concentrated potentials
are unstable is that they are ``scattered'' by the central mass,
it is more appropriate to see the central mass as introducing strong non-linearity
which, in potentials without special symmetry, creates unstable closed 
orbits.  General orbits  are then repelled in {\em phase space} by
the unstable periodic orbits destabilised by the non-linearity. The reason 
that loop orbits remain regular is that (for reasons outlined in Section~\protect\ref{tax:staper})
the closed loops remain stable in the
absence of rapid rotation. In a rapidly rotating  centrally concentrated potential,
the closed loop (or $x_{1}$)  orbits are also destabilised --- and
general loop orbits can be chaotic even though they do 
not pass close to the centre.

While the horizontal stability of periodic orbits depends only on the 
potential axis ratio in the plane, their vertical stability
crucially depend on the flatness of the halo. In general, more and 
more orbits become unstable with increasing flatness. 
For example, it 
can be seen from Fig.~\protect\ref{TM1--bz} that the fraction of unstable orbits in
the oblate models is significantly larger than the corresponding
prolate ones with the same axis ratio in the plane. 
It can be also be seen that the oblate models with 
$b/a=0.8$ also usually have a larger fraction  of unstable orbits
than the prolate models with $b/a=0.7$. 

\begin{figure}
\epsfig{file=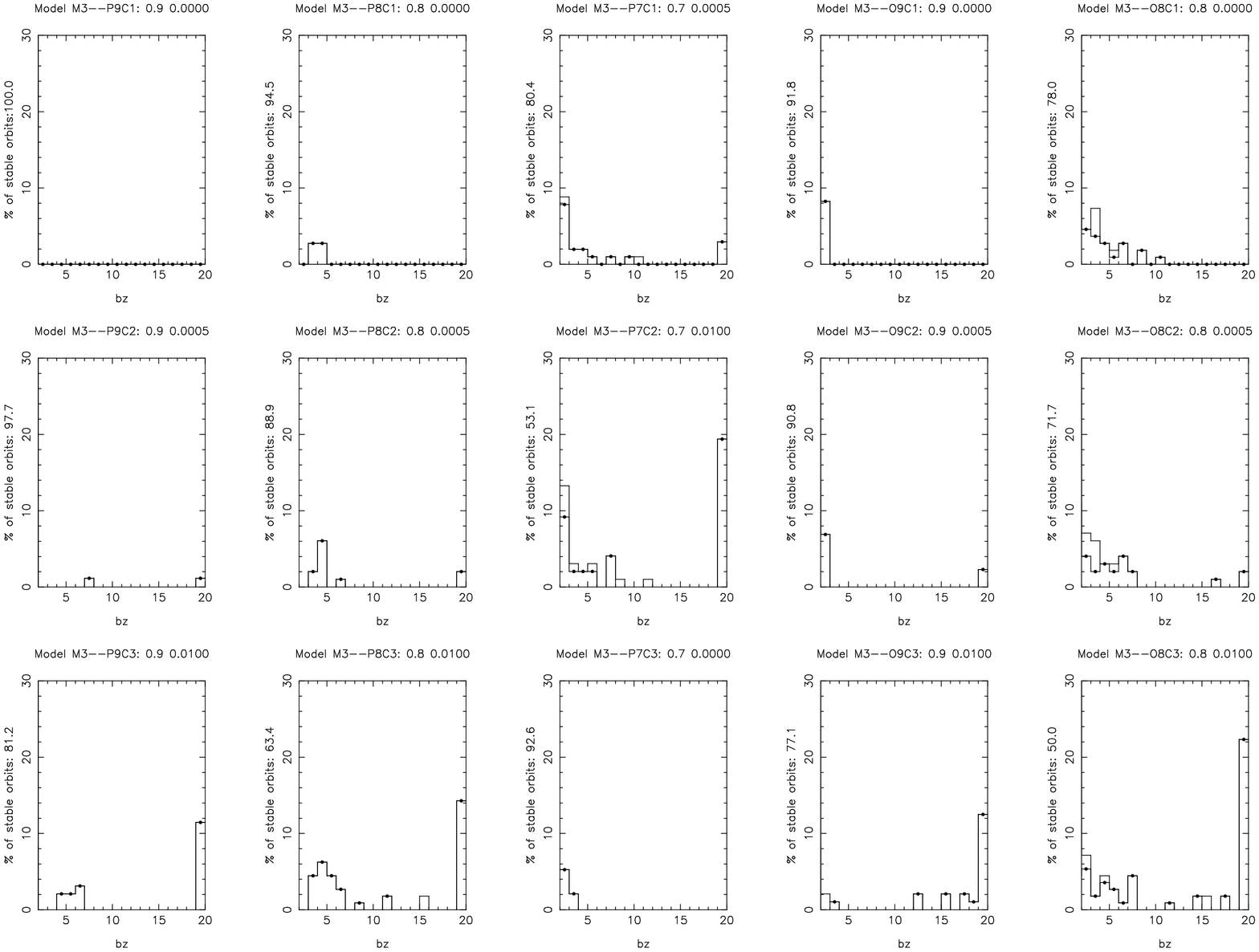,width=13.6cm,height=9.0cm,angle=-90}

\vspace{1cm}

\epsfig{file=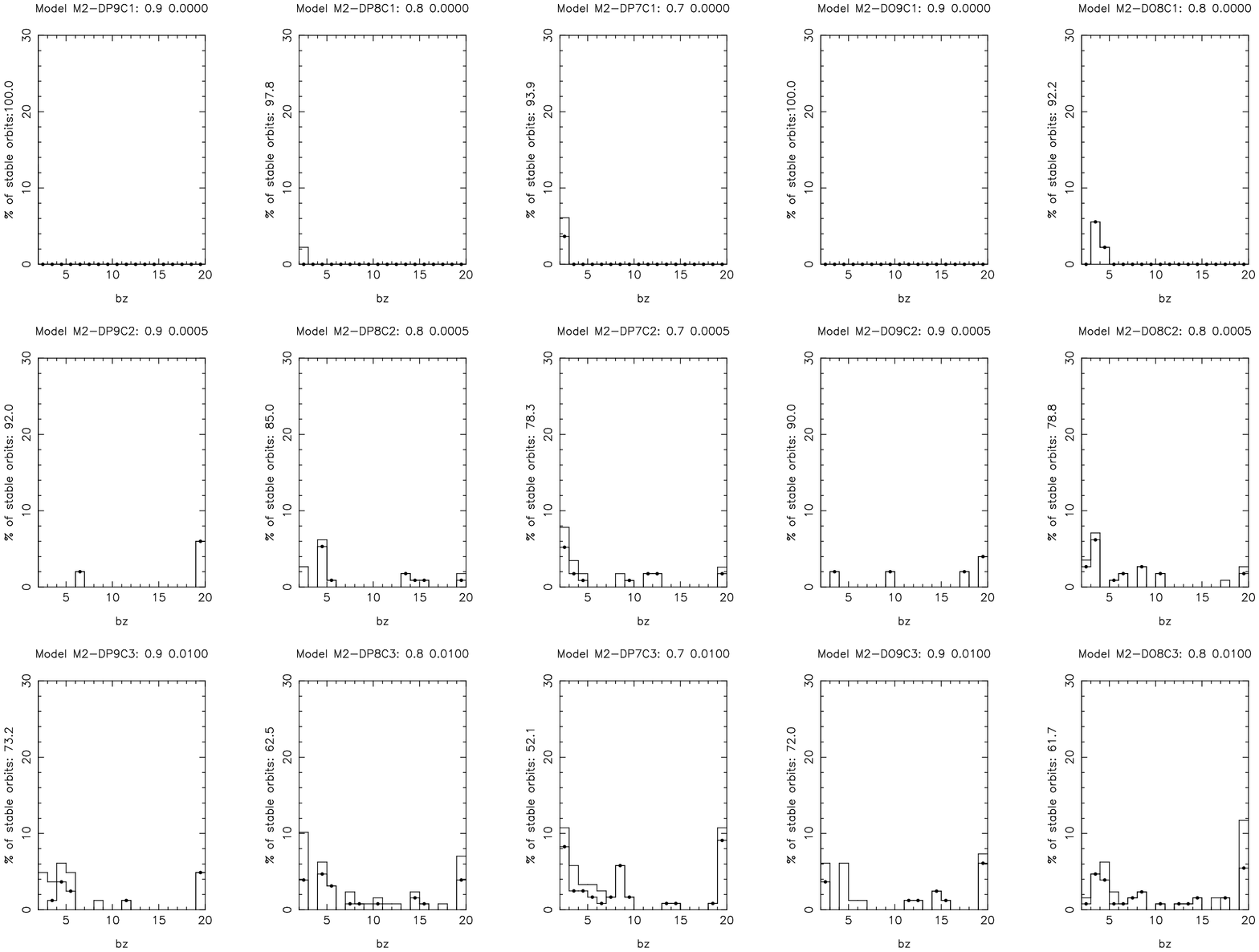,width=13.6cm,height=9.0cm,angle=-90}
\caption{\label{TM3--bz}
 Distribution of the vertical stability parameter for Model 3 (top) and
Model 2 (bottom) but with $GM_{D}=0$ (which is therefore roughly equivalent to Model 3
without a disk). Notation is the same as in Fig.~\protect\ref{TM1--bz}}
\end{figure}

The fact that the flatness of the halo is such an important factor 
in determining the stability properties of the periodic orbits suggests
that the existence of disks in the centre of dark halos may also have
a similar effect. This was indeed found to be the case and 
was more pronounced in  Model 3 which contains
a flatter disk. 
In the absence
of a disk, the halo corresponding to that model 
has a significantly smaller fraction  of $z$ unstable orbits
than when the disk is added (Fig.~\protect\ref{TM3--bz}). 
This is in contrast to the case of
horizontal stability where actually the reverse is true (since the
presence of an axisymmetric disk decreases the asymmetry of the potential
in the disk plane).

\vspace{0.3in}
\begin{table}
\begin{center}
\begin{tabular}{c|c|c|c|c|c|c|c}
\multicolumn{8}{c}{ }\\
\hline
$GM_{C}$        &  1:1  &  1:2 & 2:2 &  2:3 &  3:3 & 3:4 & 4:4  \\ 
\hline
0.0000          &  0/8  &  3/6 & 19/31 & 21/23 &  3/11 & 6/18  &  1/4  \\
\hline
0.0005          &  0/8  &  10/13 & 21/31 &  22/28  &  5/13     &  2/10  &  4/4\\      
\hline
0.0100          & 0/8   &  14/18 & 29/40 &  18/23  &  9/13     &  3/8   &  6/8  
\end{tabular}
\end{center}
\caption{\label{tax:taun} Fraction of unstable orbits of different multiplicities
in Model 1 with oblate halo and $b/a=0.8$ AND GMC=???. PLUS THE 1:1 ARE 2:2 ARE ETC!!}
\end{table}
\vspace{0.3in}

Between horizontal and vertical instability one usually finds a large fraction
of unstable periodic orbits.
The persistence
of  significant numbers of stable low order orbits however means that one
has a ``mixed'' phase space.
This is true of {\em all} orbit families as can be seen in Table~\protect\ref{tax:taun}.
Results for other models are qualitatively similar.
There are important differences between the response of such  systems 
to additional external perturbations, and that of  
integrable or  uniformly hyperbolic
 systems which are characterised by either KAM or structural
stability respectively (see Section~\protect\ref{tax:noise} for more detail). 

\section{Stability properties of ensembles of general orbits}

\subsection{Models with disk, halo and central mass}
\label{dishac}
\subsubsection{Method}

We have looked at the stability of the periodic orbits which parent (or deflect) the
general ones. We now try to get an idea of the global structure of
the phase space by studying the stability of general  orbits. 

In each model,
we have integrated an ensemble of nearly a hundred trajectories and calculated 
their maximal Liapunov exponents. The set of 12 first order differential equations --- 
representing the six differential equations of motion and their associated variational
(linearised) equations --- were numerically integrated using the variable order,
variable step size, Adams method as implemented in the NAG 
library routine D02CJF with a local tolerance of $10^{-14}$ per time step. The relative
energy change along a trajectory was smaller than $10^{-10}$.

A two parameter ``startspace'' (Schwarszchild 1993)
of  initial conditions was chosen as follows
\begin{eqnarray}
x_{ij} &=& i \times 2 \\
\label{tax:inihx}
y_{ij}&=&0.0\\
z_{ij}&=&0.02\\
\dot{x}_{ij}&=&0.0\\
\dot{y}_{ij}&=&0.02 \times j - 0.02\\ 
\dot{z}_{ij}&=&0.0,   
\label{tax:inihz}
\end{eqnarray}

where the integer $i$ was varied from $i=1$ to $i=8$,  while $j$ took values
ranging from $j=1$ to $j=12$. In total, therefore, we have 96 orbits 
started from near the disk plane.
The small initial $z$ value is included to test the vertical
stability of the orbits, which in most cases is a measure of the degree of stochasticity
of a given orbit. 
 This choice of initial conditions starting near the disk plane
is motivated by the aims outlined  at the end of Section~\protect\ref{tax:wyax}.
The choice of range and resolution in the coordinate $x$ will mean that
for Model 1 and 3 one  will be studying areas outside the core (up to eight
core radii), so that the stability results will mainly concern halo orbits
and eccentric high energy stellar trajectories.
In Model 2 one will be concerned with orbits 
inside the core and within two or three core radii. A large fraction of these
can be interpreted as stellar orbits in the dynamically hot inner areas where,
in the absence of a central mass contribution, 
motion  is likely to be on box-like orbits characteristic of the 
harmonic core. (The extent of
the effective  --- disk + halo --- 
cores of our  models can be inferred from the rotation curves where they are
represented by the rising part of these curves).  

The fact that the value of $\dot{x}_{ij}$ is always zero 
means that we are considering only orbits whose 
trajectories are normal to the $x$-$z$ plane at least once during their crossings
of that plane near $z=0$ (choosing the initial conditions on the $x$-$z$ 
plane is no restriction, since, by symmetry, all orbits must cross it at least once). 
This condition is satisfied by both box and loop orbits --- because the former can be regarded 
as independent 1-dimensional oscillations about the axial orbits while 
the latter are parented by the closed
loops, which are themselves symmetric about the $x$-axis. One can also expect many 
highly chaotic orbits to  satisfy this condition during their near random time
evolution. 
The choice of a two parameter
space of initial conditions instead of a three dimensional one also facilitates 
the representation of the results. Although there is no guarantee that all type of orbits 
will be caught in this space (for example  orbits parented by the $x$-$y$ banana
family which is asymmetric with respect to the x-axis might not be caught), a representative
sample should. In addition, this procedure avoids the non-uniqueness arising when  
the full 3-parameter space is used (where essentially the same trajectory 
may be integrated more than once: Schwarschild 1993).

\begin{figure}
\begin{center}
\epsfig{file=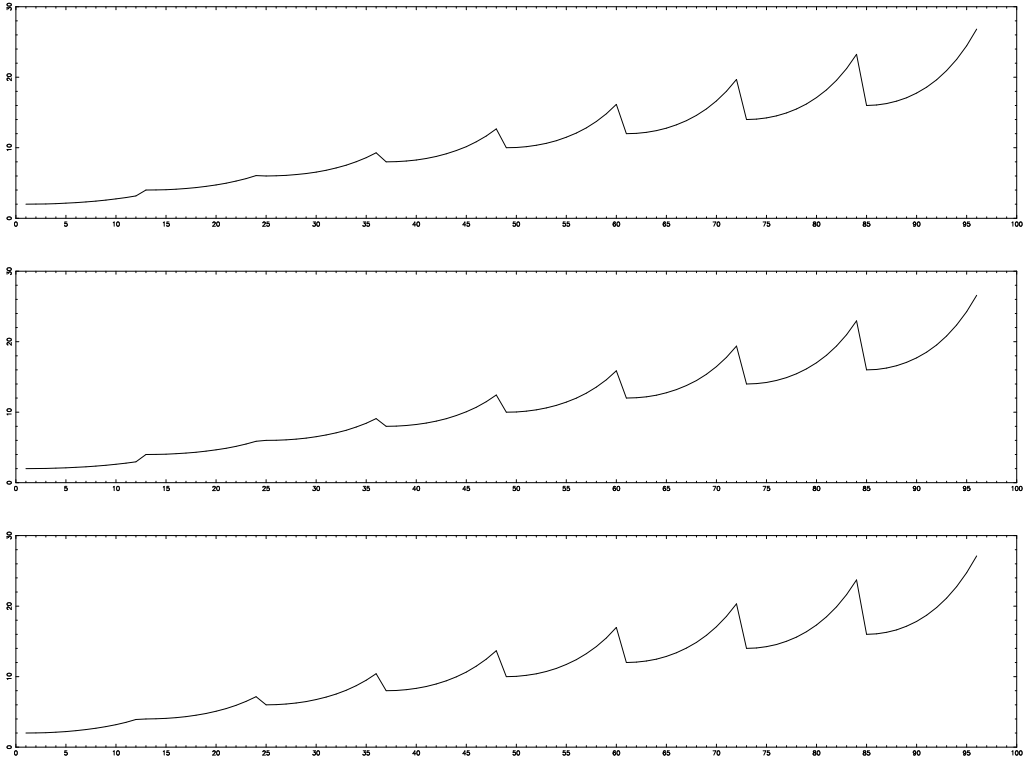,width=13.5cm,height=10.0cm,angle=-90}
\end{center}
\caption{\label{zvcic}
 Intersections of the zero velocity surfaces confining the orbits with initial
conditions given by Eq.~(\protect\ref{tax:inihx}) to~(\protect\ref{tax:inihz}) 
with the $x-y$. 
The orbits are numbered according to the rule $N_{ij}=12 (i-1) + j$. 
Top: orbits
of model 1. Centre: orbits of Model 2. Bottom: orbits of Model 3}
\end{figure}

Our startspace covers a large range of energies. However
the variation of energy with the chosen initial conditions
is slow enough  so that  orbits can be grouped together
into a few small-energy intervals 
(as was done in Fig.~\protect\ref{MAP1} and Fig.~\protect\ref{MAP1D}).   
In Fig.~\protect\ref{zvcic}  we have numbered our orbits sequentially and plotted
the orbit number $N_{ij}=12 (i-1) + j$ against the intersection of the 
zero velocity surface 
with the $x-y$ plane (assuming a  spherical halo). 
Only a few orbits
are confined within
 the same zero velocity circle, but the variation of the radius of this circle 
with the orbit numbers is seen to be slow enough so that all energy ranges are sampled 
reasonably well.

For each trajectory,  we calculate the 
numerical maximal Liapunov exponent and record its value at 50 000 Myr.
This choice, which corresponds to a few Hubble times, ensured proper convergence
for most orbits and facilitates the extraction of information from distribution
of exponents
(i.e., classifying trajectories according to their stability properties). 
We will be comparing the stability properties inferred from
calculating the Liapunov exponents over such time-scales with the actual $z$ stability of 
trajectories over smaller times-scales (comparable to a Hubble time).

A well known difficulty (e.g., Udry \& Pfenniger 1988; Merritt \& Fridman 1996) 
is the absence of a  universal criterion by which to distinguish if a 
given numerical Liapunov exponent describes a chaotic orbit or a regular one. 
We use the following
estimate for the cutoff value of the exponent (below which we call an orbit regular). 
Consider a circular orbit
of radius $R$
in the plane of an axisymmetric mass distribution. 
In the outer regions
the rotation velocity  is constant and $\sim v_{0}$, so that the differences between
the angular frequencies
of this orbit and another at a radius $R + \Delta R$ is, 
in the linear approximation, given by
\begin{equation}
\Delta \omega = \frac{v_{0}}{R^{2}} \Delta R.
\end{equation}
The  phase distance between these orbits after a time $\Delta t$ is therefore
\begin{equation} 
D=R \Delta \omega \Delta R \Delta t =\frac{v_{0}}{R} \Delta R \Delta t. 
\end{equation}
Differences in all other phase space coordinates of these two orbits are
constant, so that their contribution to the time dependent 
maximal Liapunov exponent is negligible
by comparison. This  quantity is then given by 
\begin{equation}
\lambda (t) = \frac{1}{t} 
\log  \Bl  \frac{\parallel  \frac{v_{0}}{R} \Delta R t  \parallel}
{\parallel \sqrt{6} \Delta R \parallel} \Br
\end{equation}
which for Model 1 at 10 kpc for example has a value of $\sim 1.2 \times 10^{-4}$
at 50 000 Myr, which implies an exponentiation time-scale of the order of a Hubble time.
In practice, it is found that the vast majority of regular orbits (those for
which the Liapunov exponent decreases monotonically as $\ln t/t$ even when integrated 
for very large times) have time dependent
 Liapunov exponents of the order of $\sim 5 \times 10^{-5}$ at 50 000 Myr (i.e., 
with an exponentiation time scale longer
than most estimates of the Hubble time).
 We will therefore consider orbits with a maximal exponent
less than $10^{-4}$ to be regular. The maximal exponent was calculated by the
``standard algorithm'' of Benettin et al. (1976) with a renormalisation interval of 150 Myr.

\subsubsection{Results}

We will illustrate our results using greyshade diagrams of our startspace.
 In these diagrams the (white) background
will correspond to orbits with Liapunov exponent less than or equal to $10^{-4}$,
while the (black) foreground will correspond to slots on the startspace
occupied by the initial conditions of orbits having Liapunov exponents
of $4 \times 10^{-3}$ or greater. This value implies an 
exponentiation time-scale smaller than a rotation period 
at 10 kpc in the plane of the axisymmetric Model 1. Ensembles of trajectories with such large
exponents are known to relax to time invariant distributions filling all their
available energy subspace in just a few exponentiation times (e.g., Merritt 1996).
Indeed it was found here that 
with almost all such trajectories for which this property
(ergodicity) was checked (a total of about 50 orbits), an invariant distribution
was reached in less than a Hubble time.

\begin{figure}
\epsfig{file=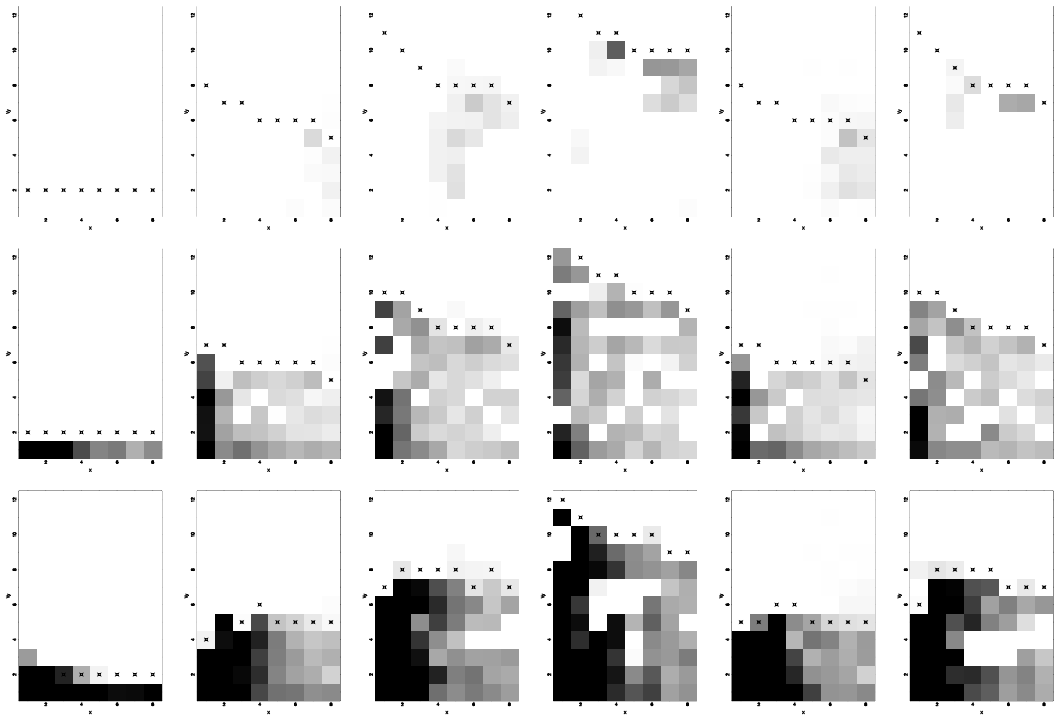,width=13.9cm,height=9.4cm,angle=-90}
\caption{\label{MOD1}
 Greyshade diagram showing the distribution of the maximal Liapunov exponents.
The diagrams corresponding to the different runs
are  the same  as in Fig.~\protect\ref{TM1--bx}
except that in the first column in the left the diagrams correspond to the case
when the halo is spherical which was not considered in the 
aforementioned figure. The initial conditions are obtained according to 
the prescription in Eq.~(\protect\ref{tax:inihx}) to 
Eq.~(\protect\ref{tax:inihz})} 
\end{figure}

Fig.~\protect\ref{MOD1} shows  the Liapunov exponent greyshade diagrams for the orbits of
Model 1 for different values of the axis ratio, triaxiality parameter and
central mass which we have considered (Section~\protect\ref{tax:mod}).
As can be seen, the results are what is to be expected from 
both the general discussion of Section~\protect\ref{tax:staper} and the analysis of periodic
orbits of Section~\protect\ref{statprop} 
In the absence of a central mass (top row) and when the halo
is spherical (far left) all orbits are regular. As one moves towards a more
asymmetric halo however (prolate model with $b/a=0.9$; second from left), 
some orbits starting with large initial $x$ coordinates
start becoming chaotic. These are  
part of the stochastic layer around the 1:2 vertical resonance
which was brought inwards with increasing asymmetry
in the halo mass distribution (as is also clear from the subsequent 
diagrams on the right). 
In the case $b/a=0.7$ the asymmetry  in the plane is large
and the horizontal 
3:4 resonance is broad enough to give rise to a significant
stochastic layer around unstable antipretzel orbits, where ``trapped'' 
chaotic motion occurs.

The greyshade diagrams of Liapunov exponents
for runs with $GM_{C}=0.0005$ are shown in the second row of Fig.~\protect\ref{MOD1}. 
As we have seen (Section~\protect\ref{tax:staper}), in the two dimensional case, this central mass
is enough for destabilising the $x$-axis orbit and to cause  most trajectories 
in the central regions to become chaotic.
Here we also see that  remarkably
modest deviations from axisymmetry
(the models with 0.9 halo potential axis ratios in the plane
are compatible with observations of present day galaxies) can have a large 
effect on the orbital structure when a modest central mass is present --- although,
as expected, the more asymmetric systems contain a larger fraction of initial conditions from which
chaotic trajectories may start. 
It is interesting to note here that chaotic behaviour is 
not limited to the central box orbits only, but 
is significant even at 16 kpc from the centre --- contrary from what may be expected 
from simple arguments (e.g., Gerhard \& Binney 1985). However, the existence in the region
once occupied by the box orbits, of a large number of trajectories which are either
regular or have small Liapunov exponents, will mean that the halo would not rapidly
(i.e., over a dynamical time or so) lose triaxiality.

The addition of a larger central mass (bottom row) increases both the exponentiation rates
of individual chaotic orbits and the fraction of such orbits. A large fraction of the 
chaotic orbits now have exponentiation time-scales of the order of a dynamical time
and, therefore, evolution is likely to take place on a time-interval of that order.
This is compatible with the results of Merritt \& Quinlan (1997) where it was found
that triaxial figures (representing elliptical galaxies in their case) do indeed
lose their asymmetry on a time-scale of a few dynamical times when the central mass
is a few percent of the total mass of the galaxy (which is the case here for the inner 
10-20 kpc). 
  Nevertheless,  as we have seen in
Section~\protect\ref{tax:staper}, some  higher order closed boxlet orbits will remain stable even for such large
central masses and asymmetries. The  white ``islands'' seen in the diagrams here correspond to
quasiperiodic orbits parented by these. In addition, 
there are, in general, fewer chaotic orbits in the central
areas because many are parented by closed loops 
created by the central mass (because of the destruction of the harmonic core and 
the more axisymmetric potential). 
Due to the reasons mentioned in Section~\protect\ref{tax:staper}, 
none of these can be  chaotic.

We have also calculated the Liapunov 
exponents for trajectories starting from the same initial conditions with
the exception that $z=0$.  
The main difference found
 was that, in the two dimensional case, significantly fewer orbits are chaotic,
but those that are, usually have larger Liapunov exponents than the
 corresponding ones in the three dimensional case.   
The fact that chaotic behavior is more widespread in the
three dimensional model is probably due to the role of the role played
by $z$ instabilities in the closed periodic 
orbit families (Section~\protect\ref{tax:staper})
--- that is, some
of the closed orbits that parented  the  regular   
islands in the two dimensional case are $z$ unstable  and therefore
cease to parent these orbits when a $z$ perturbation is applied.

\begin{figure}
\epsfig{file=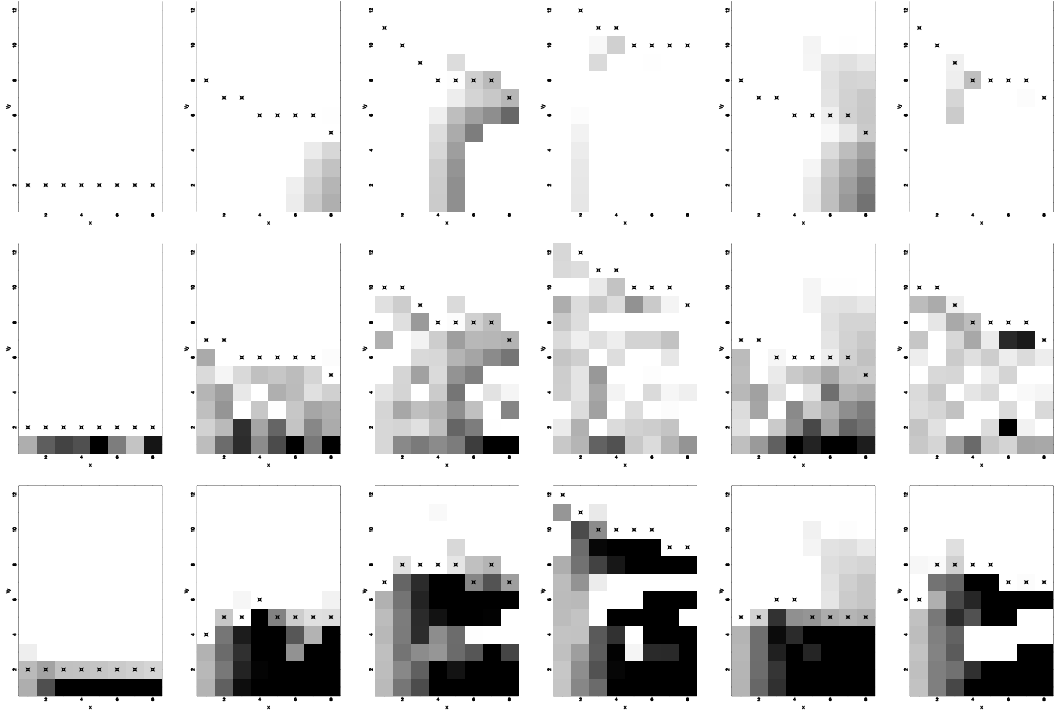,width=13.9cm,height=9.4cm,angle=-90}
\caption{\label{MOD1Z}
Greyshade diagrams illustrating the vertical stability of orbits starting near the disk
plane. The positions of the diagrams corresponding to runs with different system parameters
$GM_{C}$, T and $b/a$ are the same as in Fig.~\protect\ref{MOD1}}
\end{figure}

We have  recorded the maximum value of  $\parallel z \parallel$  along
the trajectories of the orbits for $t \leq 20 000$ Myr,
 with the $z$ coordinate being sampled at every renormalisation interval 
(sampling only until t=20 000 means we catch orbits with effective $z$ instability
over a period comparable to a Hubble time). 
These values have also been arranged in greyshade diagrams. The background
value for these diagrams was taken to be equal to 0.1 kpc. Clearly,
trajectories with larger maximum $z$ excursions 
are unlikely to be undergoing stable independent $z$
oscillations (as would happen if the vertical motion was
decoupled). The foreground was chosen as 5 kpc. The results for Model 1 are shown
in Fig.~\protect\ref{MOD1Z}. As is clear from this figure, the $z$ stability of orbits is 
usually correlated with the value of their Liapunov exponents~(Fig.~\protect\ref{MOD1}). 
There are however 
orbits that have fairly large Liapunov exponents but are confined to the disk
plane (for example those of Model M1--P7C3 at $i=7,8$ and $j=7$). These orbits
start near regular two dimensional boxlet orbits and remain trapped near the latter
for very long times. 
There are also some orbits that have fairly large $z$ excursions but small
Liapunov exponents.  These are mainly 
trapped around the vertical 1:2 resonance.

\begin{figure}
\epsfig{file=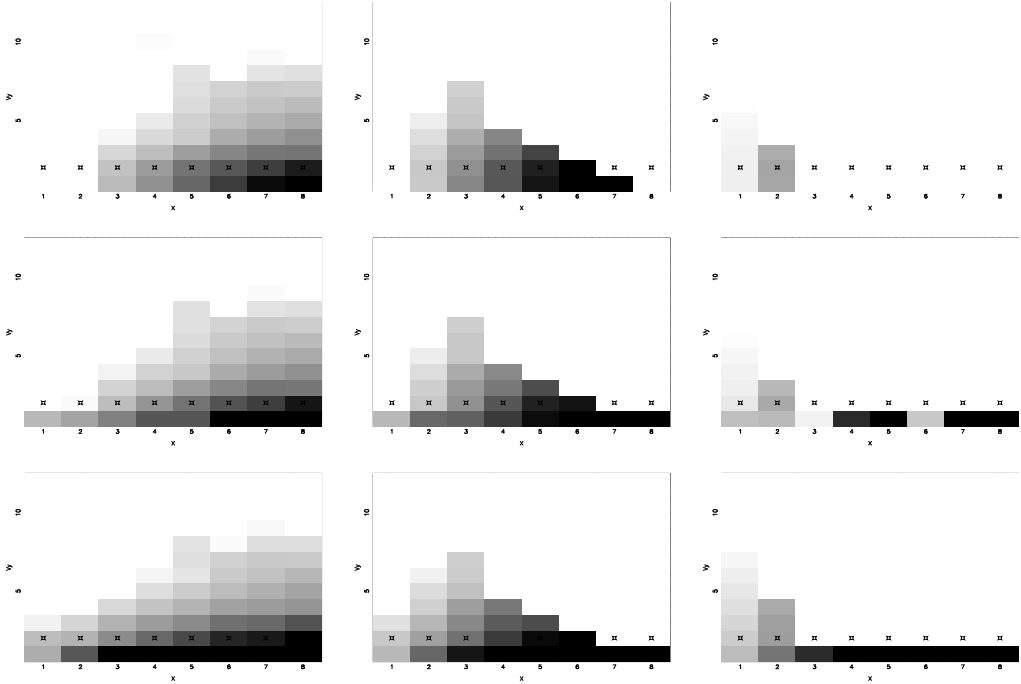,width=13.9cm,height=9.4cm,angle=-90}
\caption{\label{MODDAZ}
Greyshade diagrams illustrating the vertical stability 
of orbits starting near the disk
plane of axisymmetric models with $R_{0}^{2}=0.5$ 
with (from left to right) $c/a=0.9$, $c/a=0.8$ and $c/a=0.7$ 
and (from top to bottom) $GM_C=0$, $GM_C=0.0005$, $GM_C=0.01$ .}
\end{figure}

The vertical 1:2 resonance is also present in flat axisymmetric potentials
where it is also a source of $z$ instability for trajectories
starting near the disk plane. This resonance is too far out along the $x$ axis to be of interest
for models with large core radii but is moved inwards when the core radius is
decreased. Below $R_{0}=2$, its position along the long axis does not vary
much with core radius or the existence of
a central mass,  but depends primarilly on the flattening of the density
distribution. 
 Fig.~\protect\ref{MODDAZ} shows the maximum
$z$ excursions for ensembles of trajectories in models where $R_{0}^{2}=0.5$.
As can be seen, while the stability of trajectories
 passing sufficiently near the centre are affected by the central mass, the
main instability region around the 1:2 resonance remains unchanged. 
This is due to the fact that the central mass, by creating a strong 
nonlinearity in the potential of the central region, gives rise to a  set of 
high order vertical resonances with connected instability gaps 
 along the axial orbit. In the triaxial case, this causes widespread 
chaotic behaviour in trajectories that oscillate  about the axial orbit --- the box orbits.
 The higher order resonances are not broad however, and therefore  have little
effect away from the axial orbit itself. Therefore, in the axisymmetric
case,  all but the lowest angular momentum orbits remain 
unaffected by the addition of a central mass.

\begin{figure}
\epsfig{file=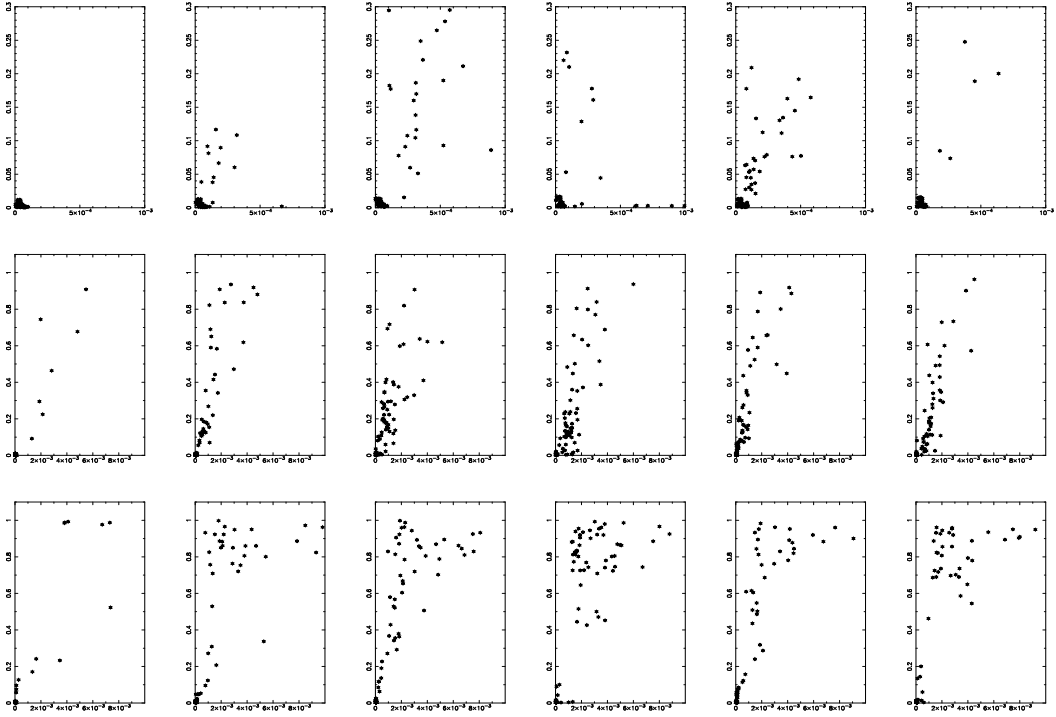,width=13.9cm,height=9.4cm,angle=-90}
\caption{\label{MOD1ZMIX}
Correlation between the values of the Liapunov exponents of trajectories of Model 1
(plotted in the abscissa)
and the ratio
of the largest $z$ excursions to the maximum 
possible $z$ excursions determined by the largest value of the $z$ coordinate on their
zero velocity surfaces. The different diagrams represent runs with the same system
parameters as the corresponding diagrams in Fig.~\protect\ref{MOD1}}
\end{figure}

Coming back to the triaxial case now, we examine in more detail the relation between the maximum
of the absolute value of the $z$ coordinates of our trajectories, and the corresponding
Liapunov exponents. 
This is shown in Fig.~\protect\ref{MOD1ZMIX}. Here we have plotted 
the values of the maximal Liapunov exponent against the ratio
of the largest $z$ excursions of the trajectories of Model 1 to  
maximum value of the $z$ coordinate on the
zero velocity surfaces of the orbits.
(i.e. the $z$ coordinate
of the zero velocity surface at $(x,y)=(0,0)$, which is the maximum possible 
$z$ value a trajectory can reach).
It is clear
that a strong correlation exists between the two quantities. For small values 
of the Liapunov exponent the relationship appears to be linear, while for larger
values most orbits reach the maximum $z$ value possible. 
This correlation is remarkable given the fact that the sampling of the $z$ coordinates
is done only up to  $t \leq 20 000$ while the
exponent is calculated until $t = 50 000$. 
While some of the scatter is due to ``trapped'' chaotic orbits, only in a few cases
did these have large Liapunov exponents. 
Thus there appears to be a strong correlation between the ergodic properties
of the trajectories over a Hubble time and the Liapunov exponents calculated over
significantly longer times. And indeed it was verified that trajectories
with the larger Liapunov exponents showed significant $z$ excursions within
a few tens or hundreds (depending on their energy) Myr,  while for trajectories
with small Liapunov exponents  a significant fraction of the 20 000 Myr period
was needed before the $z$ instability manifested itself (if at all).
We will see below however that the correlation between $z$ excursion and maximal
Liapunov exponent is much weaker in the case when rapidly rotating bars are present.

The characteristic features  distinguishing Models 2 and 3 from Model 1
are the large  halo core radii (6 kpc compared to
2 kpc in Model 1) and  the more  dominant disk.
In Model 2 the disk has a larger total mass while in Model 3 it
is more centrally concentrated.
This amounts to decreasing the nonlinearity (especially in Model 2) 
of the potential and decreasing the asymmetry in the 
central regions  (especially in Model 3).

\begin{figure}
\epsfig{file=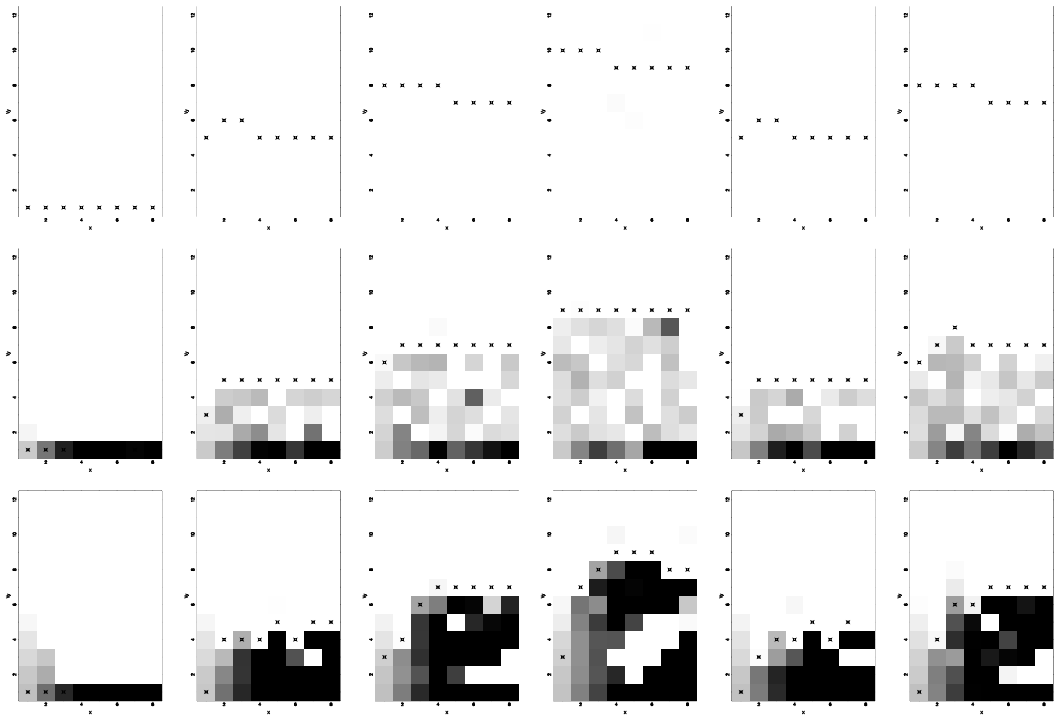,width=13.9cm,height=9.4cm,angle=-90}
\caption{\label{MOD2}
 Distribution of Liapunov exponents for orbits of Model 2. The different 
diagrams correspond to runs with the same system parameters as in 
Fig.~\protect\ref{MOD1}}
\end{figure}

In Model 2, the above  effects  will mean that the main resonances 
(both vertical and horizontal) are moved outwards.
It follows that one will not be observing the
$z$ instability around the 1:2 resonance which has moved beyond our range 
of initial conditions. 
This can be clearly seen from Fig.~\protect\ref{MOD2}, 
where  the  Liapunov exponents greyshade diagrams for Model 2 are 
shown. In Model 3, there is some instability around the 1:2
resonance but it is limited and almost all orbits near the plane are again 
stable.
As usual, this situation changes when a central mass is added.
Even the mildly asymmetric cases with $b/a=0.9$  contain  significant
fractions of chaotic trajectories --- even when the central mass is small.
So, as in Model 1, most box orbits are again replaced by chaotic ones. 
One important difference with Model 1 however is that, in runs
with a large central mass, most orbits in the inner areas which have 
definite sense of rotation are not parented by the  1:1 loops.
Many of these are chaotic; they appear to be repelled by unstable looplet 
orbits  (cf. Section~\protect\ref{statprop}). 
There is also
 an important difference in interpretation: while in Model 1 we were 
concerned mainly with trajectories which venture way beyond the small harmonic
core --- and are therefore either halo trajectories or those of stars born (or heated) 
into eccentric trajectories --- in Model 2, many trajectories we calculate are completely
confined within the harmonic core and therefore 
{\em must} correspond to stellar trajectories born on box orbits (because of the absence
of loop orbits completely confined within the core). These are subsequently
destabilised by the accretion of central mass (cf. Section~\protect\ref{tax:dispdisk}) and
are therefore candidates for forming extensive bulge-like structures out of disk material.

\subsection{The effect of discreteness noise}
\label{tax:noise}

Up to now we have assumed that stars in a galaxy
move in the mean field potential produced by the smoothed out mass
distribution. This is of course the usual assumption made in galactic
dynamics. However, it has been known for some time that, at least
in some cases, this picture may not be accurate (Pfenniger 1986; Kandrup 1994; Merritt \&
Valluri 1996). 
In considering the effect
of discreteness on the dynamics, a first approximation is just to introduce
random kicks to the  trajectory in the smoothed out potential. 
We will model  these
as perturbations to the velocity which are sampled from a Gaussian distribution. This may
not be a very accurate representation of discreteness effects 
in a real $N$-body system where one can imagine all sorts of resonance effects 
between the different degrees of freedom (El-Zant 1997),
but may be useful as a first approximation.

To examine the effects of discreteness noise, we have perturbed the velocities 
along the trajectories of Model 1 at intervals of 1 Myr. The perturbations
to each component of the  velocity were sampled from  a Gaussian distribution 
with zero mean and dispersion of $10^{-4}$ kpc/Myr,
produced using the NAG routine G05DDF.
 Although this procedure does not explicitly conserve energy
(the magnitude as well as the direction of the the velocity is changed, 
unlike the scheme of Goodman
\& Schwarschild 1981), because of the smallness of the perturbation however,
the energy is conserved to better than 1\% (usually to a few parts in a thousand)
 over a period of 50 000 Myr. That means, if one assumes
that errors in the energy grow according to a $\sqrt{t}$
diffusion law, the energy relaxation time is 
$\sim 10^{5}$ Hubble times. The velocity relaxation time resulting 
from this perturbation is of the order of
\begin{equation}
t_{r}=\frac{v_{0}^{2}}{3} \times \Bl 10^{4} \Br^{2} \sim 10^{6} \rm{Myr},
\label{difkick}
\end{equation}
which is about a hundred Hubble times. 

\begin{figure}
\epsfig{file=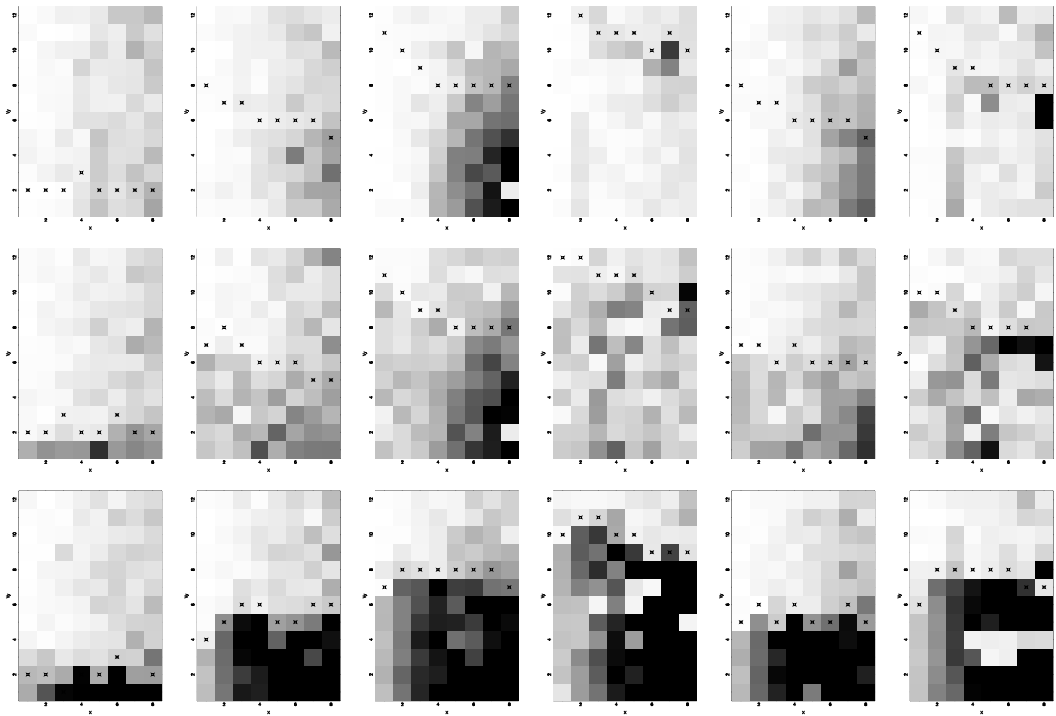,width=13.9cm,height=9.4cm,angle=-90}
\caption{\label{MOD1NN}
Greyshade diagrams showing the distribution of the maximum value of the $z$
coordinates of trajectories of Model 1 the velocity components of which are perturbed
by noise sampled from a Gaussian distribution with zero mean and dispersion $10^{-5}$.
The different diagrams correspond to runs with the same system parameters as in
Fig.~\protect\ref{MOD1}}
\end{figure}

Fig.~\protect\ref{MOD1NN} shows the greyshade diagrams for the maximum $z$ excursion of these
perturbed orbits (which are to be compared with those of Fig.~\protect\ref{MOD1}).
As is apparent from these figures, even such small perturbations as those introduced 
here can have a non-negligible effect on the orbital structure. 
 For the spherical case, the grey hue  is compatible
with  the $z$ excursion expected from normal diffusion caused by the 
  the perturbation the trajectories are subjected to.
As can be seen however, for the  triaxial models, rather
large excursions can take place. These cannot be attributed  to
standard diffusion, especially because in the flatter models, the 
$z$ force is generally stronger and trajectories tend to reside for
longer times in the central regions (which means they will require larger perturbations
to achieve the same maximum $z$ excursions than their counterparts moving
in a spherical halo). Instead, these highly unstable trajectories appear 
near the vertical resonances (especially the 1:2). Thus they are the result of
the broadening of the chaotic regions around these resonances, which is a consequence of
the external stochastic noise.

For the models with central mass one sees the above effect greatly enhanced.
In particular, many of the isolated blank areas, corresponding to stable islands, are
now filled with highly unstable orbits. This has the effect that unstable regions merge
into connected areas.
Also, orbits that were  only slightly unstable, may  
now also have very large $z$ excursions (some venturing out to vertical distances larger than
5 kpc). On the other hand it is  found that, in some cases, orbits that {\em were}
unstable can be stabilised. This effect however is less common  
(especially in the case with the models containing stronger central mass).
However it was found to be of  more relative importance when the dispersion 
of the perturbation was smaller ($10^{-5}$). In that case however the overall effect
of the perturbation was much less pronounced.

The question discussed above is that of the stability of qualitative behaviour 
against perturbations. The KAM
theorem~(e.g., Arnold 1987) guaranties actual stability against weak non-resonant external 
perturbations.  This means, for example, that stars moving in a smooth spherical 
potential are not likely to change their qualitative behaviour just because some external
Gaussian  noise is added. Roughly speaking, such noise will act to move a star from one
regular trajectory to another,  but the total displacement
will amount to the mean strength of the noise multiplied by the square
root of the number of kicks a star suffered as a result of the presence of this noise
(and will thus have an associated relaxation time given by Eq.~\protect\ref{difkick}). 
This
was indeed found to be the case for regular trajectories moving near the disk plane of
our  disk-triaxial halo models. On the other hand, very chaotic 
systems benefit from the structural stability~(or ``roughness'' as it is often called in the 
Russian mathematical literature) of average behaviour. 
When the situation   is intermediate however,
so that neither structural stability nor KAM stability are guaranteed (mixed phase space),
 non-standard diffusion processes 
can occur and intermittent behaviour can take place on a variety of scales (Zaslavsky et al
1991; Shlesinger et al. 1993; Klafter et al. 1996). 
This will mean that the phase space structure is
not stable  and that weak external noise can have  major effects not predicted  by standard
diffusion processes. It can change the qualitative orbital structure by turning regular 
or confined orbits into chaotic ones and vice-versa. 
  This state of affairs
makes it even harder to justify the assumption that galactic systems are near integrable
since, even if the smoothed density distribution produces a potential that does not support too
large a fraction of chaotic orbits which visit most of the available energy subspace, many
of these could be made to do so by the action of weak  noise or other 
weak perturbations.

\subsection{The effect of rapidly rotating barred perturbations}
\label{tax:rapbars}

To get instability  in nearly circular periodic orbits a rotating
barred perturbation must be present (Section~\protect\ref{tax:staper}). Rapidly rotating bar perturbations
are of course present in many (if not most) galaxies and they may
play an important role in galaxy evolution (e.g., Sellwood \& Wilkinson 1993). 
Here we discuss the effect of such a
perturbation in a disk galaxy with triaxial halo.

We have now more than one potential source for chaos. First, there are the axial 
orbit resonances giving rise to chaos in the region occupied mainly
by the box orbits (when no central mass is present). Second, there are
the circular orbit resonances giving rise to chaos in the $x_{1}$ loops
in the rotating barred
potential. There is also the central mass which affects the location and
width of both types of resonances and invariably increases the fraction of
chaotic orbits  present in the system. Of course these effects do not simply add up
and their interaction may be very complex. The very time dependency
of the potential, which means that not even energy is an integral of motion, 
suggests that this interaction between the various contributions will 
lead to widespread instability.

\begin{figure}
\epsfig{file=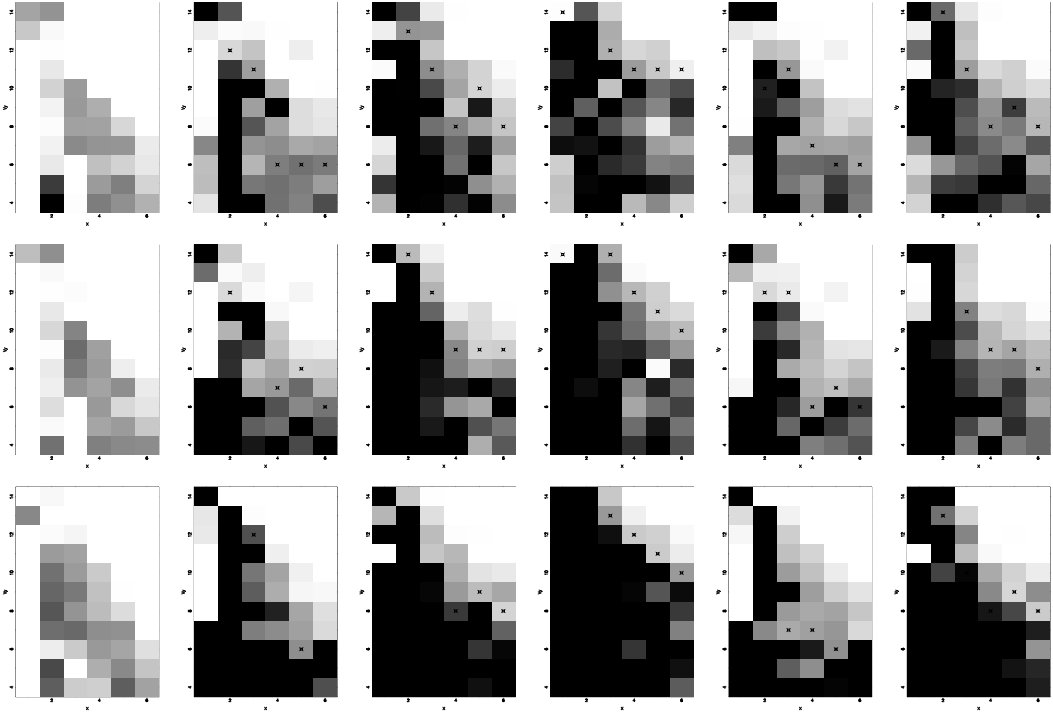,width=13.9cm,height=9.4cm,angle=-90}
\caption{\label{MOD1B}
 Greyshade diagram showing the distribution of the Liapunov exponents for
orbits of Model 1 when a rapidly rotating Ferrers bar perturbation 
with parameters described in Section~\protect\ref{tax:bars} is present. The different
diagrams correspond to runs with the same halo and central mass parameters as the 
corresponding diagrams in Fig.~\protect\ref{MOD1}}
\end{figure}

Fig.~\protect\ref{MOD1B} shows greyshade diagrams of the Liapunov exponents of  Model 1
with an added additional Ferrers bar with parameter values
  described in Section~\protect\ref{tax:bars}. 
To accommodate  the expected large fraction of chaotic orbits with
 large initial circular velocities and to concentrate the integration of orbits
to the region where the bar plays an important role, we have changed
the range of initial condition indices of Equations (\protect\ref{tax:inihx}) to
(\protect\ref{tax:inihz})  to $i=1,6$ and $j=4,14$. Also, due to the more complicated
time dependent nature of the potential studied here, it was impossible to
integrate  trajectories with a tolerance of $10^{-14}$. For most trajectories
a tolerance of $10^{-13}$ was adequate, although some required tolerances as 
high as $10^{-11}$.

 Chaotic orbits in the spherical halo potential are  mainly
those with initial $x$ position starting in the region where the bar 
dominates and especially in the radii nearing corotation. 
In this time 
independent case the above result is standard 
(because of the accumulation of the resonances
near corotation: e.g., Contopoulos 1985).  
 What is somewhat surprising is
that the fraction of chaotic orbits and their location does not change
much when a central mass is added (even when that central mass is strong).
This may be due to the relative weakness of the bar perturbation in our model to
 those in  previous studies (e.g., Hasan et al. 1993) when a central mass
was found to considerably alter the phase space structure. 
 
As can be seen however, in the presence of a
non-axisymmetric halo,
the fraction of chaotic orbits
 dramatically increases, even when no central mass  is present. 
 This effect  appears to be
directly related to the  strength of the time
dependent perturbation to the potential (since orbits
in prolate potentials with similar  axial ratios $c/a$ to the corresponding
oblate ones but with smaller axial ratio $b/a$ in the plane appear to have
larger Liapunov exponents). 
In each diagram, we have also marked with a little square the first 
of the orbits that has a definite sense of rotation.
 One can see that, while in the spherical case {\em all}
orbits have a definite sense of rotation,  
in the case when the halo is asymmetric, many of the orbits 
do not have a definite sense of rotation. 
 Although this result can be
seen as an obvious corollary of having included the stationary triaxial halo,
it can have important consequences. More precisely, it is unlikely that self
consistent bars can be built with the help of these orbits, even if the 
phase space diffusion accompanying the chaotic behaviour does not prevent the
building of such structures.

Adding a central mass increases the fraction of chaotic orbits present even further. 
This is especially true in the central regions where it is most 
effective, with the larger central mass affecting a wider area. 
This effect is again most apparent in the  more asymmetric 
models and clearly depends more on the halo potential axis ratio in the 
plane than normal to it.

\begin{figure}
\epsfig{file=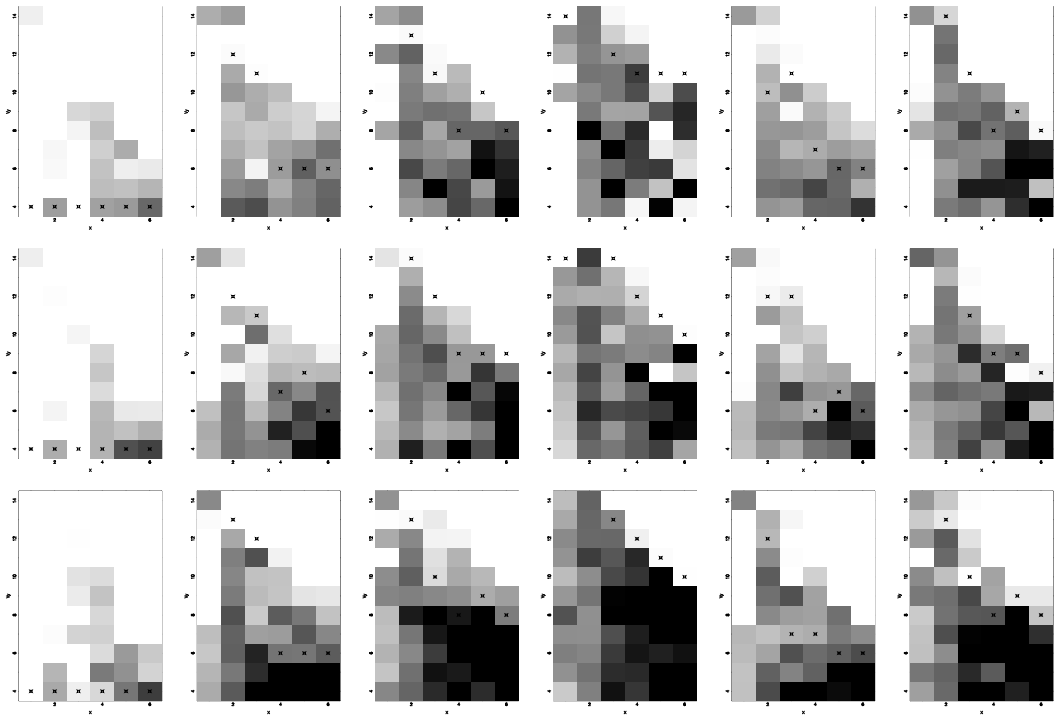,width=13.9cm,height=9.4cm,angle=-90}
\caption{\label{MOD1BZ}
  Greyshade diagram showing the distribution of the maximum $z$ excursions for
orbits of Model 1 when a rapidly rotating Ferrers bar perturbation 
with parameters described in Section~\protect\ref{tax:bars} is present. The different
diagrams correspond to runs with the same halo and central mass parameters as the 
corresponding diagrams in Fig.~\protect\ref{MOD1}}
\end{figure}

If one now looks at the diagrams showing the $z$ stability of the orbits 
of Model 1 in the presence of the Ferrers bar
(Fig.~\protect\ref{MOD1BZ}), one sees the same general trends as deduced from 
the Liapunov exponents. An interesting result which can be deduced from these 
plots however is that many orbits that have a large Liapunov exponent
are not $z$ unstable or are very mildly so. This is found to  also be  mostly true if one
samples the $z$ excursion up to the full integration time of 50 000 Myr
instead of only 20 000 as in this figure.
This effect is  probably due to the
fact that the time dependency in the potential is largest in the plane 
and therefore the $z$ stability is not as much affected as the stability
in the plane. As one would expect then, the vertical instability is simply dominated 
by the stochastic layers around the vertical
axisymmetric resonances which are concentrated near the end of the
bar, this is enhanced by the time dependency but not nearly as much as
instability in the plane.

\begin{figure}
\begin{center}
\epsfig{file=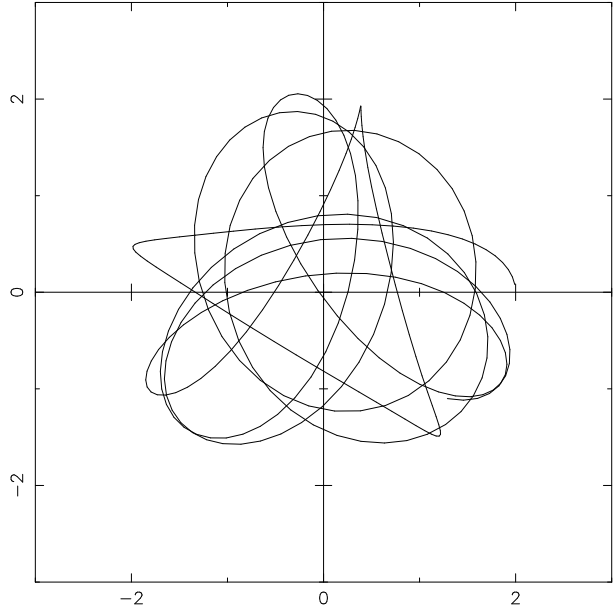,width=8.0cm,height=8.0cm,angle=-90}
\end{center}
\caption{\label{BAI2J9300MyrTrap.ps}
Highly chaotic trajectory which is trapped in the disk plane. Note the 
frequent large deflections it suffers at  successive crossings}
\end{figure}

The above means that many orbits with large Liapunov exponent are trapped
near the disk plane  because 
the diffusion rate normal to the plane is  small (or non-existent). 
Fig.~\protect\ref{BAI2J9300MyrTrap.ps}
 shows a segment of such an orbit. The most
important characteristics of the motion are the large deflection angles
as the trajectory moves in and out of the rotating potential of the bar
on eccentric motion forced by the asymmetry of the halo mass distribution. 
This type of motion results in a large Liapunov exponent. There is no equivalent 
motion in the vertical direction, since the bar is flattened
and rotates strictly in the disk plane.

\begin{figure}
\epsfig{file=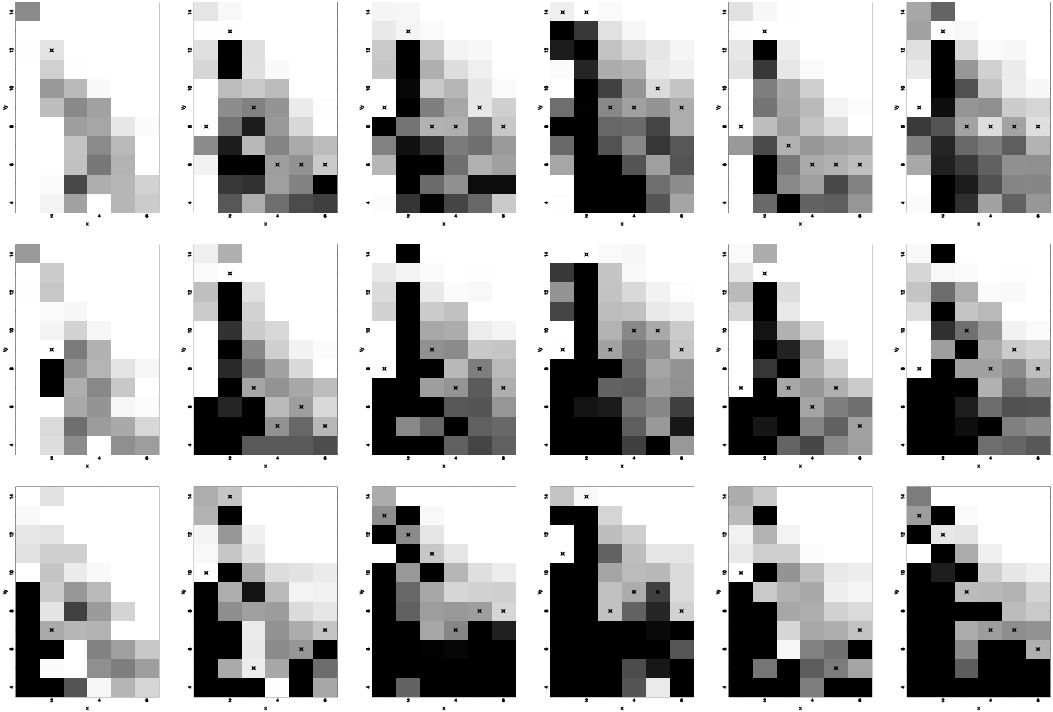,width=13.9cm,height=9.4cm,angle=-90}
\caption{\label{MOD2B}
 Same as in Fig.~\protect\ref{MOD1B} but for orbits of Model 2}
\end{figure}

The relative bar perturbation is much less (about half as) significant 
in Models 2 and 3 than in Model 1 and therefore one expects the fraction
of orbits with large Liapunov exponents to be smaller in these models than in Model 1. 
This  is easier to see in Model 2 where, in addition, by virtue of its less concentrated
density distribution,  
many low order axisymmetric resonances 
are missing  and others are deep in the core and thus 
 have a limited effect. The Liapunov exponent greyshade diagrams for that model
are shown in Fig~\protect\ref{MOD2B}.
Again there are some orbits with large Liapunov exponents that are 
confined to the  disk plane, 
this effect however was found to be  less apparent in these models 
(especially in Model 3) than in Model 1 because of their lower asymmetry
and the smaller bar contribution --- causing  the motion in the plane to be more regular. 
Clearly however, the simple relationship between the Liapunov
exponent value and the $z$ excursion found in the runs in potentials
 without bars are not valid in any of the barred
 models discussed here. It is possible that evaluating
the whole spectrum of Liapunov exponents and adding up the positive ones 
(to obtain the KS entropy: e.g., LL)
may give a better approximation to the ergodic properties of the orbits.
Chaotic two dimensional orbits must conserve some quantity at least 
approximately. 
This slow diffusion rate should correspond to a low value in one of the
 positive exponents. However this has not been checked here since we only calculate
the maximal exponent.

\begin{figure}
\epsfig{file=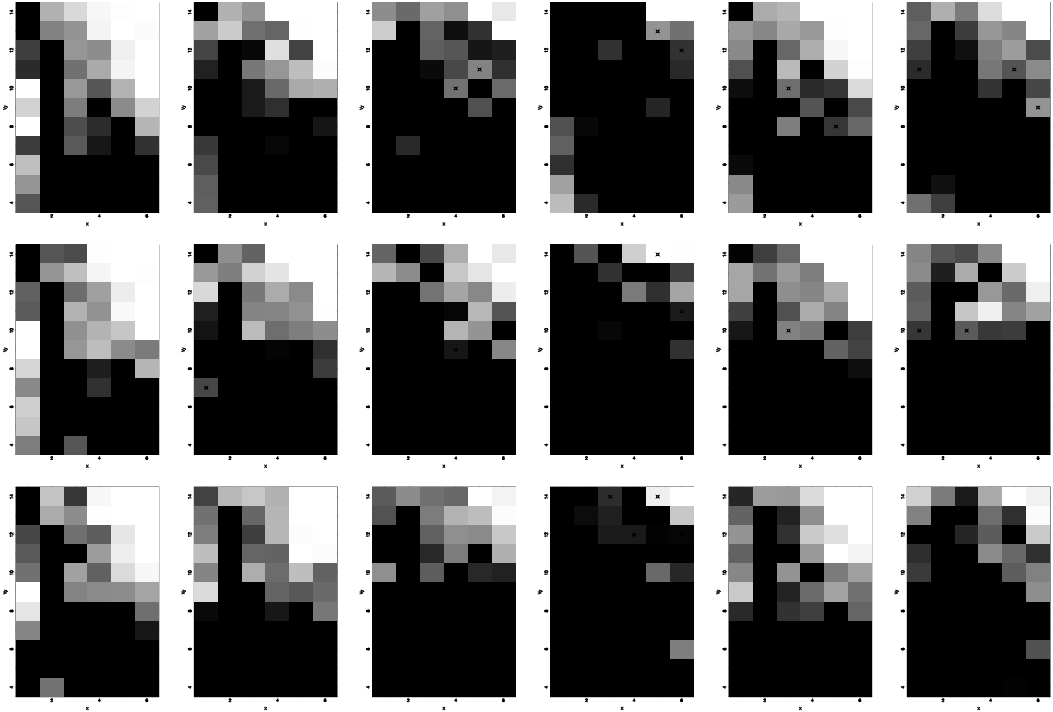,width=13.9cm,height=9.4cm,angle=-90}
\caption{\label{MOD1BR}
 Same as in Fig.~\protect\ref{MOD1} but when a homogeneous rectangular bar with the
same parameters as the Ferrers bar replaces it}
\end{figure}

The main processes described above are enhanced when one replaces the
Ferrers bar   with a 
homogeneous rectangular one. Namely, the maximal Liapunov exponents on
average become much larger but the correlation between the values
of these quantities and the maximum $z$ excursion becomes  worse.
The greyshade diagram for Model 1 with an added rectangular bar are shown 
in Fig.~\protect\ref{MOD1BR}.  Chaotic trajectories are found to be  more common and 
the exponentiation time-scales are very small (of the order of a dynamical time).  
This is of course because
of the much more asymmetric shape of the bar, which causes even larger
and more frequent large angle deflections than the ellipsoidal Ferrers
bar.  
The exceptions to this were  the runs where a spherical halo was
present and a central mass concentration is added.  
In this case, 
the potential is not time-dependent, thus there is no mechanism favouring instability
in the plane. 

Similar results to those described above were found for runs of Models
 2 and 3 with rectangular bars. However, as in the   Ferrers bar
case, the values of the Liapunov exponents for these models were 
generally found to be somewhat smaller and with  isolated regions of
stability more common. On the other hand, contrary to the Ferrers bar case, 
the discrepancy between the values of the Liapunov exponents and the $z$ excursions 
in Models 2 and 3 was 
found to be somewhat larger than in Model 1. 

\subsection{The effect of weak dissipative perturbations} 
\label{tax:disp} 

In addition to stars, the motion of which is  essentially determined by 
gravity, galaxies also contain a gaseous component.
Modelling the interstellar medium is not a trivial task. There
are various processes that are thought to take place simultaneously
in this medium. Among these are heating processes due to supernova
explosions, the degradation  of ordered motion through viscous processes 
and the inverse cooling processes which lead to heat being radiated away
(e.g., Spitzer 1978). These processes appear to act in a complex and
highly non-uniform manner, leading to several phases with very 
different properties (e.g., Norman \& Ikeuchi 1989; Clifford 1983).

One important characteristic that has to be taken into account when attempting
to model the interstellar medium is its highly clumpy and apparently
non-uniform nature. This property renders standard hydrodynamical treatments
based on the continuum approximation and a perfect gas equation inadequate,
since the assumption of local thermodynamic equilibrium stemming from 
a certain separation of ``fast'' and ``slow'' processes 
is no longer satisfied in a straightforward manner. In view of the above difficulty, an 
alternative approach based on the dynamics of
a collection of locally interacting ``sticky'' particles has been tried by
many authors (e.g., Combes \& Gerin 1985). In this scenario, particles move 
influenced only by gravity until they come close enough together when they
collide inelastically. Although such  a procedure might seem at first sight
artificial and somewhat trivial, it can actually be fairly rigorously
justified under certain conditions if some refinements are introduced.
The clumpy and apparently scale invariant nature of the gaseous interstellar
medium has also inspired attempts to treat it as a fractal object (Pfenniger
\& Combes 1994). This method still awaits detailed application to realistic situations. 

Some of the
issues mentioned above are discussed in a little more detail in Section~\protect\ref{tax:disin}.
We just mention here that there are indeed fast and slow processes which 
can be separated. These are the collision time of individual gas clouds,
 which is of order of 1
to 10 Myr, and their dynamical time which is much longer --- being of the
order of  60 to 600 Myr. This feature enables one to reconcile the 
two approaches (continuum approximation and sticky particle techniques) described above.
(Details of the grounds on which such an approach may be justified are given 
by Scalo \& Struck-Marcel 1984). 
In this new formulation, one considers the hydrodynamics of 
collections of gas clouds which interact with each other over time-scales
which are short compared to the dynamical time. A new set of hydrodynamic
equations  more appropriate to this situation is thus obtained. In these
equations a ``fluid element'' is therefore a region small enough so that
the macroscopic gravitational field can be considered  roughly constant  
while large enough to contain a fair sample of gas clouds. 

According to
Combes (1991), the total mass of molecular Hydrogen in the Milky Way, for
example, is about $2-3 \times 10^{9}$ solar masses concentrated in clouds
of  mass greater than $10^{5}$ solar masses.   At about 10 kpc
these are concentrated in a region of a hundred pc from the plane of the
disk. Therefore, the volume mentioned above would be of the order of, say,
500 by 500 by 100 pc, but of course will be smaller in the central areas
where the concentration of molecular hydrogen increases significantly (Combes 1991). 

Assuming that the hydrodynamic effects (e.g., pressure, viscosity etc.) are
small (i.e., gas clouds move primarily under the influence of gravitational forces)
and that any evolutionary effects are slow, one can imagine the full hydrodynamic
equations to be perturbations to the Collisionless Boltzmann Equation.
 This approach is similar in nature 
to the representation in Section~\protect\ref{tax:noise}  of discreteness noise in the $N$-body
problem as a perturbation to single particle trajectories. Here we will follow
Pfenniger \& Norman (1990; hereafter PN)  and assume that in the context discussed above,
the  main hydrodynamic effect can be approximated to first order as a velocity
dependent viscous force, the value of which is fairly small in comparison to 
the value of the mean gravitational field. In PN, dissipative perturbations
of the form
\begin{equation}
F_{fric}= - \gamma v {\bf v},
\label{tax:difor}
\end{equation}
where  $\gamma$ is a constant which determines the strength of the dissipation,
were used. 
This form is similar to the form  obtained if one  assumes that the 
individual clouds are composed of infinitely compressible isothermal
gas. Then in each collision  the acceleration
during an encounter of two clouds is given  (e.g., Quinn 1991) by 
\begin{equation}
a_{en} \sim u^{2}/r,
\end{equation}
where $u$ is the absolute value of the relative velocity of the clouds and
$r$ is their separation. 

One can, to first order, estimate $\gamma$ using the usual ``mean free path''
approximation  of (linear) viscosity:
\begin{equation}
\gamma= n \lambda v_{rms},
\end{equation} 
where $n$ is the volume number density of gas clouds, $\lambda$ the mean free path and
$v_{rms}$ is the average random velocity. Supposing that the volume density is
of the order of a few hundred or so per cubic kpc and the random velocity of the order
of a few parts per hundred kpc/Myr, so that the mean free path is also of that
order if the mean free time is about 1-10 Myr. This gives a value $\gamma  \sim 0.01-0.1$ . 
Following PN we  use  more conservative values
for $\gamma$ and see that even these can have highly non-trivial effects.

\subsubsection{\bf Dissipation in disk-halo systems}
\label{tax:dispdisk}

Using a value of $\gamma=0.005$ we have integrated the equations of motion 
of dissipative particles moving in potentials with disk and triaxial halo contributions.
 The results are easier to interpret (and the figures
in general tidier) in the case when a halo alone is present, although the situation
is qualitatively the same when an additional disk contribution
is also present (as long as significant asymmetry remains of course). 
We  stick here to the case $GM_{D}=0$ and the 
other parameters being those of Model 2 with $c/a=0.8$ and $b/a=0.9$.
 All trajectories studied start on the $x$-axis
with the
local rotation velocity in the azimuthally averaged potential.

\begin{figure}
\begin{center}
\epsfig{file=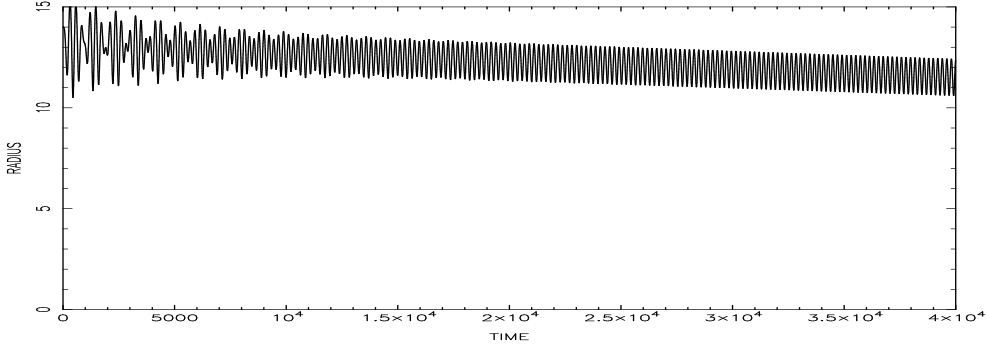,width=13.0cm,height=4.5cm,angle=-90}
\end{center}
\caption{\label{dispq}
 Evolution of radial coordinate of trajectory starting from an initial value of 14 kpc}
\end{figure}

The quantity ${\bf v}$ in Eq.~(\protect\ref{tax:difor}) is taken to be the 
vector difference 
between the velocity of the test particle and that of a particle moving with the
local circular velocity.  Such motion cannot exist because in a
non-axisymmetric potential the orbits with the smallest radial departure from
circular motion are ovally distorted closed loop orbits. Because the dissipation 
rate is small however, far from the core where these orbits are fairly
close to circular, the difference in velocities is fairly small and 
trajectories can oscillate around the closed loops for very long times (Fig.~\protect\ref{dispq}). 
This means that, effectively, our dissipative point particles representing
the fluid elements of gaseous disks,
settle into
regular quasiperiodic ``limit cycles''.  In the case where  we would have calculated
$\bf v$ in terms of the difference between a particle's trajectory and the 
local periodic loop orbit, we would obtain a true limit cycle as the
asymptotic {\em attractor} (an excellent review of  concepts related
to chaos in dissipative systems is  given by Eckmann \& Ruelle 1985).

\begin{figure}
\begin{center}
\epsfig{file=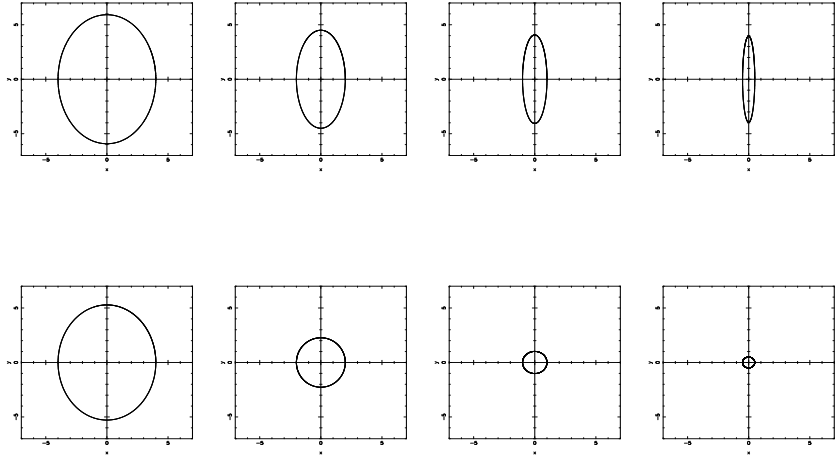,width=11.0cm,height=6.0cm,angle=-90}
\end{center}
\caption{\label{PEDISP}
 Sequence of closed loop orbits. Top: When no central mass is present. Bottom:
In the presence of a central mass $GM_{C}=0.01$}
\end{figure}

The situation is rather different however as one moves nearer the 
halo core. In this case,
the closed loop orbits become more and more eccentric and cannot parent any
trajectories that spend all their time inside the core (Fig.~\protect\ref{PEDISP}).
 One can then 
say that the above attractor ceases to exist, leaving only the center as
a stable attractor for trajectories inside the core.

\begin{figure}
\begin{center}
\epsfig{file=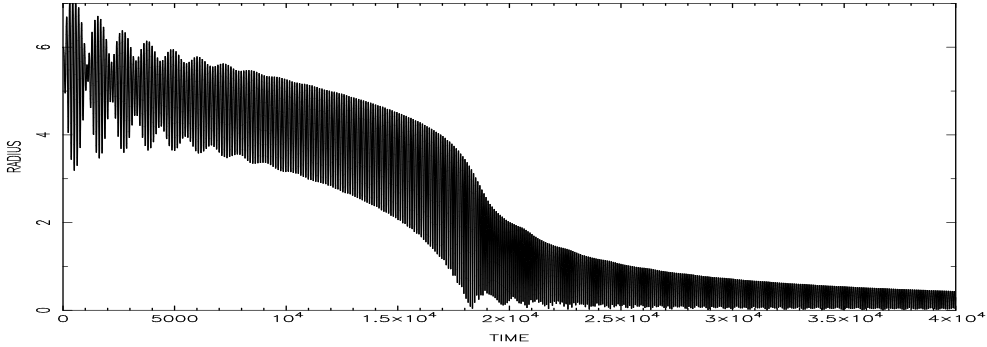,width=13.0cm,height=4.5cm,angle=-90}
\end{center}
\caption{\label{dispa}
 Evolution of radial coordinate of trajectory starting from an initial value of 6 kpc}
\end{figure}

In Fig.~\protect\ref{dispa} we plot the radial coordinate of a trajectory 
starting at 6 kpc. 
As can be seen, as one moves closer and closer towards the centre,
the dispersion in the radial coordinate of the particle increases as the local
loop orbits become more and more eccentric. At the separatrix 
 beyond which
no  loop general loop orbits can exist (which occurs at about 4 kpc),
there is a sharp transition in the dissipation rate and the particle quickly moves towards the centre.

\begin{figure}
\epsfig{file=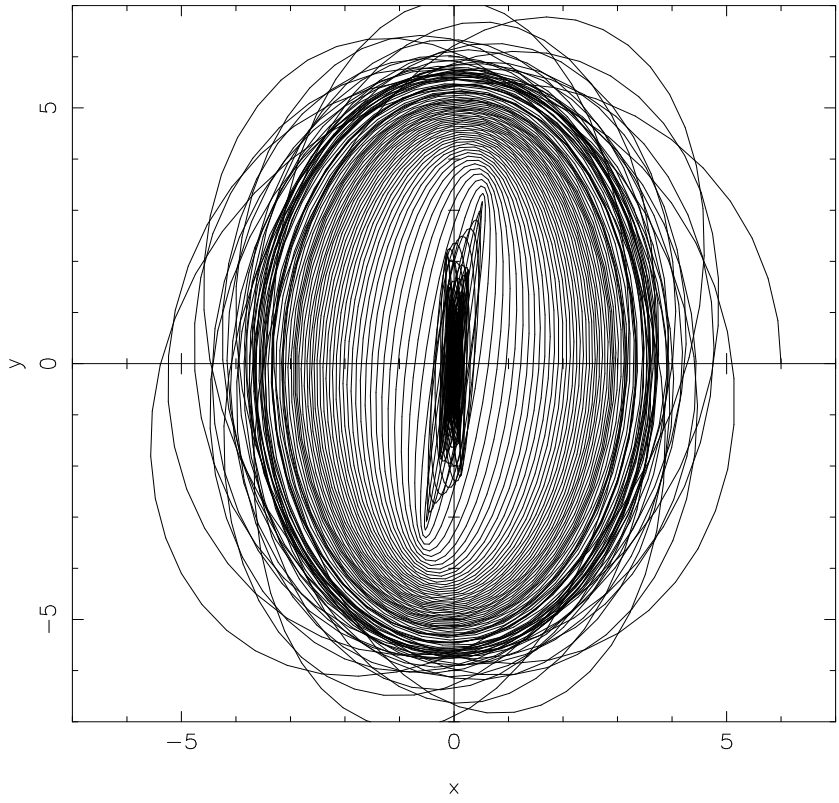,width=11.0cm,height=10.5cm,angle=-90}

\vspace{1cm}

\epsfig{file=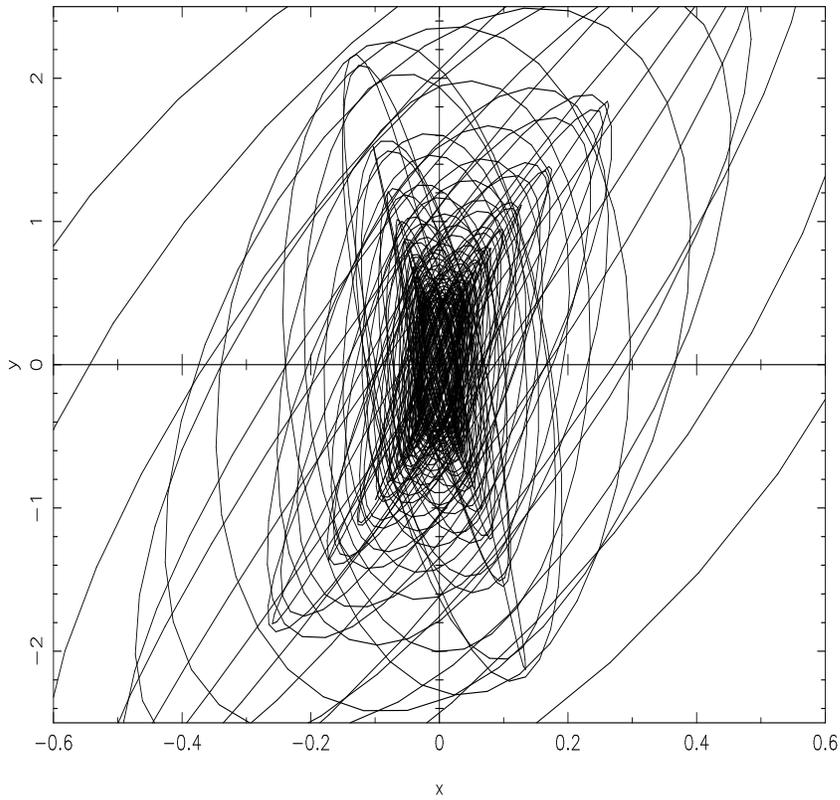,width=11.0cm,height=10.5cm,angle=-90}
\caption{ \label{dispaxy}
 Trajectory in the $x-y$ plane of orbit with radial coordinate represented in Fig.~\protect\ref{dispa}
and a more detailed view of the segment of the trajectory from 1750 Myr 
to end of the run at 50 000 Myr (bottom)}
\end{figure}

Fig.~\protect\ref{dispaxy} displays the spatial evolution of this trajectory,
 which shows it to follow a
sequence of closed loop orbits followed by what essentially are sections of box orbits.
This latter behaviour, which starts roughly at about 16 000 Myr and increases
the dissipation rate dramatically, is shown in the plot on the right hand side
of this figure. 
This process of course will lead to the growth of central mass concentrations,
which as we have seen, can be important in determining the orbital structure.
However it appears that it slows down significantly as the central mass
increases. The central mass ruins the harmonic nature of the potential
core which also becomes  more symmetric in the inner regions so that the closed
loop orbits, which now exist in the central areas, 
are close to circular (Fig.~\protect\ref{PEDISP}). The extent of the region 
where the potential is strongly affected by the presence of the central mass
will of course depend depend on its strength.
If the central mass is strong enough, this will be the whole harmonic core
and nearly circular loop orbits could exist everywhere in the central areas.

\begin{figure}
\begin{center}
\epsfig{file=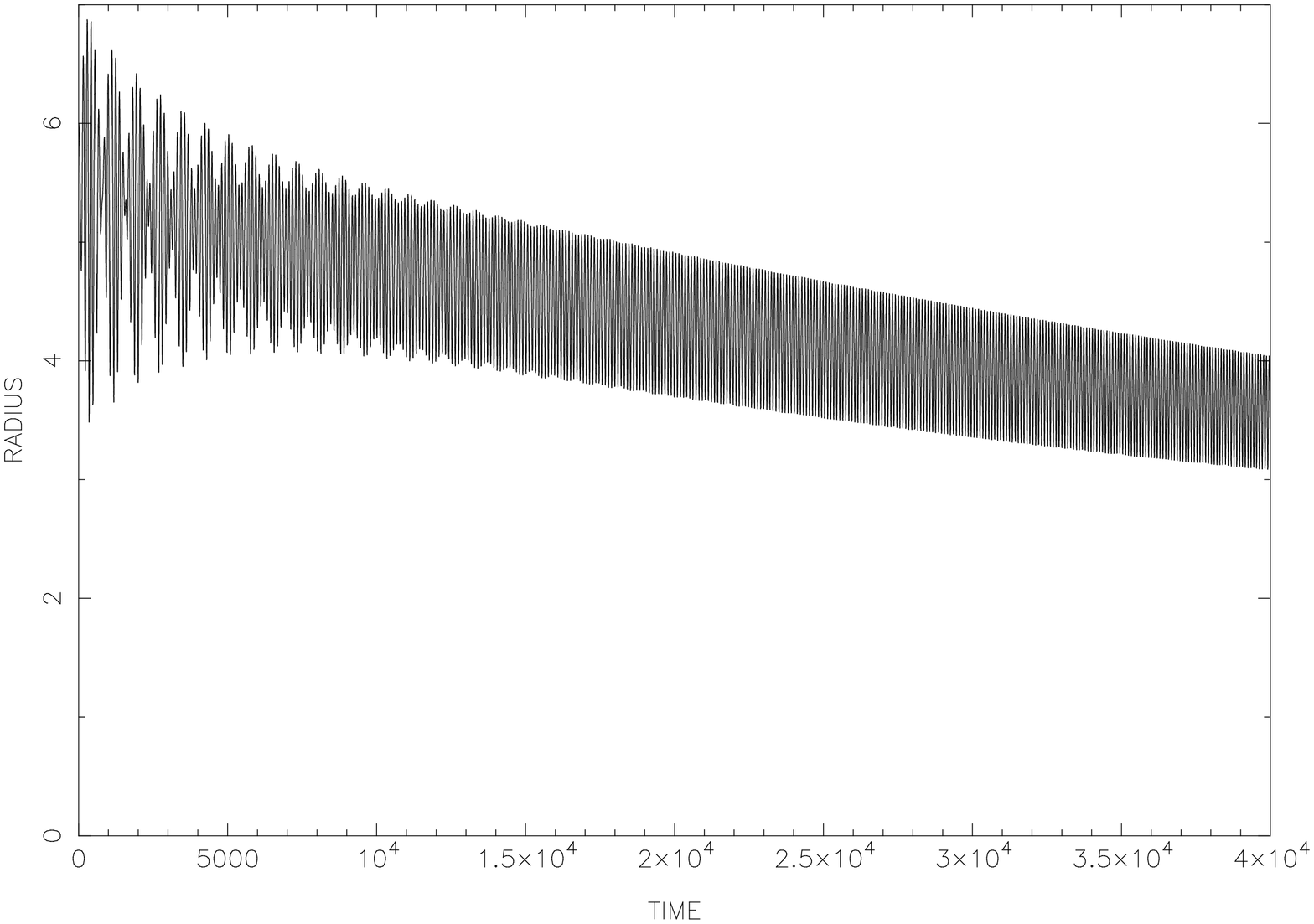,width=13.0cm,height=4.5cm,angle=-90}

\vspace{1.1cm}

\epsfig{file=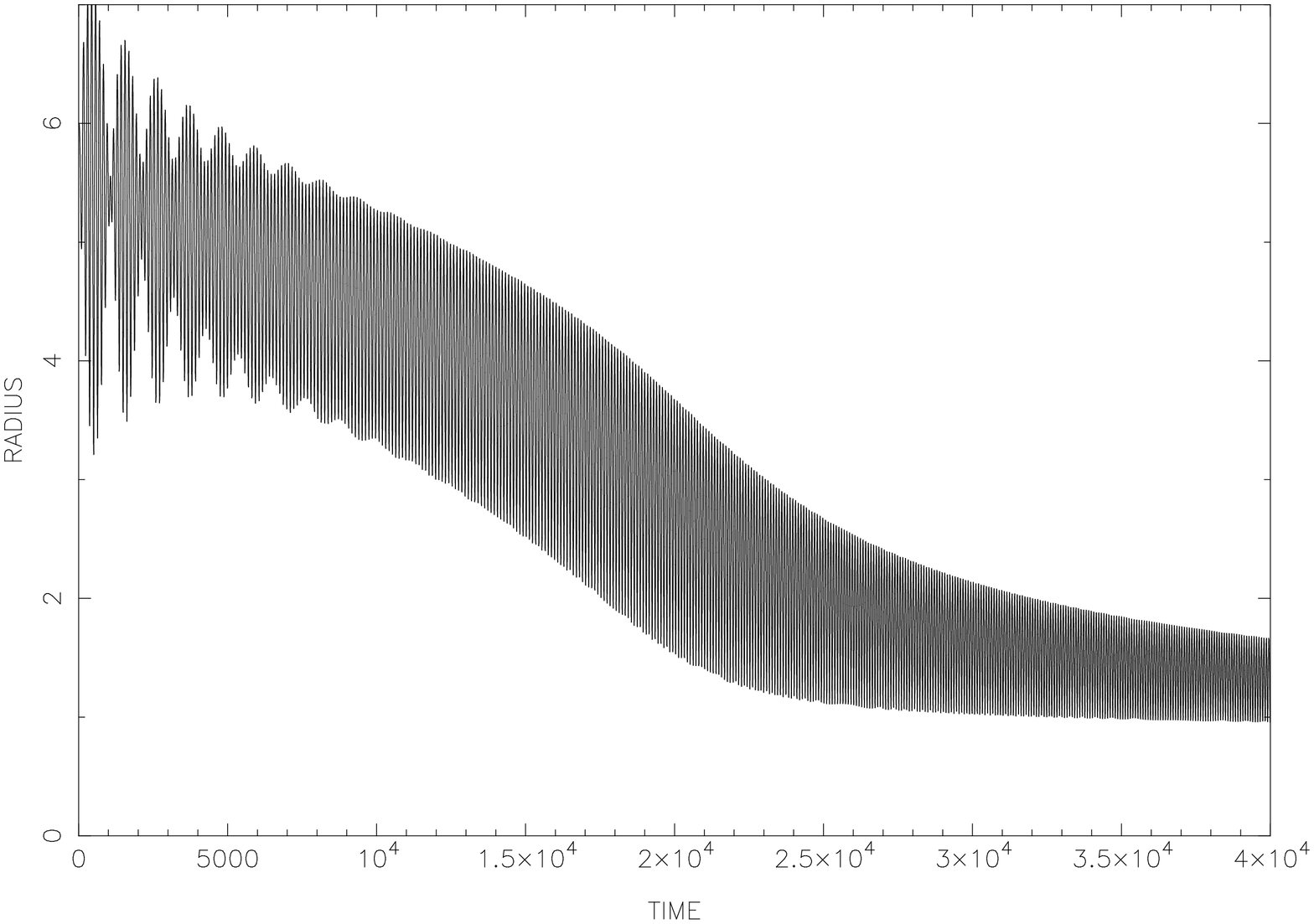,width=13.0cm,height=4.5cm,angle=-90}

\vspace{1.1cm}

\epsfig{file=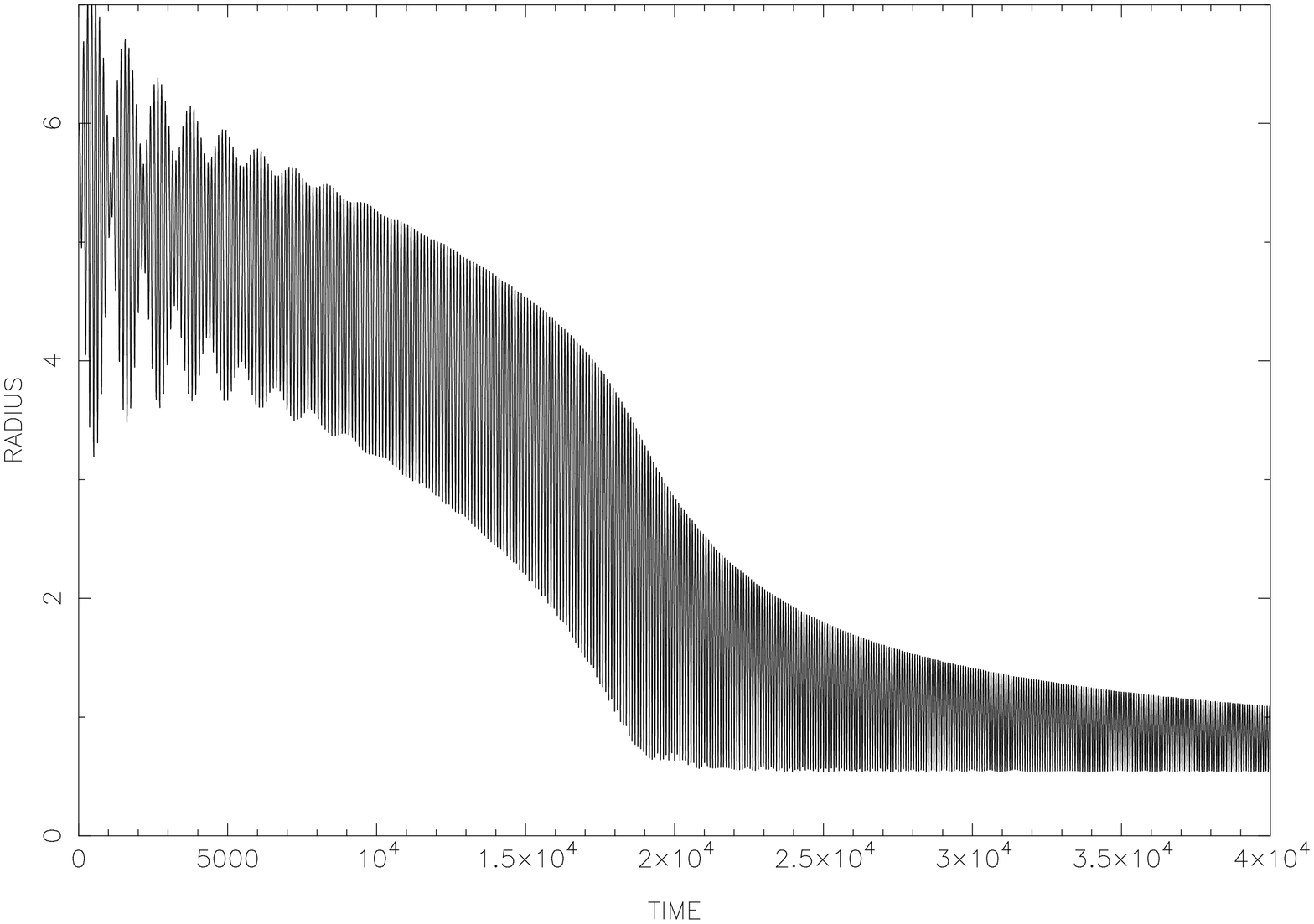,width=13.0cm,height=4.5cm,angle=-90}

\vspace{1.1cm}

\epsfig{file=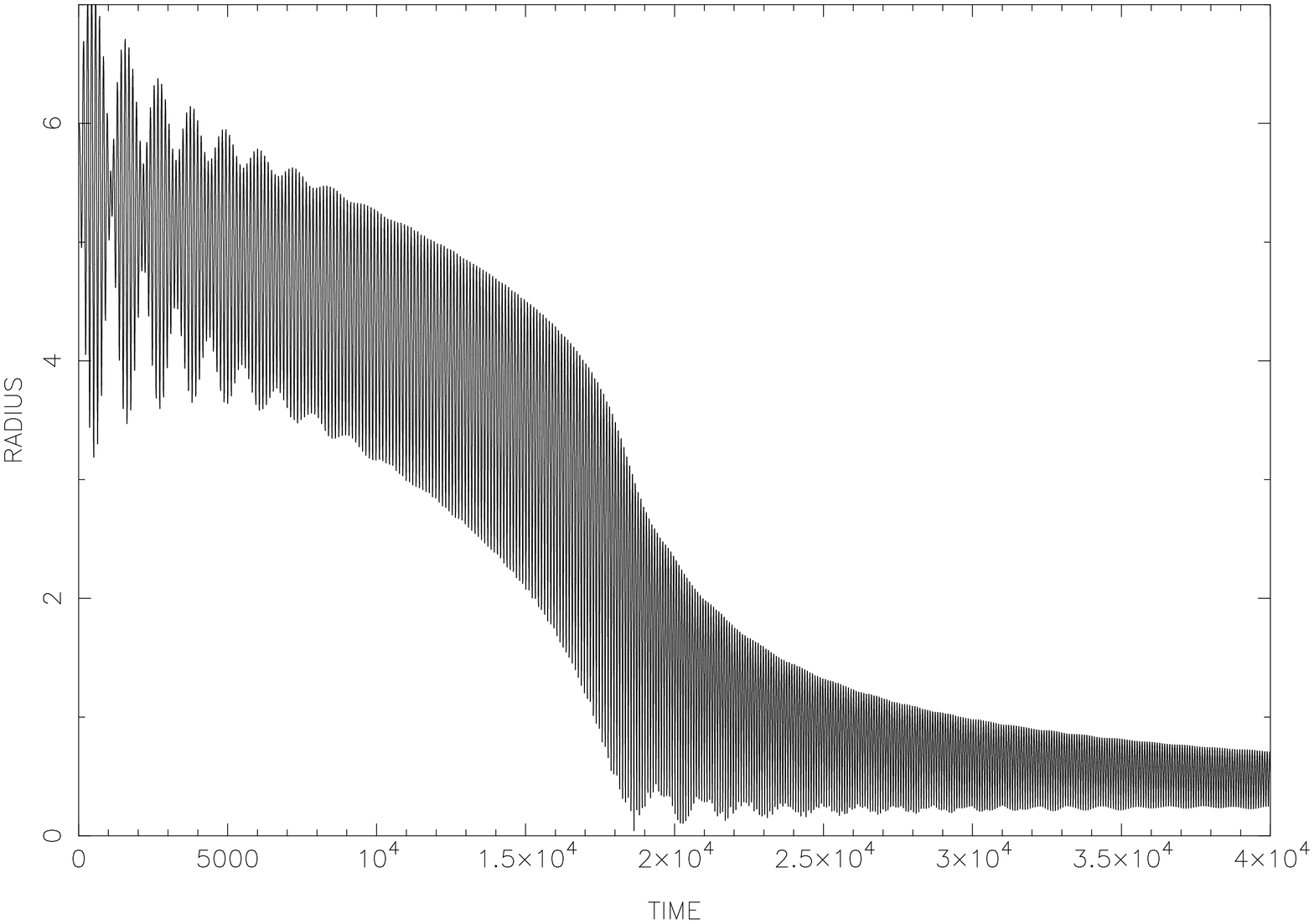,width=13.0cm,height=4.5cm,angle=-90}

\end{center}
\caption{\label{dispc}
 Same as in Fig.~\protect\ref{dispa} but with a central mass of (from top to
bottom)
$GM_{C}=0.01$, $GM_{C}=0.0005$, $GM_{C}=0.0001$, $GM_{C}=0.00001$}
\end{figure}

The top diagram in Fig.~\protect\ref{dispc} shows the evolution of the radial coordinate
for a trajectory 
starting from the same initial conditions as in
Fig.~\protect\ref{dispa} and exactly the same potential except that a central mass of $GM_{c}=0.01$
is present. Clearly, the behaviour previously observed --- that is the accelerated 
dissipation rate --- is no longer present. 
For very small central masses (anything
greater than $GM_{C} \sim 0.00001$), loop orbits exist deep inside the core. 
Subsequent plots of Fig.~\protect\ref{dispc} 
clearly show that there is a certain region, between the halo core radius and the centre,
where the trajectories are extremely 
eccentric and the dissipation is very large. 
The scope of this region increases with decreasing central mass.
 Orbits completely confined in this annulus cannot 
be parented by the closed loops, 
 therefore they  can either
oscillate around the variety of boxes and boxlets available or move chaotically.
This causes the large  increase in the dissipation rate. If the central mass is strong 
enough,
closed loops exist everywhere and so this region does not exist --- which means
that  the dissipation rate is always small. {\em Thus there is a threshold 
for the central mass concentration beyond
which accretion completely stops --- and does not appear to be above a few tenths of
a percent of the relevant total mass} (e.g., within 10-20 kpc).

It is also possible for dissipative orbits to be $z$ unstable. However, given
the fact that the instability requires significant central mass and
 that in this case trajectories settle down on closed loops, few orbits
are $z$ unstable unless the dissipation rate is very small. Nevertheless, we expect 
the process of accretion of material to the central region  (in a realistic
situation) to be accompanied by shocks which may cause
 much of the gas to be transformed into stars (e.g., Pfenniger 1993). These 
will be born on eccentric trajectories which will usually be vertically 
unstable as we have seen in Section~\protect\ref{dishac}. As the central mass increases, one
expects this to take place at  progressively larger radii    
(as the accretion  from the innermost region stops and  trajectories 
with higher energy  are destabilised by the growing central concentration).
 Once the harmonic core is completely destroyed,
the accretion ceases completely, and all box orbits within the core are also
destroyed. One therefore expects the formation
of a bulge-like structure, the extent of which is similar to that of the
original harmonic core, and which is formed ``inside out'' 
from layers of unstable trajectories. Since however the inflow of mass is gradual 
and saturates before a situation is reached  whereas most box orbits are unstable
over short time-scales (cf., Section~\protect\ref{dishac} Merritt \& Quinlan 1997)
the above processes are expected to be slow. In particular, a significant bulge may
form before the halo loses triaxiality (if at all).

\subsubsection{\bf Dissipation in systems with rotating bars}

When the halo is axisymmetric, the effect of dissipation is
 analogous to the models studied by PN --- in regions  where stable
$x_{1}$ periodic  orbits exist, the dissipative trajectories mimic these,
however in the case the $x_{1}$ are unstable, the dissipative trajectories 
are quickly brought inwards, a process that is accompanied by (modest) $z$ instability.
The addition of a central mass brings about large instability strips
in the $x_{1}$ family, especially in the central areas (PN). (This
 is in contrast to the case when no bar is present, when the
central mass actually {\em stabilizes} the closed loop orbits!) 
The effect of orbit dissipation  therefore  becomes much more dramatic
with an attracting point  at the centre quickly reached.
 Varying the  flattening  of the halo
has the only effect of suppressing the $z$ instabilities even further (since the vertical
resonances are moved in).

\begin{figure}
\begin{center}
\epsfig{file=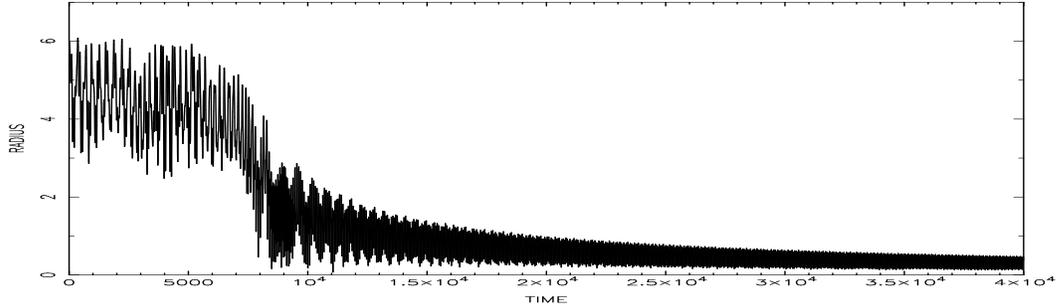,width=13.9cm,height=4.0cm,angle=-90}
\end{center}
\caption{\label{dispr}
 Trajectory started from edge of the bar in Model 2 with axis rations $b/a=0.8$
and $c/a=0.7$}
\end{figure}

\vspace{1.5cm}

\begin{figure}
\epsfig{file=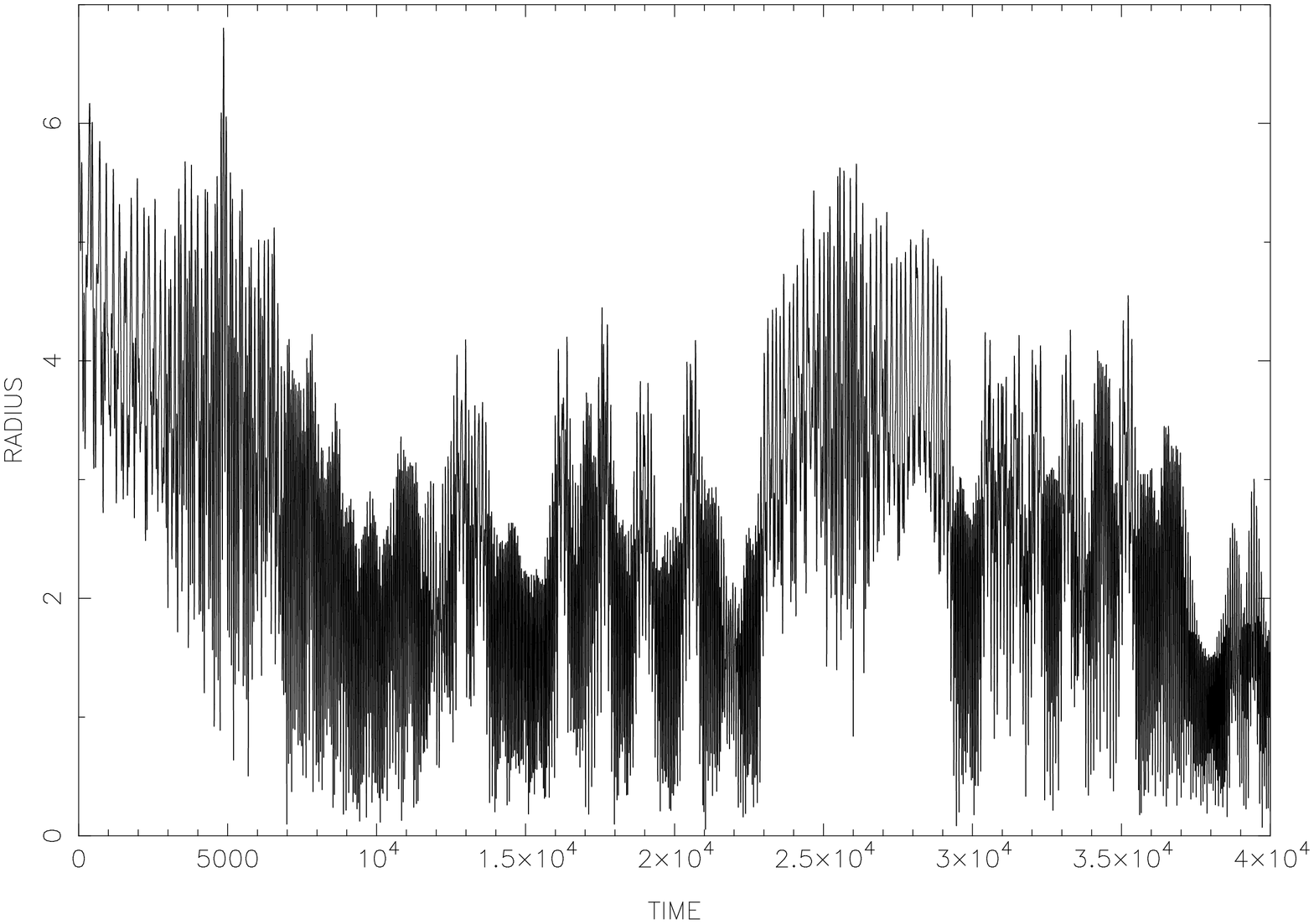,width=6.7cm,height=4.0cm,angle=-90}
\hspace{0.2cm}
\epsfig{file=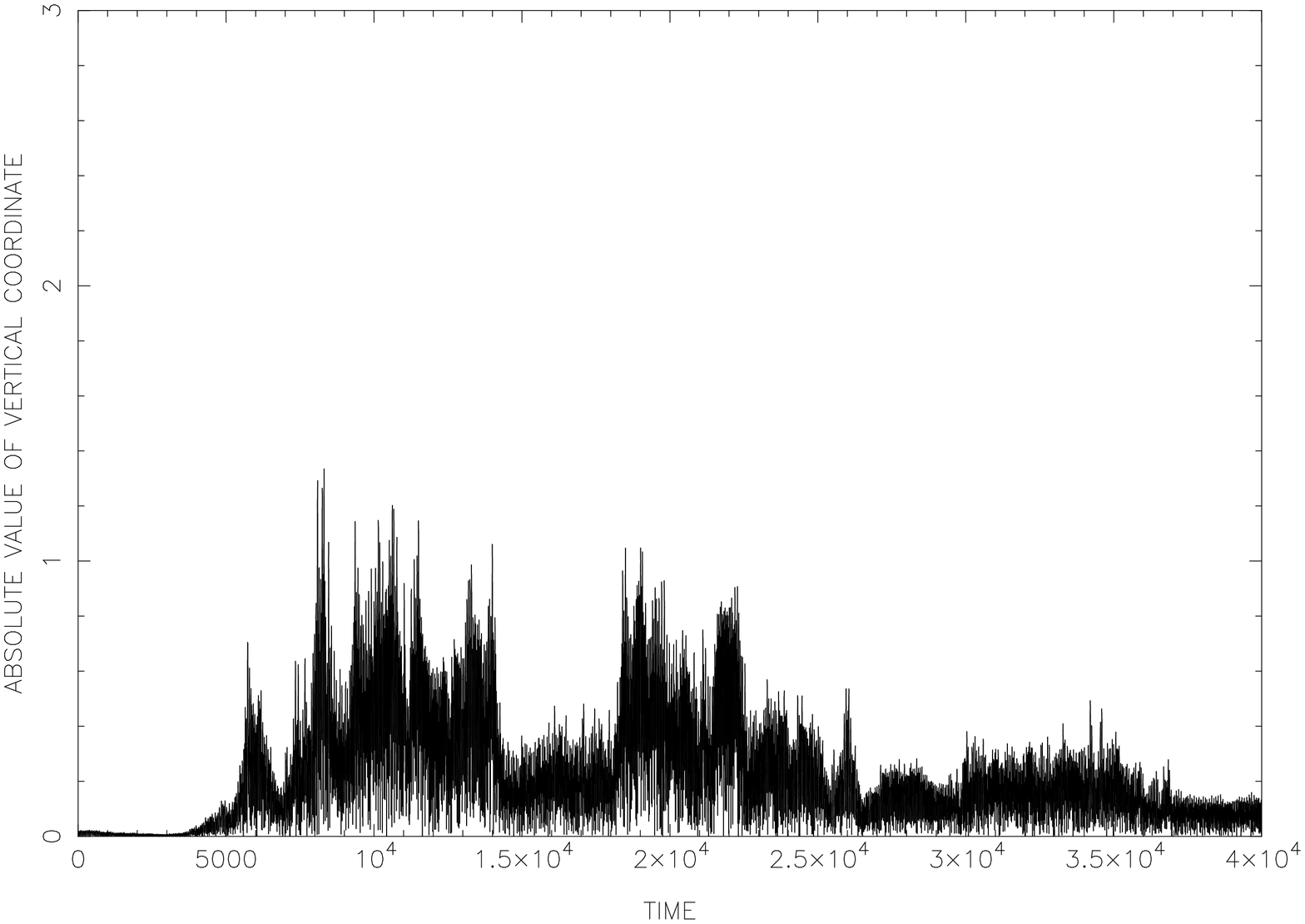,width=6.7cm,height=4.0cm,angle=-90}
\caption{ \label{dispw}
Left: Same as in Fig.~\protect\ref{dispr} but with $\gamma=0.0005$. 
Right: Corresponding evolution 
of the absolute value of the $z$ coordinate}
\end{figure}

\vspace{1.5cm}

\begin{figure}
\epsfig{file=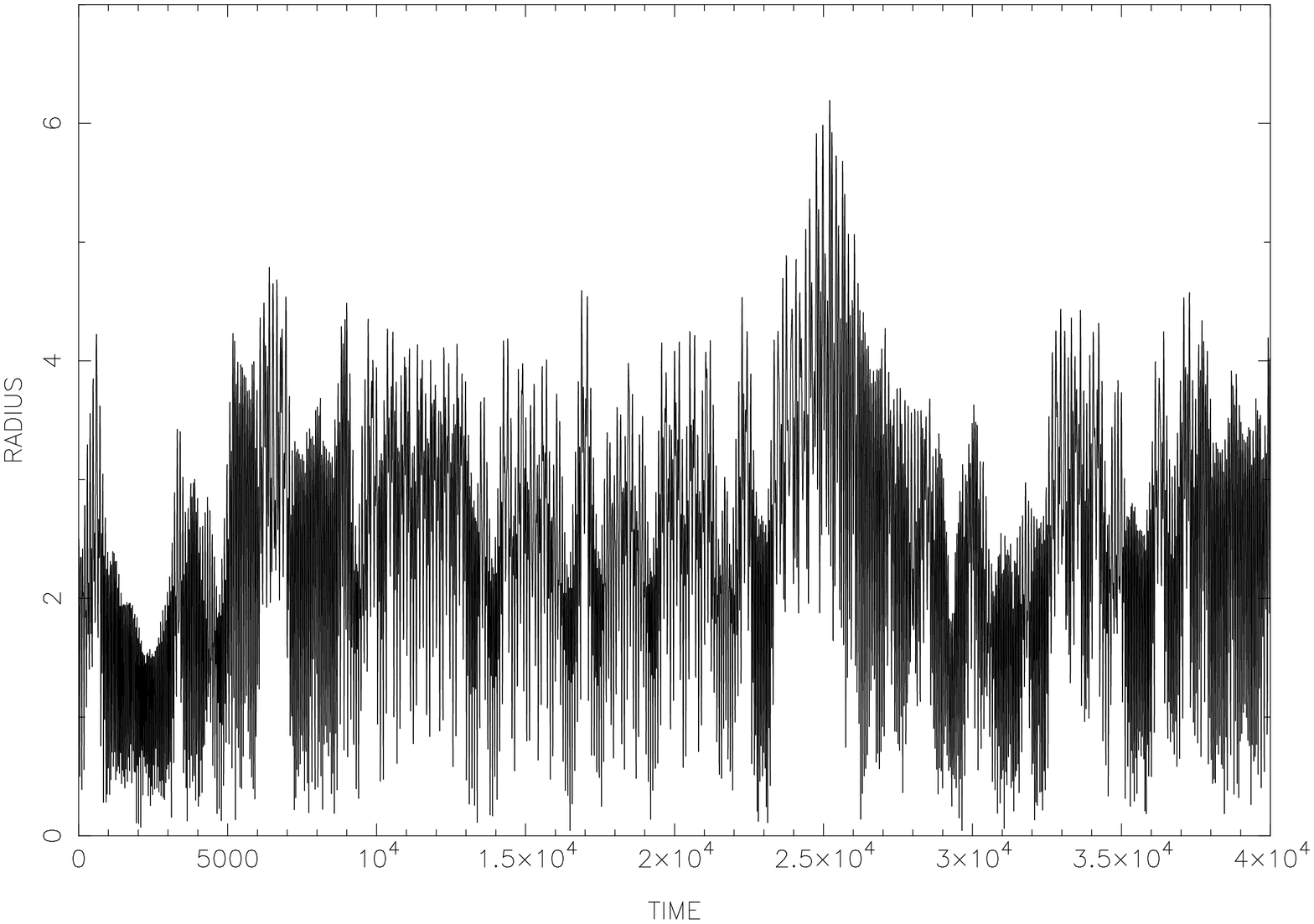,width=6.7cm,height=4.0cm,angle=-90}
\hspace{0.2cm}
\epsfig{file=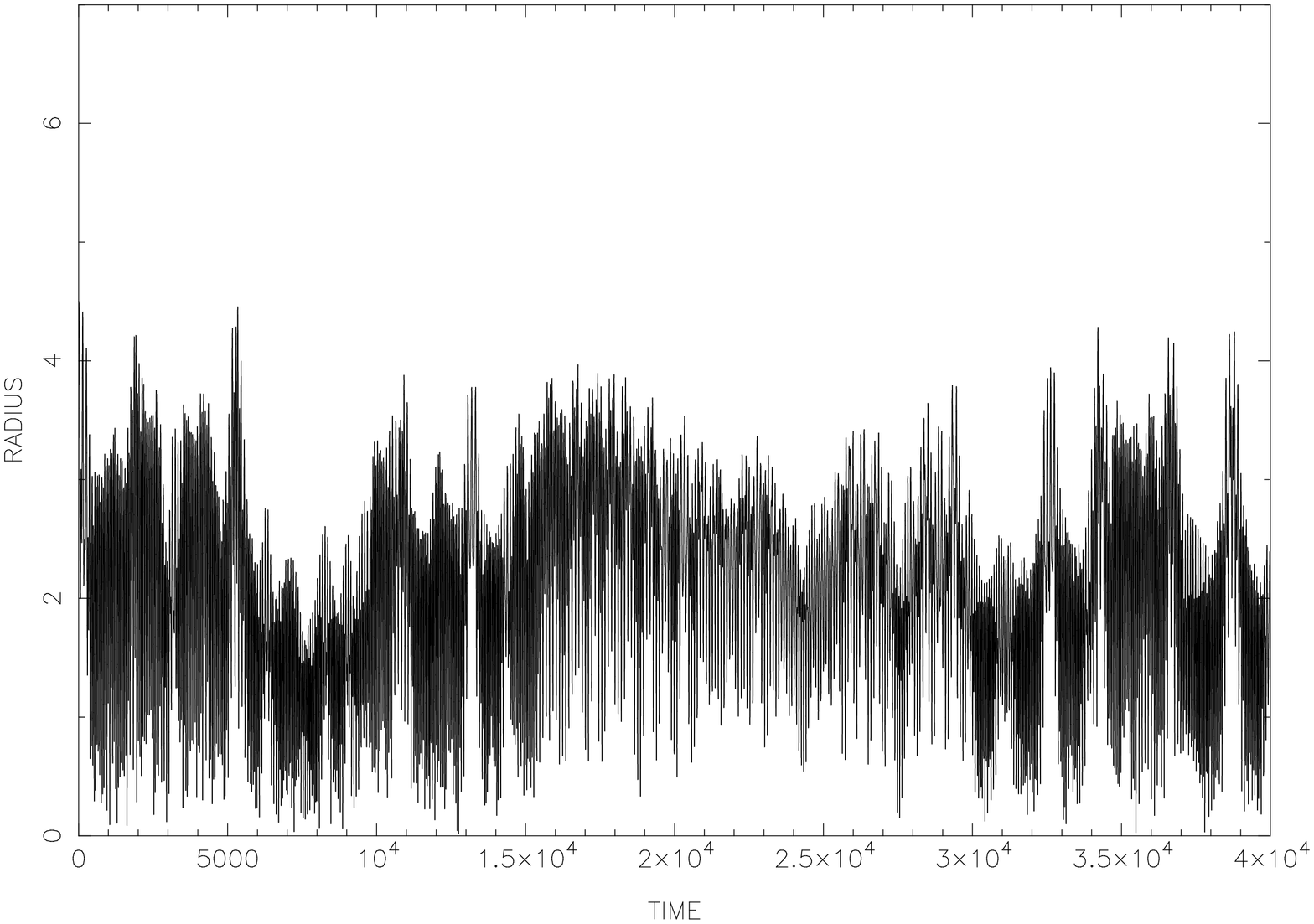,width=6.7cm,height=4.0cm,angle=-90}
\caption{ \label{dispu}
 Evolution of radial  coordinates for trajectories starting at a 
only 2.5 kpc (left) and 4.5 kpc from centre of the mass distribution but appearing
 to end up in the same attractor as the trajectory in Fig.~\protect\ref{dispw}.
Here also $\gamma=0.0005$}
\end{figure}

We focus on the case where the halo is non-axisymmetric. 
We will concentrate  on trajectories of  Model 2 which has its rotation
curve rising over the extent of the bar's major axis (an
observed property of bars: e.g., Pfenniger 1984a) and a large axisymmetric
disk contribution.  
Indeed, this model appeared in the Hamiltonian limit to
have a more regular phase space than the other two models --- thus suggesting
that it may be possible for a  bar to survive long enough so that its effects on the gas dynamics
be important (see Section~\protect\ref{tax:rapbars} above).
We fix the halo potential axis ratio 
in the disk plane at $b/a=0.8$ while normal to the plane of the
disk we take  the axis ratio to be $c/a=0.7$. 

Fig.~\protect\ref{dispr} shows the time evolution
of the radial coordinate of a particle  started from  the edge of the bar with
the velocity of a circular orbit in the azimuthally averaged potential.
As can be seen from that figure, even trajectories 
starting  this far out from the centre of the
potential can be transported there. It can be noted however that after the
initial rapid decay in the radial coordinate, the trajectory settles down 
to a state where this coordinate oscillates about a value of roughly 0.5 kpc  with little
further decay. It is not obvious  whether this is a regular limit cycle or a strange
attractor. 
In principle this can be checked  by calculating the Liapunov exponents for 
example, though the difference here is of no great practical consequence. 
However, when 
 the dissipation was decreased  (by lowering the value of  $\gamma$
to $\gamma=0.0005$), it was found that some  end  states 
can be seen to be  highly chaotic even by simple inspection. Fig.~\protect\ref{dispw} displays
the radial coordinate time series of such an orbit. This behaviour is also usually
accompanied by non-negligible excursions in the vertical direction which
 are suppressed by high dissipation rates. It is also interesting to note that 
trajectories  starting from as small an initial radial coordinate as 2.5 kpc 
appear to end up in this attractor, at least as far as
the radial coordinate behaviour is concerned (Fig.~\protect\ref{dispu}).

In the presence of a stronger bar and in the absence of dissipation
(e.g., $GM_{B}=0.04$), it was found that many of the 
trajectories starting near the end of bar escape in the Hamiltonian limit. 
In the presence of significant dissipation however most of these trajectories end
up in stable attractors {\em outside} the bar.

\subsection{Gas dynamics}
\label{tax:disin}

The dissipative point particle approach used above may give some indication
about the effect of dissipation but is certainly not suitable for detailed work.
The best numerical approach to modelling  clumpy structures such as gas cloud 
systems in galaxies is the kinetic approach. 

The gas in galaxies moves mainly under the influence of gravity. However
when gas clouds, or groups  of gas clouds, come within a certain distance of
each other,  non-gravitational forces such as viscous and pressure forces act. Pressure forces
act to scatter the clouds off each other, while viscous forces act to cause
dissipation (only the latter  were considered in the point particle case). 
Thus we have the standard weak fluctuation dissipation behaviour described 
by the Fokker-Planck Equation. The Fokker Planck equation is
a partial differential equation of seven  variables  and is thus not easily solved 
directly in the general case. However, one can set up numerical experiments that mimick
the behaviour expected of the evolution ruled by this equation. 
A simple algorithm displaying the Fokker Planck kinetic behaviour was
developed by Hoogerbrugge \&  Koelman  (1992). We describe it here in language
 adapted to our situation (instead of that used in conjunction with laboratory
systems for which the method was developed).

In the aforementioned algorithm, particles move in the smoothed out galaxy
potential for a time interval $\Delta t$ after which a search is made to identify 
the nearby neighbours $j$ of each of the particles. At that point, the  velocity
of the $i^{th}$ particle is updated according to
\begin{equation}
\dot{\bf x}_{i}=\dot{\bf x}_{i} + \sum \Omega_{ij} {\bf e}_{ij},
\end{equation}
where 
${\bf e}_{ij}=({\bf x}_{i}-{\bf x}_{j})/\parallel {\bf x}_{i}-{\bf x}_{j} \parallel$.
Linear and angular momentum conservation require that  $\Omega_{ij}=\Omega_{ji}$,
and that it is a scalar function of the relative positions and velocities
of the particles. A particularly simple function satisfying these conditions
is a direct modification of the one used by the original authors, but facilitating
 direct comparison with our dissipative point particle simulations and conditions
in the interstellar medium~(Section~\protect\ref{tax:dispdisk})
\begin{equation}
\Omega_{ij}= \Theta_{ij}  \Bl \Pi_{ij} - \gamma_{p} 
 \parallel \dot{\bf x}_{i} - \dot{\bf x}_{j} \parallel^{2} {\bf e}_{ij} \Br,
\end{equation}
where $\Theta=1$ when the separation between the two particles is less than
a certain radius $r_{int}$ and zero otherwise. $\gamma_{p}$ determines the strength
of the dissipation. $\Pi$ is sampled from a Gaussian distribution with mean $\Pi_{0}$
and dispersion $\delta \Pi$. Given that these collisional processes are fast compared 
to a dynamical time, we expect
that hydrodynamical behaviour is recovered on large scales. Indeed, it can be rigorously
shown that this is what the algorithm reproduces when the above condition is satisfied
 (Espagnol 1995). In this case we have a stable local thermodynamic
equilibrium if the  ``generalized Einstein relation'' (e.g., Klimontovich 1994)
is satisfied. In our case this means
\begin{equation}
\Pi_{0}=\gamma_{p}  \parallel \dot{\bf x}_{i} - \dot{\bf x}_{j} \parallel
\Bl \langle   \parallel \dot{\bf x }_{j} \parallel^{2} \rangle - 
\Bl \langle \dot{\bf x}_{j} \rangle \Br^{2} \Br,
\label{tax:ein}
\end{equation}
where the local mean  squared velocity  is averaged over the 
neighbours interacting with particle $i$. 

In practice the right hand side of~(\protect\ref{tax:ein}) should be multiplied
by a factor smaller than $1$ to have predominantly dissipative behaviour.
In the simulations we now describe this was taken as 0.9 . Also, since we use a 
relatively small number of particles (500), we just use the heuristic relation
\begin{equation}
\Pi_{0}=\gamma_{p} \parallel \dot{\bf x}_{i} - \dot{\bf x}_{j} \parallel^{2},
\end{equation} 
which makes the fluctuation term of similar form to the dissipation term.
In this case the dissipation term of Eq.~\protect\ref{tax:difor} is related to
$\gamma_{p}$ simply by
\begin{equation}
\gamma \times \Delta t \sim  (1 - 0.9) \gamma_{p} \times \sqrt{N_{int}}. 
\end{equation}
The interaction range (which scales as $\sim \sqrt{N}$ in a flat system)
was also taken to be rather large to make up for the 
small number of particles: $r_{int}=0.5$ kpc . 
There were   roughly 
$N_{int}=4-8$ particles within that radius for each particle. $\Delta t$ was taken to be
5 Myr. We therefore take a value of $\gamma_{p}=0.12$ to match the value $\gamma=0.005$
usually adopted in Section~\protect\ref{tax:disp}. Finally the dispersion of the Gaussian
distribution was taken as $\delta \Pi =0.2 \Pi_{0}$.

\begin{figure}
\epsfig{file=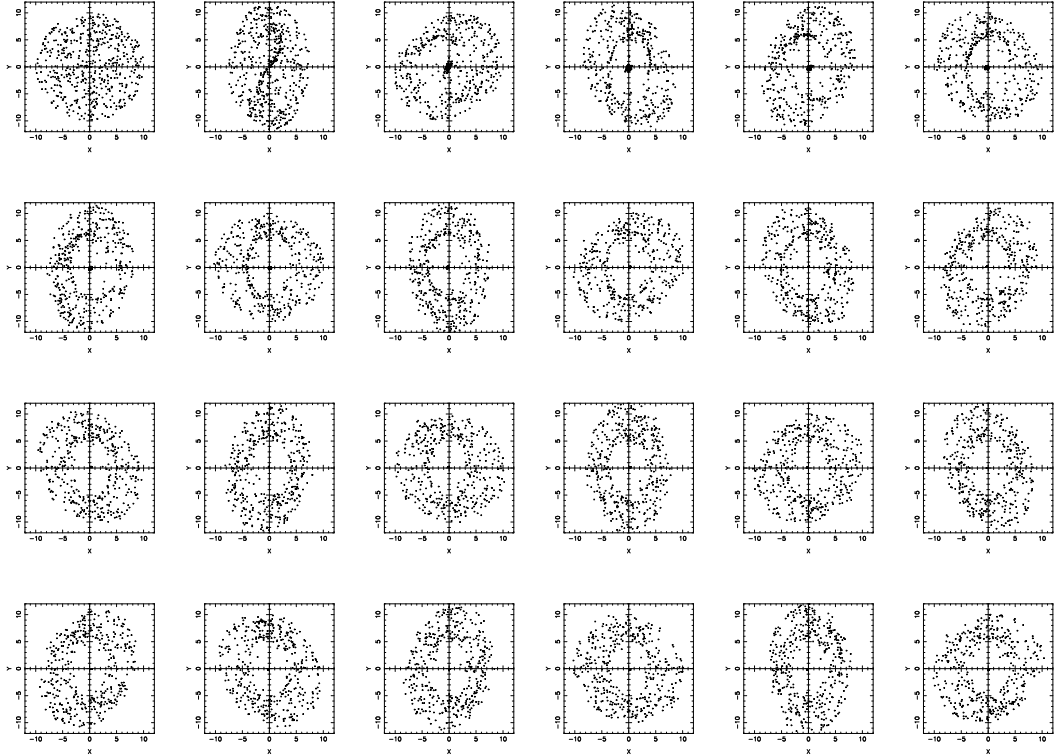,width=13.9cm,height=10.0cm,angle=-90}
\caption{\label{stick1}
$x-y$ projection of spatial evolution of Fokker-Planck type sticky particle system
in  model corresponding to the plots in 
Figs.~\protect\ref{dispq} and \protect\ref{dispa}. Snapshots are shown at intervals of 500 Myr 
(in sequence from left to right and from top down)  }
\end{figure}

\begin{figure}
\epsfig{file=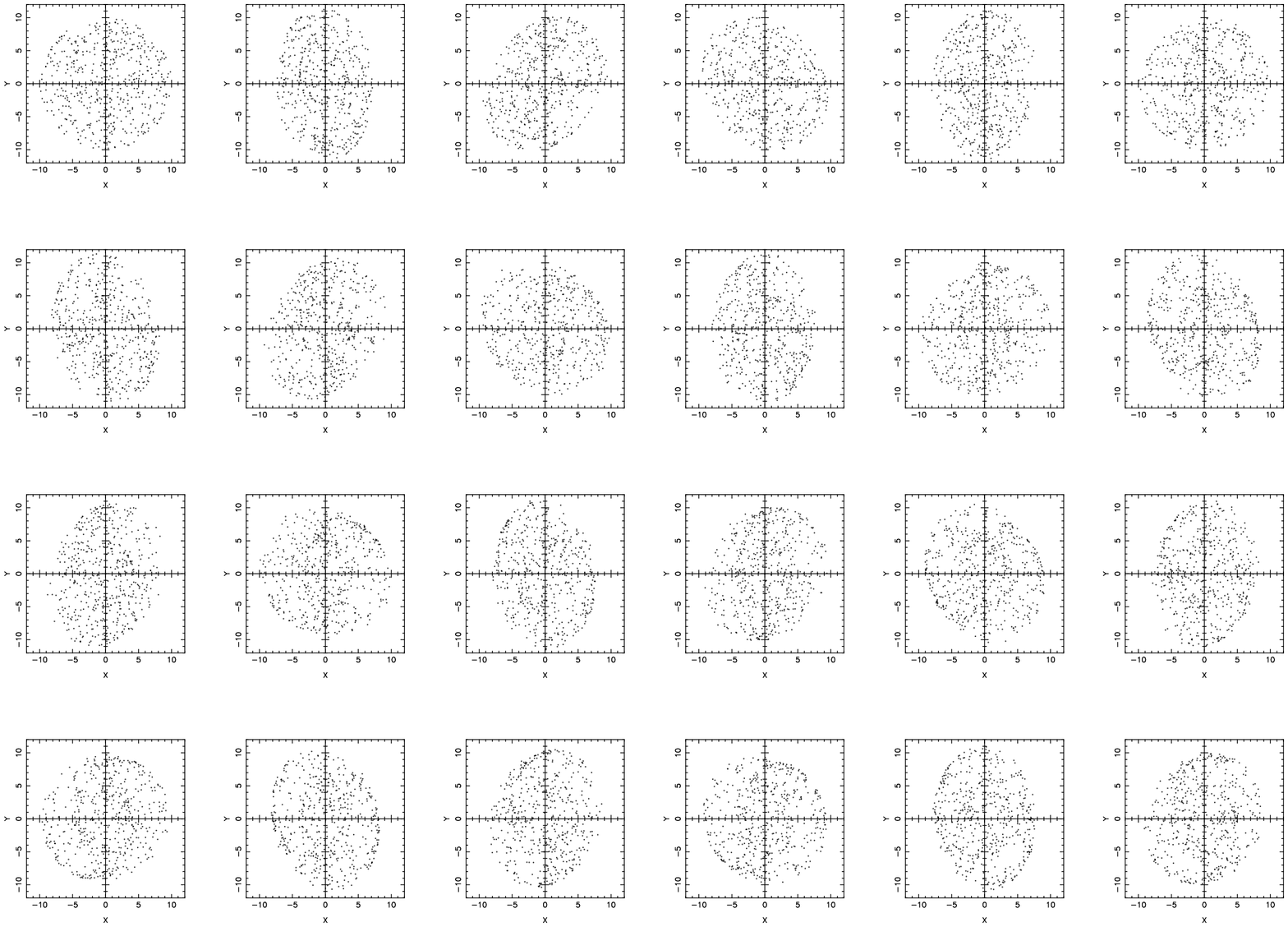,width=13.9cm,height=10.0cm,angle=-90}
\caption{\label{stick2}
Spatial evolution of sticky particle system of Fig.~\protect\ref{stick1} 
when a central mass is added. 
Snapshots are shown at intervals of 500 Myr 
(in sequence from left to right and from top down) }
\end{figure}

The particles are placed at random in the inner 10 kpc of a given model.
A hundred of them have $z=0$ and the rest are in groups of a hundred with
$z=-0.2,-0.1,0.1,0.2$. All particles are given the local rotation velocities in the
plane. 
Fig.~\protect\ref{stick1} shows the results for the halo models in which the dissipative 
point particle runs of Figs.~\protect\ref{dispq} and \protect\ref{dispa} were integrated. Clearly,
one recovers the same conclusions as before: trajectories inside the 
core radius of the halo quickly dissipate and spiral down towards the centre, while
those in the outer areas find stable loop orbits to oscillate around and are relatively
long lived.
Fig.~\protect\ref{stick2}  shows the case when a central mass $GM_{C}=0.01$ is present.
Again, just like in the case of single dissipative point particles, no dissipation towards
the centre takes place~(cf. Fig.~\protect\ref{dispc}).

Modelled in a self consistent way (with the gravitational field of the gas taken
into account), this process will lead to a {\em self regulating mechanism} for
gas accession to the central regions of a galaxy.
In a general situation the particles should also be assigned different sizes, masses and
 ``internal energies'' which they can exchange with  each other
during collisions. This might enable one to model ``physical'' processes
such as fragmentation of gas clouds, coalescence, star formation
and energy feedback into the interstellar medium, which are now believed to have an 
important effect on the dynamics and which should be included in any detailed modelling.
However, it is clear that to a first approximation, the main effect of a triaxial
potential with a harmonic core is to provide {\em an efficient self regulating
 mechanism for accretion of
gaseous material within that core to the central region}. This is not only important for
the formation of central masses which are so crucial  in determining  the dynamics, 
but perhaps to other phenomena like the fueling of active galactic nuclei.
It provides an alternative mechanism whereby non-axisymmetric perturbations
can lead to gas  accretion into the inner areas of galaxies, now that the correlation
between the presence of galactic  bars and of AGN in galaxies
is thought to be weak (e.g. Mulchaey \& Regan  1997). The fact the accretion is takes place over a Gyr
and is self regulating makes it compatible  with the lifetime of AGN and with the
observation of Merritt \& Quinlan (1997) that observationally detected black holes 
 have smaller masses than the critical mass required to destroy axisymmetry
--- the more obvious  mechanism for self regulation.

\section{Conclusions and possible consequences}
\label{tax:concpos}

\subsection{Summary of results}

Triaxial structures are expected from the instability of gravitational
collapse to non-spherical perturbations and the ineffectiveness of 
violent (or collisionless) relaxation in washing out completely the 
initial state.
Triaxial dark matter halos  are predicted by hierarchical
theories of galaxy formation  and although there is some doubt if they
could survive dissipational collapse, it seems that this would bring about 
at most a rearranging of shape but not complete loss of triaxiality.
Here we have mainly concentrated on models with  constant density core halos.
These are favoured by observations but are not predicted by cosmological
simulations which nevertheless are in agreement with other observed features
(Section~\protect\ref{tax:halo}).

We have studied the orbital structure near the disk plane
of disk galaxy models with triaxial halos.  This was done by examining the
vertical stability and calculating the maximal Liapunov exponent of 
general orbits and studying the stability of periodic orbits that parent them.
These effects were, as could be expected, in general correlated. 
There are three factors which mainly determine the orbital structure of our 
models: the non-linearity of the force field, the asymmetry in the density
distribution (Section~\protect\ref{tax:staper}) and the presence of a rotating perturbation 
(Section~\protect\ref{tax:rapbars}). 
The nonlinearity is determined by the central concentration
of the density distribution.
In models with core radii of $R_{0}=2$ kpc significant instability is found around
the 1:2 resonance of the $x$-axial orbit, which is moved inwards when the asymmetry is greatest. For
models with very small core radius, widespread instability occurs even in the axisymmetric
case. For models with larger core radius or which were disk dominated in the central
areas, the major low order resonances were too far out to be significant even when
significant asymmetry is present. 

 One way of enhancing nonlinearity is by adding a central mass concentration.   
Then, as is usually the case (e.g., Gerhard \& Binney 1985),
almost all box orbits are destabilised 
and replaced by either boxlets or chaotic orbits.  
The effect of the central 
mass can be understood in terms of the stability properties of the closed periodic orbits. 
Although the lower order resonances (e.g., the 1:2 resonance)
appear at somewhat smaller radii when a central mass is present, the main effect 
causing the instability seems
to be the creation and broadening of higher  order resonances.
When
the central mass is strong enough, the instability gaps on the $x$ axis orbit merge and,
even for  relatively small central masses of about $10^{-5}$
the total galaxy mass at 20 kpc 
the axis orbits (which parent the boxes) are
destabilized and are replaced by higher order (KAM islands known as)
``boxlets''  (Section~\protect\ref{tax:staper}). Increasing the 
central mass destabilizes these and the remaining stable orbits are mainly those
of still higher order in the self replicating phase space hierarchy (e.g., LL).
This leads to widespread chaotic behaviour
 in the region of the phase space once occupied by the box orbits.
In addition, it was found that the destabilisation of the box 
orbits takes place out to many core radii. 
 Therefore high energy stars born on eccentric orbits may be also be unstable.
For central masses of up to $0.05 \%$ the total galaxy mass at 20~kpc many
trajectories in the region once occupied by the box orbits are either regular
or have small Liapunov exponents (and therefore also small diffusion rates)
and thus any resulting  evolutionary effects are expected to be slow.
For central  masses of $\sim 1\%$ the galaxy mass (at 20 kpc) most orbits
in the aforementioned region are strongly chaotic and quickly relax to 
an invariant distribution
(Section~\protect\ref{dishac}).

Not all unstable periodic orbits are ``centrophylic'' boxlets. In fact many
higher order looplets (in particular the 2:2 loops) which have a definite sense 
of rotation and do not pass near the centre, may be unstable in the presence
of a central mass concentration. As might be expected, the fraction of looplet 
orbits is greater in more axisymmetric potentials, and, in general, the fraction
of vertically unstable orbits was found to be larger in flatter potentials.
Thus, the presence of a disk may, in some situations, increase the fraction of 
unstable general orbits. For although there are
fewer box orbits to start with, many orbits with a definite sense of rotation
(which are parented by higher order looplets)
 are unstable in the central areas and fewer 
orbits are parented by stable boxlets.

Even for highly asymmetric systems with strong central mass concentrations, 
and in the region of phase space once occupied by the box orbits, there remains 
a significant fraction of stable periodic orbits (albeit parenting  a small fraction
of general orbits).  This will mean that even though
there is widespread chaotic behaviour, trajectories in non-integrable but smooth galaxy
potentials do not approximate strongly chaotic uniformly hyperbolic systems.
 Such   systems therefore are not structurally stable and thus 
their qualitative behaviour can be affected by small perturbations (Section~\protect\ref{tax:noise}). 
This was indeed found to be the case. It was found that perturbations that take many Hubble 
times to change the energy can have a significant effect on the orbital structure. 
For smaller perturbations, the effect was more or less equally likely to {\em stabilize}
the trajectories as to destabilize them. For stronger perturbations however the latter
effect was much more predominant as these destroyed the intricate island structures.
The fact that weak discreteness noise can have such an effect on the dynamics
is important for both the estimation of the importance of discreteness effects in
real systems and the faithfulness to which $N$-body simulations can reproduce 
the dynamics of galaxies. In particular, the triaxiality of the halos produced 
by $N$-body simulation should be taken as a conservative  estimate since these
simulations employ far fewer particles than present in halos mainly composed
of elementary particles.

In the absence of rapidly rotating bar perturbations, the nearly circular closed 
loops are stable. This is  due to the absence of resonances, which arises
from the fact that the angular frequency and the $R$ and $z$ oscillation 
frequencies have  nearly constant ratios in the type of models chosen 
(this is  analogous to the situation  when   response is linear, in which case
 {\em each} of the 
frequencies is constant).
The addition of a rotating  bar  breaks this symmetry and therefore 
markedly increases the fraction of chaotic orbits present --- since now widespread
chaotic behaviour can occur in {\em both} regions once occupied by the main
orbit families (boxes and loops).
The presence of the rotating perturbation within the triaxial halo potential
means that the potential is time dependent and that no transformation of
coordinates can change that. This will lead to very complicated behaviour near the 
disk plane, with trajectories suffering large angle scattering. The result is that
trajectories are much more irregular in the disk plane where the time dependence
is greatest. These trajectories are sometimes not  $z$ unstable even though they 
may have a very large maximal Liapunov exponent. Thus some additional ``invariant''
may be at least approximately conserved. In this case therefore the maximal exponent
does not provide complete information on the phase space transport properties.
This was especially true if the bar has a rectangular  shape.

 Central masses can be produced if small dissipative perturbations are given to the loop orbits 
in the inner areas near the halo core radius (where the 1:1 resonance
occurs). Trajectories are then observed to spiral inwards in a few
rotation times, even though the magnitude of the dissipation would require 
many Hubble times for this effect to be noticeable 
in the outer areas of the potential. This behaviour is easy to understand
since closed loop orbits (which parent the general loop orbits) become more 
and more eccentric as one moves nearer to the center of the potential and do not
exist at all deep inside the core, at the point where this happens (the ``separatrix'') 
dissipation rates are greatly enhanced. Beyond this, the trajectories do not find any
closed loop orbits to oscillate around and quickly spiral to the centre. However 
once a strong central mass has formed, closed loop orbits are stabilised in the central 
area and dissipative trajectories may settle around them in stable limit cycles instead
of spiraling to the centre. The build up of the central mass thus saturates. 
This provides a self regulating mechanism for the formation of central masses in galaxies,
and may explain why black holes are observationally found to have masses smaller than
those required to destroy the triaxial equilibrium (Merritt \& Quinlan 1997).
Since a central mass of $0.05 \%$ of the total galactic mass at 20 kpc
appears to be enough to essentially stop further gas inflow. And thus from the discussion 
earlier in this section one expects the evolution caused by the central mass to be slow. 
The time-scale for the formation of the central mass is about 1 Gyr.

In the presence
of time-dependent forcing (that is a rotating bar), dissipative  trajectories were found to
settle into long lived chaotic states where their radial coordinates oscillated around definite
values. Such strange attractor states may explain some of the observed ring structure in
galaxies (Gu et al. 1996)  --- provided of course that a bar can survive long enough in
a triaxial halo potential. Trajectories on some of these attractors may
also have large $z$ excursions and large radial motion through the galactic disk.

\subsection{Possible consequences}

The effects described above may have the following  consequences. First there will 
be a redistribution of the gas accompanied by the fuelling of a central mass. 
In general therefore, gas  motion in nonrotating  non-axisymmetric
potentials may be linked to the fuelling of central black holes or trigger bursts of star
formation as the gas moving on self crossing trajectories is shocked and dissipates
towards the centre as in the case of barred potentials 
(Beckman et al. 1991; Pfenniger 1993). This provides an alternative mechanism
now that the correlation between the presence of galactic bars and of AGN's is thought
to be weak (Mulchaney \& Regan  1997).

In the absence of a central mass, and within the region delimited by the harmonic core 
(approximately equal to $R_{0}$ if the halo is dominant but 
decreases with increasing symmetry),
only box orbits can exist and almost all of these
are unstable to perturbations out of the disk plane in the presence of a central mass
concentration. 
Therefore, initially any stars born  
 from a settling disk in this region will become unstable when a central mass forms. 
The vertical instability will lead to the formation of bulge like structures 
even if this  system of stars, which is dynamically hot, is initially confined near the
disk plane. As the central mass builds up, the harmonic nature of the potential 
is destroyed in its vicinity and  
closed loop orbits are stabilised at progressively larger radii within the core.
Eventually, when the central mass is large enough, the harmonic core is 
completely destroyed.
At that point the gas accretion also stops, since the gas can now 
settle on the newly created closed loops {\em anywhere} inside the core 
region (Fig.~\protect\ref{dispc}).
One then expects the end result of this process to be the formation
of a bulge-like structure, the extent of which is similar to that of the
original harmonic core and is formed from layers of unstable trajectories 
forming from inside out (see the discussion at the end of Section~ref{tax:dispdisk}).

It then follows that haloes with larger cores will be expected to develop centrally
condensed ``bulgy'' disks. Once fully evolved
 these will then be expected to dominate the rotation
curve in the inner areas. On the other hand, for halos with smaller core radii,
this effect will be less significant and the halo would dominate the rotation curve at all
radii. This suggests an explanation as to why in galaxies with widely different
disk-halo contribution rotation curves one still gets these to be approximately
(but not exactly:  Persic et al. 1996) flat over most of the detected radius of a galaxy. 
If this is the case then it will have to be that more massive haloes have larger
core radii, since it is observed that galaxies in which are disk dominated near the 
centre have larger terminal rotation velocities.  Since the effective harmonic
core radius in the disk plane for a given $R_{0}$ is also a function of the asymmetry in that
plane, one may also suggest that larger mass haloes may have a larger asymmetry.

We have seen that, in the presence of bars rotating in triaxial halos,
almost all orbits become chaotic. In addition, these orbits do not have 
a definite sense of rotation. It is not clear if such a bar  could form in the first 
place because
of the large random motion (due to the triaxiality). However,
if it is possible to still have a bar unstable disk inside a triaxial halo, it is 
likely to evolve a random bulge like structure extremely quickly as the bar is destroyed.
It may be interesting to note here that  observations from the Hubble deep field survey 
(van den Bergh et al. 1996)
 suggest that barred 
galaxies are rare at high redshifts. 
It is also found that the fraction of late type galaxies is much lower than in standard catalogues.
 This may be expected if disks
form inside strongly triaxial haloes where star formation may be caused by gas
particles moving in the central regions on self-intersecting trajectories.
In the outer regions, gas particles
may move around (non-self-intersecting) closed loop orbits  so that  
the star formation rate is much smaller. In galaxies with small halo core radii 
(which according to our scenario would end up as late type galaxies), the more violent effects are 
limited to a small central region and the evolution will generally proceed at a much slower rate ---
which means less of them are observed at high redshift.

Evidently, as the above processes unfold, the halo gradually loses triaxiality
(because of the destruction of the box orbits which are crucial for maintaining the triaxial
structure, and the settling of stars onto loop orbits which are alligned perpendicular
to the halo). 
This also can bring an end to the gas inflow and the scattering of stars into a bulge.
This process is not expected to be very fast however, since for a weak
central mass there is usually a large fraction of regular and 
weakly chaotic (ie., with small Liapunov exponents and phase space diffusion 
diffusion rates) orbits remaining
in the region once occupied by the box orbits. The threshold central mass
needed for the majority of trajectories in that region to be highly
chaotic (i.e., to have an exponentiation time-scale of a dynamical time
or so) appears to be much larger than the mass needed to stop the accretion.
Thus, such central concentrations would probably never be reached because
of the self regulating nature of the accretion mechanism, which may explain why 
central concentrations as high as those apparently required for rapid evolution
towards axisymmetry in elliptical galaxies are rarely observed (Merritt \& Quinlan 1997).
Therefore,
one may expect there to be a period where the effects described above act on the
stellar trajectories before the halo becomes completely axisymmetric (at which point
it is then possible to develop a barred disk).

Although no precise determination of the time-scales of the processes described
in this section can be obtained without full $N$-body simulation, including 
hydrodynamic effects, one
can, to first order, perhaps give the following order of magnitude
estimates. Since the time-scale over which originally planar trajectories 1
are destabilised corresponds to a few exponentiation time-scales,
we expect that, after a significant central mass has formed (which takes in
$\sim 500$ Myr), a significant fraction of the orbits within the 
inner few kpc will be destabilised over a few dynamical times (less than a
Gyr). In the outer areas this will correspond to a few Gyr. Complete loss 
of triaxiality and the evolution towards a final state should take about
a Hubble time.

\section*{Acknowledgments}
Thanks go to Martin Hendry for reading the manuscript. 
The routine calculating the forces due to order 2 Ferrers bars 
was kindly made available by Daniel Pfenniger. This work has benefited
from many discussions with the late Professor R. J. Tayler.

\appendix
\section{Equidensity contours}

\begin{figure}
\begin{flushleft}

\epsfig{file=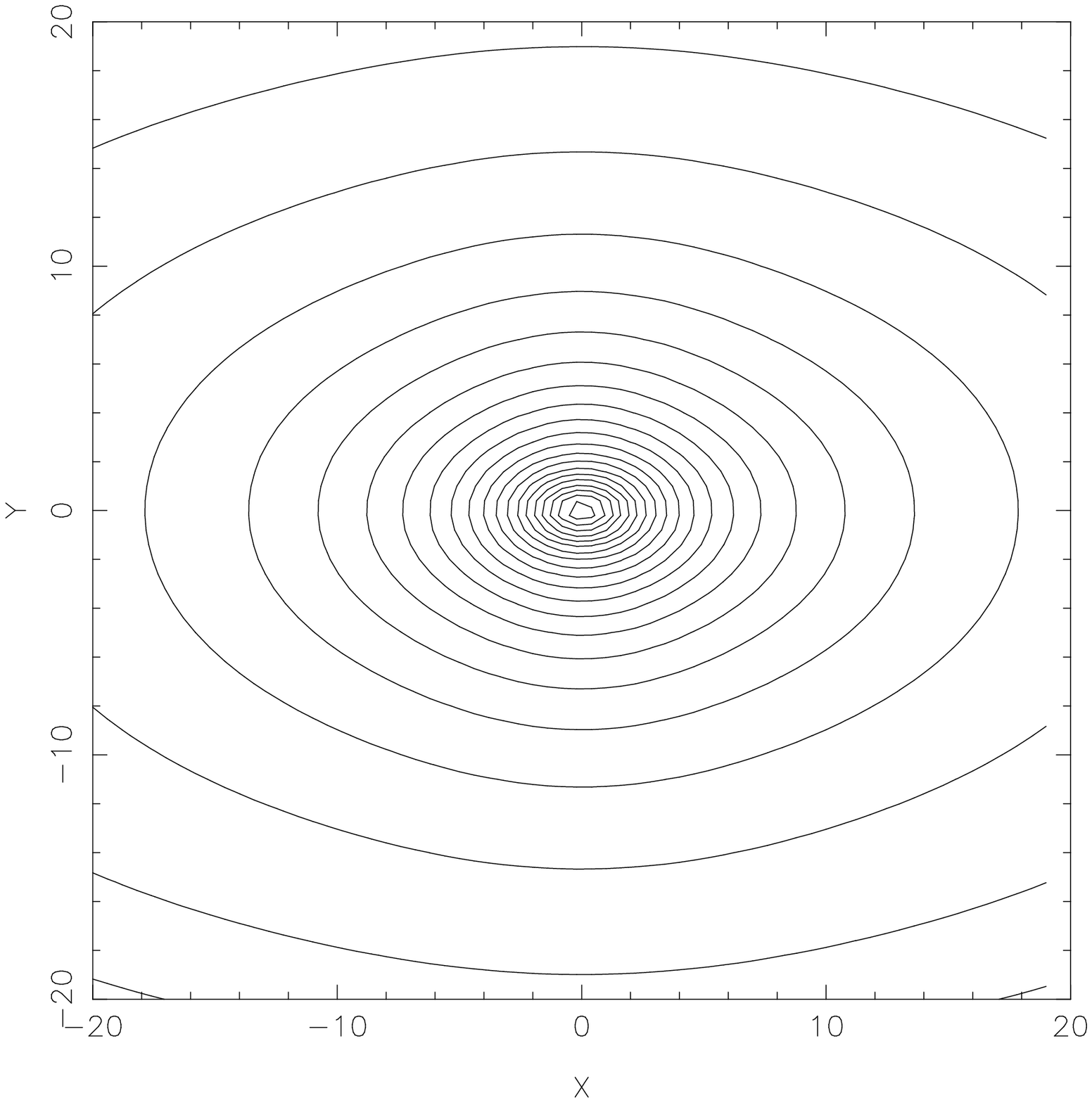,width=6.5cm,height=6.5cm,angle=-90}
\hspace{0.5cm}
\epsfig{file=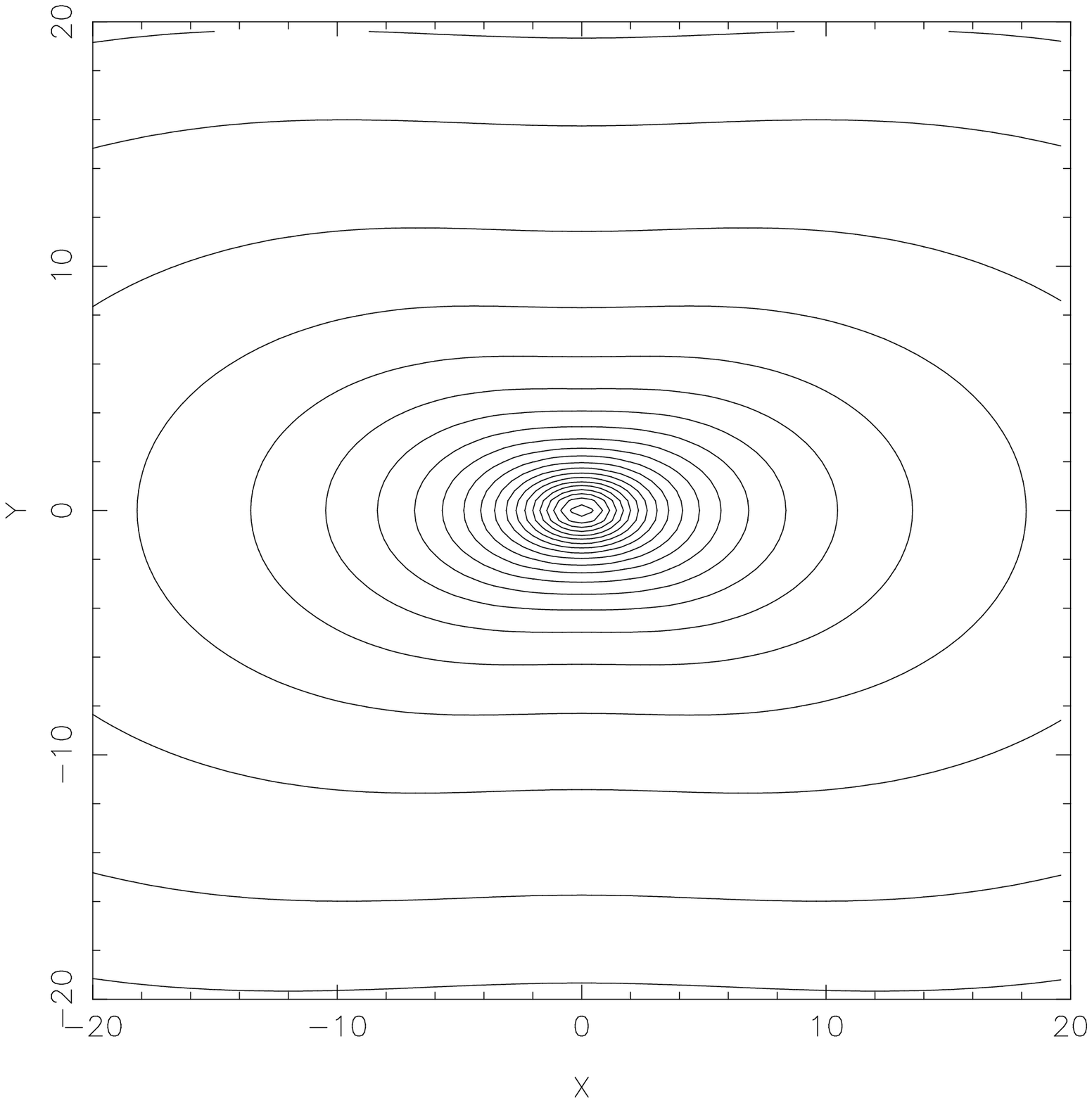,width=6.5cm,height=6.5cm,angle=-90}

\end{flushleft}
\caption{\label{EQMOD1D764}
Equidensity contours in $x-y$ of Model 1 with prolate halo ($T=0.85$) and
$c/a=0.7$ (left) and when the disk contribution is omitted}
\end{figure}

It is well known (e.g., BT Fig. 2-8) 
that the logarithmic potential can have wild dents in its density
distribution when the axis ratio in one plane is not very near unity. 
In addition, 
the density distribution can even be negative. This happens because the potential
was chosen for convenience and does not arise from any presupposed realistic density distribution (but actually the reverse is done --- the density distribution 
arises from the potential: This point is discussed in  more detail by Pfenniger 1984b
who also gives references to the relevant mathematical literature). 
Nevertheless, the situation is less serious for the three dimensional triaxial
potential (as opposed to the flattened axisymmetric one). With two parameters to
adjust one can obtain fairly realistic and well behaved density distributions. 
For example, the diagram on the left hand side of
Fig.~\protect\ref{EQMOD1D764}  shows the equidensity contours for the same model parameters 
as in the
aforementioned BT figure but when $c/a=0.64$. Clearly, although some denting remains, the density distribution exhibits a much better behaviour. Moreover, it was checked that
it is always positive. These two properties are connected and it is a general
result that triaxial logarithmic haloes for which the axis ratios $b/a$ and
$c/a$ are not too different have fairly well behaved density distribution even
for significantly asymmetric systems. To see how this happens, we use
apply the Poisson's equation  to the potential in Eq.~(\protect\ref{tax:log})
and obtain the relation
\begin{equation}
(p+q-1) x^{2} + p(1+q-p) y^{2} + q(1+p-q) z^{2} =\frac{ 4 \pi G \rho}
{v^{2}_{0}} R^{2}_{e}
- R^{2}_{0} P
\label{tax:dentd}
\end{equation}
where $R^{2}_{e}=R^{2}_{0} + x^{2} + p y^{2} + q z^{2}$ and $P=1+p+q$.

\begin{figure}
\begin{flushleft}




\end{flushleft}

\caption{\label{M3O8}
 Three dimensional equidensity contours of some of the systems studied.
Top left: Model 3 with oblate halo and $c/a=0.8$. Top right: Model 2 with 
oblate halo and $c/a=0.8$. Bottom left: Model 3 with prolate halo and $c/a=0.8$.
Bottom right: Model 1 with prolate halo and $c/a=0.7$. 
THIS FIGURE WAS TOO LARGE TO BE INCLUDED}
\end{figure}

If the right hand side and the terms in brackets on the left hand side
are positive then this equation  can be written in the form 
of an equation of an ellipsoid with {\em variable axis ratio} 
\begin{equation}
\frac{x^{2}}{A^{2}} + \frac{y^{2}}{B^{2}} +\frac{z^{2}}{C^{2}} = 1
\end{equation}
with
\begin{eqnarray}
A^{2}=\frac{ \frac{ 4 \pi G \rho_{e}}{v^{2}_{0}} R^{2}_{0} +R^{2}_{e}- R^{2}_{0} P
}{p+q-1}
\label{tax:axden1}\\
B^{2}=\frac{ \frac{ 4 \pi G \rho_{e}}{v^{2}_{0}} R^{2}_{0} +R^{2}_{e}- R^{2}_{0} P
}{p(1+q-p)}
\label{tax:axden2}\\
C^{2}=\frac{ \frac{ 4 \pi G \rho_{e}}{v^{2}_{0}} R^{2}_{0} +R^{2}_{e}- R^{2}_{0} P
}{q(1+p-q)},
\label{tax:axden3}
\end{eqnarray}
where $\rho_{e}$ is the value of the density along an equipotential.
It is easy to see that
\begin{equation}
(p+q-1) x^{2} + p(1+q-p) y^{2} + q P z^{2}>0
\end{equation}
is the condition for the right hand side of~(\protect\ref{tax:dentd}) to be positive and that
\begin{equation}
R^{2}_{0} P +(p+q-1) x^{2} + p (1+q-p) y^{2} + q(1+p-q) z^{2}>0
\end{equation}
is the condition for the density to be positive. Therefore  the conditions
for the density to be well behaved --- i.e. not being constant on hyperbolic concave isocontours and being positive everywhere --- is that
\begin{eqnarray*}
1+p+q>0\\
p+q-1>0\\
p(1+q-p)>0\\
q(1+p-q)>0\\
\end{eqnarray*}
The first two of these conditions are always satisfied while the other two imply that
\begin{equation}
\parallel p-q \parallel <1.
\label{conax}
\end{equation}
This means that one can have a fairly realistic but very flat density distribution
using the logarithmic potential as long as the asymmetry is not too different
in the principal coordinate planes. However it is clear 
from~(\protect\ref{tax:axden1}),~(\protect\ref{tax:axden2}) and~(\protect\ref{tax:axden3}) that  the equal
value surfaces of the right hand side of~(\protect\ref{tax:dentd}) and the equidensity surfaces
have axis ratios that differ by factors of 
\begin{equation}
\sqrt{ \frac{p(1+q-p)}{1+q-1} }
\end{equation}
for $(B/A)$ and 
\begin{equation}
\sqrt{ \frac{q(1+p-q)}{1+q-1} }
\end{equation}
for $(C/A)$. Therefore for $p$ and $q$ differing significantly from one,
the density distribution may be severely dented even when the condition 
(\protect\ref{conax}) is satisfied. This is because the axis ratio will vary
significantly along the equidensities. Nevertheless, the equidensities will
not be as severely distorted as to be locally hyperbolic 
(and therefore open) and the density will remain positive.

From the above discussion  we expect that the density contours
would be better behaved if the potential axis ratios in the plane and normal
to it are not too different.
For example, the 
halo equidensity contours even for such a rather extreme potential axis ratio as $b/a=0.7$
are found to have reasonable shape if $c/a \sim 0.64$. 
This can be seen from Fig.~\protect\ref{EQMOD1D764} where have plotted
the equidensity contours in the $x-y$ plane of Model 1 with halo 
potential axis ratio $b/a=0.7$ with and without the disk contribution (the latter case
should be compared to Fig. 2-8 of BT where the contours are potted for the case $c/a=1$,
in which case they are found to be extremely dented and the density can be negative).

We have also checked the shapes of the three dimensional density distributions
by producing some three dimensional contours of this quantity. 
The relevant contours for several 
Model parameters are shown in Fig.~\protect\ref{M3O8}


\begin{thebibliography}{ANOSOV 99}

\bibitem{}
Aarseth S.J., Binney J.J., 1978, MNRAS 185,22
\bibitem{}
Arnold V.I., 1987, In
Mackay R.S., Weiss J.D. (eds) 
 Hamiltonian dynamical systems. J.W. Arrowsmith ltd., Bristol
\bibitem{}
Aubry S., 1995, Physica D86,284
\bibitem{}
Barbanis B., Contopoulos G., 1995, A\&A 294,33
\bibitem{}
Beckman J.E., Varela A.M., Munoz-Tunon C., Vilchez J.M., Cepa J., 1991, A\&A 245, 436
\bibitem{}
Benettin G., Galgani L., Strelcyn J.M., 1976, Phys. Rev. A14, 2338 
\bibitem{}
Binney J.J. \& Tremaine S., 1987,
Galactic dynamics. Princeton Univ. Press, Princeton (BT)
\bibitem{}
Blitz L., Spergel D.N., 1991, ApJ 370,205
\bibitem{}
Botteno R., 1996,  A\&A 306,345
\bibitem{}
Burkert A., 1997, To appear in: Klapdor-Kleingrothaus H.V, Ramachers Y. (eds)
Aspects of Dark Matter in Astro-and Particle Physics (astro-ph/9703057) 
\bibitem{}
Clifford P.J., 1983, The role of Neutral Hydrogen clouds in the interstellar medium.
PhD thesis, Univ. of Sussex
\bibitem{}
Cole S., Lacey C., 1996, MNRAS 281,716
\bibitem{}
Combes F., 1991, ARAA 29,195
\bibitem{}
Combes F., Gerin  M., A\&A 150,327
\bibitem{}
Contopoulos G., 1985, In: Buchler J.R., Perdang J.M., Spiegel E.A. (eds.) Chaos
in Astrophysics. Reidel Publishing Company, Dordrecht
\bibitem{}
Debattista V.P., Sellwood J.A, 1996, To appear in the proceedings of the Dark
Matter 1996 conference, Sesto, Italy. Rutgers Astrophysics Preprint Series 194
(astro-ph/9610009)
\bibitem{}
Dubinsky J., 1994, ApJ 431,617
\bibitem{}
Dubinsky J.,  Carlberg R., 1991, ApJ 378,496
\bibitem{}
Eckmann J.P., Ruelle D., 1985, Rev. Mod. Phys. 57, 617
\bibitem{}
El-Zant A.A., 1997, A \& A in press
\bibitem{}
Espagnol P., 1995, Phys. Rev. E52,1734
\bibitem{}
Evans N.W., Collett J.L.,1994, ApJ 420,L70
\bibitem{}
Flores R.A., Primack J.R., Blumenthal G.R., Faber S.M., 1993, ApJ 412,443
\bibitem{}
Flores R.A., Primack J.R., 1994, ApJ 427,L1
\bibitem{}
Franx M., de Zeeuw T., 1992, ApJ 392,L47
\bibitem{}
Franx M., Illingworth G., de Zeeuw T., 1991, ApJ 383,112
\bibitem{}
Freeman K.C., 1993, In: Thuan T.X., Balkowski C.,
Trân Thanh Vân J. (eds) Physics of nearby Galaxies: nature or Nurture?  
Editions Frontières, Paris  
\bibitem{}
Friedli D., Benz  W., 1993, A\&A 268, 65
\bibitem{}
Gerhard O.E., Binney J., 1985, MNRAS 216,467
\bibitem{}
Goodman J., Scchwarzschild M., 1981, ApJ 245,1087
\bibitem{}
Gu Q.S., Liao X.H., Huang J.H. Qu, Q.Y., Su H., 1996, A\&A, in press
\bibitem{}
Gutzwiller M.C., 1990, Chaos in Classical and Quantum Mechanics.
Sringer Verlag, New York
\bibitem{}
Hasan H., Pfenniger D., Norman C.A., 1993, ApJ 409,91
\bibitem{}
H\'enon M., 1973, A\&A 28,415
\bibitem{}
Hofner P., Sparke L.S., 1994, ApJ 428,466 
\bibitem{}
Hoogerbrugge P.J., Koelman J.M.V.A., 1992, Europhys. Lett. 19,155
\bibitem{}
Kandrup H.E., 1994, invited talk at seventh Marcel Grossman meeting July 1994 
(astro-ph/9410091)
\bibitem{}
Katz N., Gunn J.E., 1991, ApJ 377,365
\bibitem{}
Klafter J., Shlesinger M.F., Zumofen G., 1996, 
Phyiscs Today Vol. 49 No. 2 P. 33 
\bibitem{}
Klimontovich Yu. L., 1994, Phys. Uspekhi 37,737
\bibitem{}
Kuijken K., Tremaine S., 1991, In: Sundelius B. (ed) Dynamics of disk galaxies.
G\"oteborg Univ. Press., G\"oteborg
\bibitem{}
Kuijken K., Tremaine S., 1994, ApJ 421,178
\bibitem{}
Kuiken K., Fisher D., Merrifield M.R., 1996, MNRAS 283,543
\bibitem{}
Lewis J.R., Freeman KC, 1989,  AsJ, 97,139 
\bibitem{}
Lichtenberg A.J., Lieberman M.A., 1983, Regular and stochastic motion.
Springer, New York (LL)
\bibitem{}
Lin C.C., Mestel L., Shu F.F., 1965, ApJ 142,1431
\bibitem{}
Macmillan M.D., 1958, The theory of the potential. Dover, New York
\bibitem{}
Merritt D., 1996,
 In:
Chaos in gravitational $N$-body systems, Muzzio J.C. (ed). Kluwer
\bibitem{}
Merritt D., Fridman T., 1996, ApJ 440,136
\bibitem{}
Merritt D., Quinlan G., 1997, ApJ Submitted, astro-ph/9709106 
\bibitem{}
Merritt D., Valluri M., 1996, ApJ 471,82
\bibitem{}
Miyamoto M., Nagai R., 1975, PASJ 27,533
\bibitem{}
Mulchaey J.S., Regan M.W., 1997, ApJ Letters in press (astro-ph/9704094) 
\bibitem{}
Navarro J.F., 1996, astro-ph/9610188
\bibitem{}
Norman C.A., Ikeuchi S., 1989, ApJ 372,383
\bibitem{}
Norman C.A., May A., van Albada T.S., 1985, ApJ 296,20
\bibitem{}
Norman C.A., Sellwood J.A., Hasan H., 1996, ApJ 462,114
\bibitem{}
Parker T., Chua L.O., 1989, Practical Numerical Algorithms for Chaotic Systems.
Springer Verlag, New York  
\bibitem{}
Persic P., Salucci P., Stel F., 1996, MNRAS, in press
\bibitem{}
Pfenniger D., 1984a, A\&A 134,384
\bibitem{}
Pfenniger D., 1984b, A\&A, 141,171
\bibitem{}
Pfenniger D., 1986, A\&A 165,74
\bibitem{}
Pfenniger 1993, In: Thuan T.X., Balkowski C.,
Trân Thanh Vân J. (eds) Physics of nearby Galaxies: nature or Nurture?  
Editions Frontières, Paris 
\bibitem{}
Pfenniger D., Combes F., 1994, A\&A 285,94
\bibitem{}
Pfenniger D., Norman C.A., 1990, ApJ 363,391
\bibitem{}
Pfenniger D., Friedli D., 1991, A\&A 252,75
\bibitem{}
Powell M.J.D., 1968, Atomic Energy Research Establishment Reports: 5947
\bibitem{}
Quinn T., 1991, Particle simulations of polar rings. In: Warped disks and 
inclined rings around galaxies. Cambridge University Press, Cambridge
\bibitem{}
Rix H.-W., 1995, nvited Review at the IAU Symposium 169, "Unsolved Problems of the Milky Way",
The Hague, August, 1994 (astro-ph/9501068)
\bibitem{}
Rix H.-W., Zaritsky D., 1995, ApJ 447,82
\bibitem{}
Rix H.-W., Franx M., Fisher D., Illingworth G., 1992, ApJ L5
\bibitem{}
Roberts M.S., Haynes M.P., 1994, ARAA 32,115
\bibitem{}
Rubin V.C., Graham J.A., Kenney J.D.P., 1992, ApJ 394,29
\bibitem{}
Sackett P., Sparke L., 1990, ApJ 361,408
\bibitem{}
Sackett P., Rix H.W., Jarvis B., Freeman K., 1994 ApJ 436,629
\bibitem{}
Scalo J.M., Struck-Marcell C., 1984, ApJ 276,60
\bibitem{}
Schmidt M., 1985, IAU Symposia 106,75
\bibitem{}
Schwarzschild M., 1993, ApJ 409,563
\bibitem{}
Sellwood J.A., Wilkinson A., 1993, Rep. on Prog. in Phys. 56,173
\bibitem{}
Shlesinger M.F., Zaslavsky G.M., Klafter J., 1993, Nature 363,31
\bibitem{}
Spitzer L.,  1978, Physical processes in the interstellar medium. Wiley, New York
\bibitem{}
Tully R.B., Verheijen M.A.W., 1997, Contibution to ESO workshop November 1996
(astro-ph/9702221) 
\bibitem{}
Udry S., Pfenniger D., 1988, A\&A 198,135
\bibitem{} 
van den Bergh S., Abraham R.G., Ellis R.S. et al., 1996, AsJ 112,359
\bibitem{}
van der Kruit P.C., 1986, A\&A 157,230
\bibitem{}
van der Kruit P.C., 1989, Kinematics and Mass Distributions in Galaxies.
 In: Gilmore G., King. I.R., van der Kruit P., Buser R. (eds.), 1989, 
The Milky Way as a galaxy. Geneva Observatory, Sauverny
\bibitem{}
Verdes-Montenegro L., Bosma A., Athanassoula, 1996, A\&A 306,1011
\bibitem{}
Warren M.S., Quinn P.J., Salmon J.K., Zurek W.H., 1992, ApJ 399,405
\bibitem{}
Zaslavsky G.M., 1991, Sagdeev R.Z., Chernikov A.A., Usikov D.A.,
Weak chaos and quasi-regular patterns. Cambridge University Press, Cambridge

\end{thebibliography}
\end{document}